 \ifx\compilefullpaper\undefined  
 \documentclass[pra,longbibliography,twocolumn,showpacs,nofootinbib,
                superscriptaddress,notitlepage]{revtex4-2}
 
 \pdfoutput=1 
 
 \usepackage{mathtools}
 \usepackage{amssymb}
 \usepackage{bm}
 \usepackage{amsthm}
 \usepackage{mathrsfs}
 \usepackage{stmaryrd}
 \usepackage{latexsym}
 
 \usepackage[table]{xcolor}   
 \renewcommand{\arraystretch}{1.2}
 \usepackage{booktabs}
 \usepackage{array,makecell}
 \usepackage{boldline}
 
 \usepackage[utf8]{inputenc}  
 \usepackage[protrusion=true,expansion=true]{microtype}
 \usepackage{ragged2e}
 \usepackage{setspace}
 \usepackage{lipsum}
 \usepackage{ulem}
 
 \usepackage{graphicx}
 \usepackage{float}
 \usepackage{verbatim}
 \usepackage[caption=false]{subfig}
 
 \usepackage{enumitem} 
 
 \usepackage{dsfont}
 \usepackage{url}      
 \usepackage{keyval}
 
 \usepackage[noend]{algpseudocode} 
 \usepackage{algorithm}

 \definecolor{cardinal}{rgb}{0.827,0,0}

 \newcommand{\jan}[1]{\textcolor[RGB]{48,212,16}{#1}}
 
 \usepackage[colorlinks=true,hyperindex,breaklinks,
             linkcolor=blue,urlcolor=blue,citecolor=blue]{hyperref}
 \usepackage[capitalise]{cleveref}
 
\begin{document}
 
 \title{Fast and accurate AI-based pre-decoders for surface codes}
 
\author{Christopher Chamberland}
\thanks{These authors were the main contributors to this work.}
\email{cchamberland@nvidia.com}
\affiliation{NVIDIA Corporation, USA}

\author{Jan Olle$^\ast$}
\email{jolleaguiler@nvidia.com} 
\affiliation{NVIDIA Corporation, USA}

\author{Muyuan Li$^\ast$}
\email{muyuanl@nvidia.com} 
\affiliation{NVIDIA Corporation, USA}

\author{Scott Thornton}
\affiliation{NVIDIA Corporation, USA}

\author{Igor Baratta}
\affiliation{NVIDIA Corporation, USA}

\begin{abstract}
Fast, scalable decoding architectures that operate in a block-wise parallel fashion across space and time are essential for real-time fault-tolerant quantum computing. We introduce a scalable AI-based pre-decoder for the surface code that performs local, parallel error correction with low decoding runtimes, removing the majority of physical errors before passing residual syndromes to a downstream global decoder. This modular architecture is backend-agnostic and composes with arbitrary global decoding algorithms designed for surface codes, and our implementation is completely open source. Integrated with uncorrelated PyMatching, the pipeline achieves end-to-end decoding runtimes of order $\mathcal{O}(1 \mu\text{s})$ per round at large code distances on NVIDIA GB300 GPUs while reducing logical error rates (LERs) relative to global decoding alone. In a block-wise parallel decoding scheme with access to multiple GPUs, the decoding runtime can be reduced to well below $\mathcal{O}(1 \mu\text{s})$ per round. We observe further LER improvements by training a larger model, outperforming correlated PyMatching up to distance-13. We additionally introduce a noise-learning architecture that infers decoding weights directly from experimentally accessible syndrome statistics without requiring an explicit circuit-level noise model. We show that purely data-driven graph weight estimation can nearly match uncorrelated PyMatching and exceed correlated PyMatching in certain regimes, enabling highly-optimized decoding when hardware noise models are unknown or time-varying, as well as training pre-decoders with realistic noise models. Together, these results establish a practical, modular, and high-throughput decoding framework suitable for large-distance surface-code implementations.
\par\medskip
\noindent\centering\small
\textbf{Code:} \href{https://github.com/nvidia/ising-decoding}{GitHub}
\qquad
\textbf{Models:} \href{https://huggingface.co/collections/nvidia/nvidia-ising}{Hugging Face}
\par
\end{abstract}
 
\maketitle

\section{Introduction}
\label{sec:Intro}

 \begin{figure*}
     \centering
 \subfloat[\label{fig:visualization_3dEX} ]{\includegraphics[width=.8\textwidth]{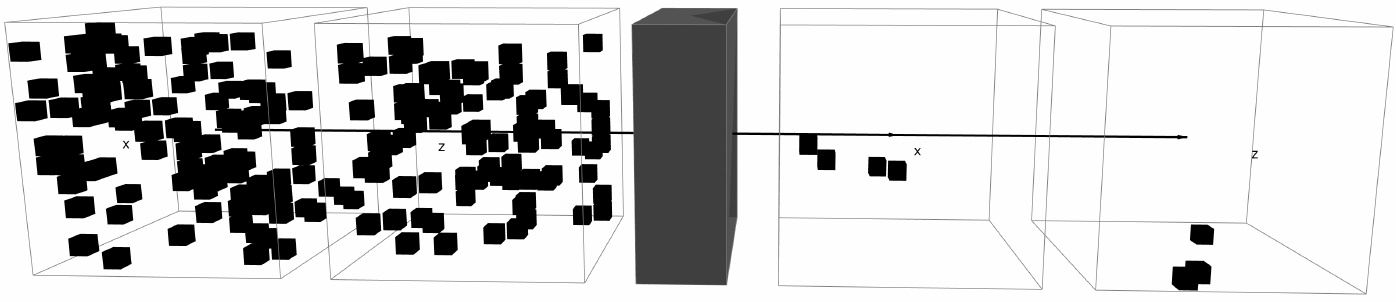}}
 \caption{ Example showing the syndrome density being reduced by the pre-decoder for both $X$-type and $Z$-type stabilizers. The residual syndromes are passed on to a global decoder to perform final corrections.  }
 \label{fig:visualization_3d}
 \end{figure*}

Quantum error correction (QEC) is a fundamental requirement for building large-scale fault-tolerant quantum computers (FTQC) \cite{Shor95,KnillLaflamme2000}. QEC decoders are classical algorithms that infer physical errors—or, equivalently, the values of logical observables—from syndrome measurement data and, in some schemes, additional information such as flag-qubit outcomes \cite{ChaoFlag1,Chamberland2018flagfaulttolerant,ChaoFlag3,Chamberland2019Magic1,chamberland2020very}. As shown in Refs.~\cite{TerhalBacklog,ChambsLocalNN22}, decoder runtimes must be sufficiently high to prevent an exponential backlog of unprocessed syndrome data during the execution of a quantum algorithm. In what follows, \textbf{runtime} will be referred to as the time taken for the decoder to process a block of syndrome measurement rounds. For many hardware platforms, sliding-window decoding imposes runtime requirements on the order of $\mathcal{O}(1 \mu\text{s})$ per syndrome measurement round \cite{ChambsLocalNN22}, a regime that is challenging for current state-of-the-art classical hardware. Parallel block-wise decoding architectures can partially alleviate this constraint by decoding commit and cleanup windows concurrently, provided sufficient classical resources are available \cite{CampbellParallelV1,AlibabaParallel}. Nevertheless, the runtime of a quantum algorithm remains fundamentally constrained by the time required to decode a block of $d_m$ syndrome measurement rounds for a distance-$d$ code, even when $d_m \ll d$ \cite{Chamberland22,PCTels}. Minimizing decoding runtimes at the block level is therefore of central importance for scalable FTQC.

 A variety of AI-based QEC decoders have been proposed with the goals of achieving low decoding runtimes and improved logical error rates (LERs) \cite{ChamberlandRonagh,Baireuther_2019,TransformerGoogle,AlphaQubit2,AlphaQubit1China}. However, many such approaches encounter scalability challenges, both in the amount of training data required as the code distance increases and in their compatibility with parallel block-wise decoding architectures in time and in \textbf{space}. Spatial parallelism is particularly critical for fault-tolerant logical operations based on \textbf{lattice surgery} \cite{fowler2018low,Litinski19,Chamberland22,Chamberland22b}, where merged code patches can have effective distances $d_{\text{eff}} \gg 100$. In this regime, meeting real-time decoding requirements may necessitate spatially parallel block-wise decoding across large patches \cite{CampbellParallelV1}. As a result, decoders that are not compatible with parallelism in space risk becoming bottlenecks for logical operations, even if they perform well at moderate code distances for memory settings.

AI-based pre-decoders have been developed explicitly to address speed and scalability to very large code distances \cite{Gicev2023scalablefast,ChambsLocalNN22,Australia3DConvPred,LATTEPredecoder}. A non-AI-based decoder that uses Belief Propagation as the pre-decoder was also explored in \cite{caune2023beliefpropagationpartialdecoder}. Since pre-decoders are trained on labeled data and operate locally, such pre-decoders are naturally compatible with parallel block-wise decoding in both space and time. Moreover, their locality allows models trained at a modest distance $d_1$ to generalize to much larger distances $d_2 \gg d_1$. In a typical pipeline, the pre-decoder processes syndrome data locally, performs corrections and passes residual syndromes and logical information to a global decoder, which performs the final correction. An example of the residual syndromes passed to a global decoder after the application of a pre-decoder is shown in \cref{fig:visualization_3d}. While this hybrid approach leverages the strengths of both learned and algorithmic decoders, prior to this work it has not been demonstrated that a pre-decoder combined with a state-of-the-art global decoder can simultaneously achieve total decoding runtimes on the order of $\mathcal{O}(1 \mu\text{s})$ per round \textit{and} lower logical error rates than the global decoder alone.

In this work, we introduce a new AI-based pre-decoder architecture for the rotated surface code \cite{DennisSurface,fowler2012surface,TomitaRotatedSurface2014}. We develop new methods for processing labeled training data that explicitly address both spacelike and timelike failure mechanisms. These methods substantially improve pre-decoder performance and enable end-to-end decoding runtimes on the order of $\mathcal{O}(1 \mu\text{s})$ per syndrome measurement round, including both pre-decoding and subsequent global decoding using PyMatching \cite{HiggottPyMatch}. We demonstrate these results at code distances $d=21$ and $d=31$, where the combined pre-decoder + uncorrelated PyMatching pipeline achieves lower logical error rates than uncorrelated PyMatching alone, while simultaneously reducing total decoding runtime. Moreover, the relative improvement in total decoding time compared to PyMatching increases with code distance. For a correlated PyMatching global decoder, we train a larger model which outperforms correlated PyMatching alone and achieves lower runtimes at up to distances 13. Larger models can be trained to achieve LERs which are lower than correlated PyMatching for distances $d \leq 13$. The low runtimes arise from a combination of significant reductions in effective syndrome density produced by the pre-decoder and efficient deployment on state-of-the-art NVIDIA GB300 GPUs. When applying our pre-decoder in a temporal parallel block-wise decoding scheme, runtimes well below $1 \mu\text{s}$ can be achieved with access to enough GPUs. 

In standard implementations of PyMatching, edge weights in the matching graph are derived from an assumed circuit-level noise model to optimize logical error rate (LER) performance. However, the application of a pre-decoder modifies the syndrome statistics in ways that are not captured by the original noise model, leading to suboptimal matching weights. More broadly, there are many practical settings in which the full circuit-level noise model is either unknown or subject to drift over time, while syndrome data from the underlying hardware remains accessible. This motivates the need for methods that infer effective decoding parameters directly from observed data.

To address these challenges, we introduce an AI-based noise-learning architecture that infers near-optimal edge weights for both uncorrelated and correlated PyMatching using syndrome statistics alone, without requiring explicit knowledge of the underlying noise model. We demonstrate that applying this protocol to raw syndrome data yields edge weights that achieve nearly identical LERs for uncorrelated matching and improved LERs for correlated matching compared to those obtained from the known noise model.

When applying the noise-learning architecture to syndrome statistics produced by the pre-decoder, we do not observe further improvements in LER. This behavior is consistent with the structured nature of the residual errors output by the pre-decoder, which already encode much of the relevant information for downstream decoding and thus limit the extent to which additional gains can be realized through weight re-optimization.

This work is organized as follows. In \cref{sec:SurfaceCodeReview}, we review key properties of the rotated surface code relevant to the development of our pre-decoder. The pre-decoder architecture is presented in \cref{sec:PreDecArch}. After motivating its use in \cref{subsec:Motivation}, we describe the neural network architecture and associated simulation and data-processing techniques in \cref{subsec:NNArchHyperParam}. In \cref{sec:EffectivePreDecNoiseModel}, we introduce our noise-learning framework based on syndrome statistics. Numerical results for both the pre-decoder and noise-learning models are presented in \cref{sec:Numerics}. In particular, \cref{subsec:SynDensLER} analyzes syndrome density reduction and the resulting logical error rates (LERs) when combining the pre-decoder with uncorrelated PyMatching, while \cref{subsec:SynDensLERCorrMatch} extends these results to correlated PyMatching using a larger model. Runtime performance is examined in \cref{subsec:GPURuntimes}, where we report per-round decoding times for the pre-decoder on NVIDIA GB300 GPUs, as well as total runtimes for the combined pre-decoder and PyMatching pipeline. In \cref{subsec:TimeLikeParallel}, we demonstrate how per-round decoding times can be further reduced by increasing the number of GPUs within a temporal parallel, block-wise decoding scheme. In \cref{subsec:NoiseLearnImprove}, we evaluate the noise-learning model on syndrome data generated from a circuit-level noise model, comparing LERs obtained using learned edge weights against those derived from the known noise model. The impact of larger batch sizes on reducing resource requirements for real-time decoding is explored in \cref{sec:BatchingImprove}. Finally, \cref{sec:Conclusion} summarizes our results and outlines directions for future work.

\section{Summary of contributions}
  \label{sec:Contributions}

  The main contributions of this work are as follows.

  \begin{enumerate}[leftmargin=*]

  \item \textbf{Pre-decoder architecture with spacelike and timelike corrections.} We introduce a fully convolutional 3D neural network
  pre-decoder for the rotated surface code that jointly predicts spacelike (data-qubit) and timelike (measurement) corrections across the
  full space--time syndrome volume (\cref{sec:PreDecArch}). The architecture is backend-agnostic: it composes with any global decoder
  designed for surface codes, not only PyMatching, and can be adapted to different noise models, code distances, and runtime budgets by
  adjusting model depth, width, and training configuration. We develop new data-processing techniques---including a protocol for isolating
  timelike failure components (\cref{Algo:TimelikeOutputGen}), a fault-deferral scheme that prevents artificial timelike detection events
  (\cref{Algo:DataGenOptimize}), and a timelike homological equivalence protocol (\cref{Algo:TimelikeHomologicalEquivZ})---that
  substantially improve training label quality and pre-decoder performance.

  \item \textbf{Simultaneous LER improvement and end-to-end runtime reduction.} We demonstrate that combining our pre-decoder with
  uncorrelated PyMatching achieves both lower logical error rates and lower total decoding runtime than uncorrelated PyMatching alone at
  code distances $d \ge 21$ near the surface-code threshold (\cref{subsec:SynDensLER,subsec:GPURuntimes}). To our knowledge, this is the
  first demonstration that an AI-based pre-decoder can simultaneously improve both metrics relative to a state-of-the-art global decoder.
  The relative improvements in both LER and runtime grow with increasing code distance. By training a larger model with residual connections
  (\cref{fig:Model8Representation}), we further show LER improvements over correlated PyMatching at distances up to $d=13$
  (\cref{subsec:SynDensLERCorrMatch}).

  \item \textbf{GPU deployment and benchmarking of decoder runtimes.} We benchmark five pre-decoder architectures on NVIDIA GB300 GPUs at FP8
  precision, systematically exploring tradeoffs between model width, depth, kernel size, inference runtime, and LER performance
  (\cref{subsec:GPURuntimes}). The combined pre-decoder + PyMatching pipeline achieves total speedups of up to $3.4\times$ over
  uncorrelated PyMatching and $3.5\times$ over correlated PyMatching at $d=31$ and $p=0.006$
  (\cref{tab:Summary_Speedup,tab:runtimes_mwpm_bs1_correlated_total_speedup}). When deployed in a temporal parallel block-wise decoding
  scheme with multiple GPUs, per-round pre-decoder runtimes fall well below $1\,\mu\text{s}$ (\cref{subsec:TimeLikeParallel}).

  \item \textbf{Noise-learning architecture from syndrome statistics.} We introduce an AI-based architecture that infers near-optimal edge
  and hyperedge weights for both uncorrelated and correlated PyMatching directly from experimentally accessible syndrome statistics,
  without requiring knowledge of the underlying circuit-level noise model (\cref{sec:EffectivePreDecNoiseModel}). The architecture exploits
   distance-independent probability formulas for all 18 edge types and 43 hyperedge type compositions, enabling a model trained at a single
   code distance to generalize to arbitrary distances. Applied to raw syndrome data, the learned weights nearly match uncorrelated
  PyMatching performance and improve correlated PyMatching LERs relative to weights derived from the known noise model
  (\cref{subsec:NoiseLearnImprove}).

  \item \textbf{Resource reduction through batching.} We show that increasing the GPU batch size within a parallel block-wise decoding
  scheme can reduce the number of parallel classical resources $N_{\text{par}}$ required for real-time decoding by up to $12.5\times$, a
  consideration that becomes critical when decoding lattice-surgery operations across very large merged patches
  (\cref{sec:BatchingImprove}).

  \end{enumerate}
 
 \section{Brief review of the surface code}
 \label{sec:SurfaceCodeReview}
 
 \begin{figure}
     \centering
 \subfloat[\label{fig:SurfaceCodeExampV1} ]{\includegraphics[width=.4\textwidth]{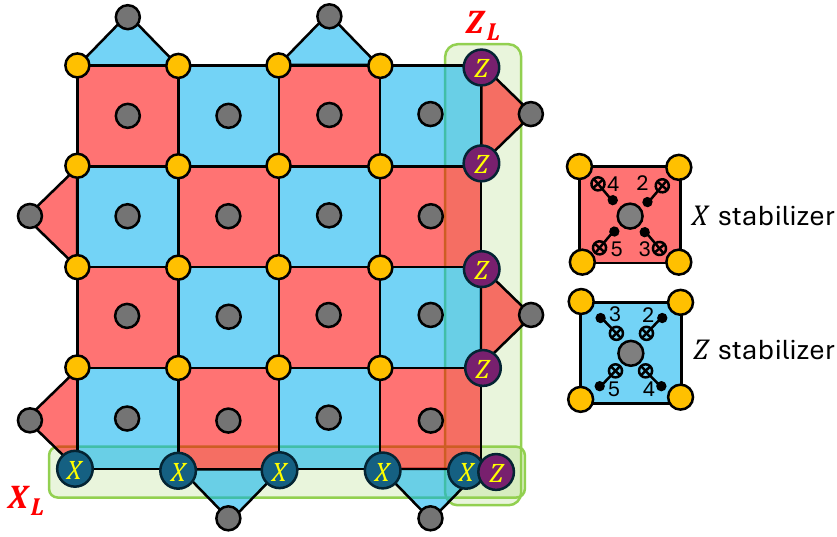}}
 \caption{ Example of a surface code patch for $d=5$. Data qubits correspond to yellow vertices, whereas ancillas used to measure the stabilizers correspond to grey vertices. $X$ ($Z$) stabilizers are represented by red (blue) plaquettes. Minimum-weight representatives for logical $X_L$ ($Z_L$) observables are shown as horizontal (vertical) strings. We provide a gate scheduling such that weight-two errors arising from a single fault propagate perpendicular to its corresponding logical observable.  }
 \label{fig:SurfaceCodeExamp}
 \end{figure}
 
Throughout this work, we train our models using the surface code \cite{DennisSurface,fowler2012surface}. However, the methods introduced in \cref{sec:PreDecArch} are not specific to the surface code and can be adapted to other topological QEC codes. To make the presentation as self-contained as possible, we begin with a brief review of the surface code and establish the notation used throughout the paper.
 
The surface code is a two-dimensional topological quantum error-correcting code whose stabilizers can be measured using nearest-neighbor interactions and which exhibits a threshold of approximately $0.7 \%$ for a circuit-level depolarizing noise model. Moreover, universal fault-tolerant quantum computation can be implemented using only nearest-neighbor interactions via lattice surgery \cite{fowler2018low,Litinski19,Litinski2018latticesurgery,Chamberland22,Chamberland22b}. As a result, despite the development of many alternative codes with attractive theoretical properties, the surface code remains a leading candidate for near- and mid-term quantum computing architectures, particularly those with limited qubit connectivity.
 
The surface code is characterized by the parameters $\left[\!\left[ d_x d_z, k, \text{min}(d_x,d_z) \right]\!\right]$, where $k=1$ is the number of encoded logical qubits and $d_x$ ($d_z$) denotes the minimum weight of logical $X$ ($Z$) operators. In this work, we focus on square patches with $d_x = d_z = d$, although the methods presented in \cref{sec:PreDecArch} naturally extend to rectangular patches with arbitrary $d_x$ and $d_z$. An example of a $d=5$ surface code patch is shown in \cref{fig:SurfaceCodeExamp}. For the chosen patch orientation, minimum-weight representatives of the logical operators $X_L$ and $Z_L$ correspond to horizontal and vertical strings, respectively. \cref{fig:SurfaceCodeExamp} also illustrates a valid gate scheduling for measuring $X$- and $Z$-type stabilizers, chosen such that a weight-two error arising from a single fault propagates perpendicular to the corresponding logical operator. The numbers shown beside the CNOT gates indicate the time steps at which the gates are applied, with time steps 1 and 6 reserved for ancilla state preparation and measurement.

We define the error syndrome as the set of stabilizer measurement outcomes. To distinguish spacelike from timelike errors, stabilizer measurements are repeated over multiple rounds. The number of required measurement rounds depends on the desired suppression of timelike logical failures, which is particularly relevant for lattice-surgery-based protocols (see, for example, Appendix C of Ref.~\cite{Chamberland22} and the extended discussion in Ref.~\cite{PCTels}). Throughout this work, the error syndrome is understood to include stabilizer measurement outcomes from all syndrome measurement rounds. We denote the measured syndromes in round $k$ for $X$- and $Z$-type stabilizers as $\text{SynX}^{(k)}$ and $\text{SynZ}^{(k)}$, respectively, and define the full syndrome as
\begin{align}
    \text{Syn} = (\text{SynX}^{(1)},\text{SynZ}^{(1)}, \cdots, \text{SynX}^{(d_m)},\text{SynZ}^{(d_m)})
\end{align}
A decoding algorithm processes $\text{Syn}$ to infer a likely error configuration. Two widely used decoders for the surface code are minimum-weight perfect matching (MWPM) \cite{HiggottPyMatch} and Union Find (UF) \cite{DelfosseUnionFind}. Importantly, the runtime of both decoders depends on the syndrome density $s$. For $d_m$ measurement rounds and $S(d)=d^2-1$ stabilizers per round, we define
\begin{align}
    s = |\text{Syn}| / (d_m S(d))
\end{align}
where $|\text{Syn}|$ denotes the number of non-trivial detection events. The decoding complexity of MWPM scales as $\mathcal{O}(s^3)$ \cite{Edmonds_1965}, while UF scales as $\mathcal{O}(s)$. Although UF offers faster runtimes, MWPM typically achieves lower logical error rates \cite{DelfosseUnionFind}. In contrast, AI-based decoders have a fixed complexity independent of $s$.
 
As shown in Refs.~\cite{TerhalBacklog,ChambsLocalNN22}, when decoding a sequence of syndrome measurement rounds using a sliding-window approach, an exponential backlog arises if the decoding time per round, $T_{\text{DEC}}$, exceeds the time required to measure the stabilizers, $T_s$. In Ref.~\cite{ChambsLocalNN22}, the wait time for updating the Pauli frame as a function of circuit depth was derived as
\begin{align}
T^{b_j} = \frac{c^j r}{T_s^{j-1}} + T_l\Big[ \frac{T_s^{1-j}(c^j - T_s^j)}{c - T_s} \Big],
\label{eq:TbjSlidingWindow}
\end{align}
where $T_l$ denotes the runtimes associated with transmitting measured stabilizers to the classical processing device. Equation~\eqref{eq:TbjSlidingWindow} assumes a linear-time decoder, $T_{\text{DEC}}(r) = c r$, where $c$ is a constant that depends on the code distance $d$ and $r$ is the number of syndrome measurement rounds.
 
To mitigate the exponential backlog when $T_{\text{DEC}} > T_s$, Refs.~\cite{CampbellParallelV1,AlibabaParallel} introduced a parallel window decoding strategy. Instead of decoding windows of size $d_m$ sequentially with buffer regions of equal size, the syndrome measurement history is partitioned into commit regions of size $d_m$ with buffer regions of equal size placed both before and after each commit region. All commit regions are decoded in parallel, and the remaining cleanup regions can likewise be partitioned into blocks that are decoded concurrently. Ref.~\cite{CampbellParallelV1} showed that the exponential backlog can be avoided provided the number of parallel decoding resources $N_{\text{par}}$ satisfies
\begin{align}
N_{\text{par}} \ge \frac{2 T_{\text{DEC}}}{(T_l + T_s)(n_{\text{com}} + n_W)},
\label{eq:NparSats}
\end{align}
where $n_{\text{com}}$ is the number of syndrome measurement rounds in the commit region and $n_W$ is the number of rounds in each buffer region. Nevertheless, even in this parallelized setting, overall algorithm runtime remains strongly dependent on $T_{\text{DEC}}$. In \cref{sec:PreDecArch}, we introduce a pre-decoding architecture that achieves both fast execution on GPUs and substantial reductions in syndrome density $s$, thereby minimizing $T_{\text{DEC}}$ when combined with a global algorithmic decoder such as MWPM or Union Find.
 
 \section{Pre-decoder architecture}
 \label{sec:PreDecArch}
 
\begin{figure}
     \centering
 \subfloat[\label{fig:PreDecOverviewV1} ]{\includegraphics[width=.4\textwidth]{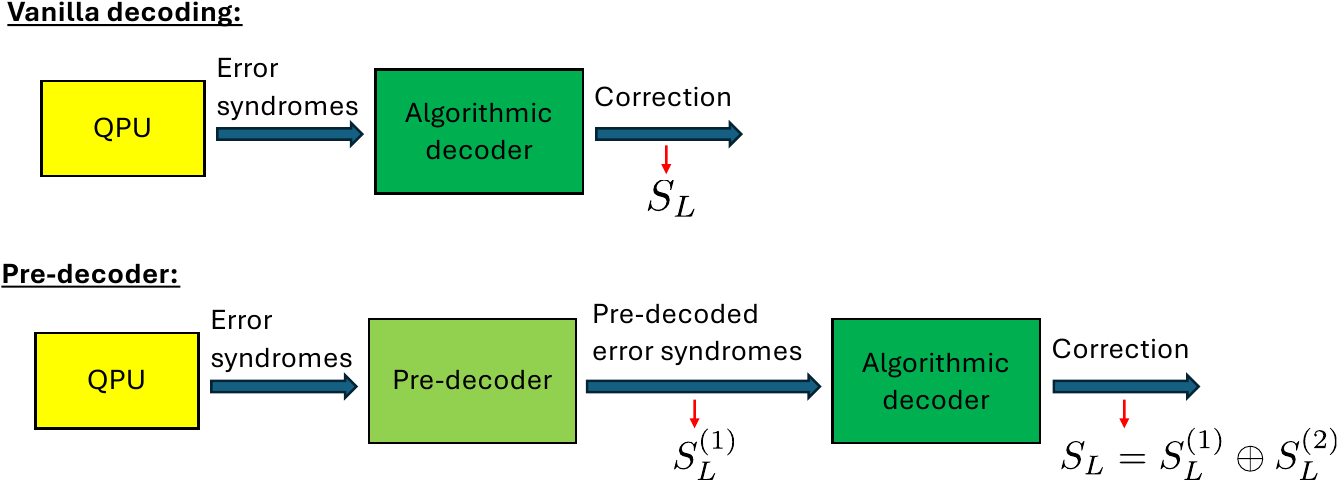}}
 \caption{In a vanilla decoding algorithm, an algorithmic decoder receives the error syndromes from the QPU and performs corrections to determine the signs $S_L$ of the relevant logical observables. When using a pre-decoder, the pre-decoder receives the error syndrome from the QPU and applies spacelike and timelike corrections across all syndrome measurement rounds that were used as inputs. Such corrections produce the signs $S_L^{(1)}$ of the logical observables. The new error syndrome obtained from the corrections are then passed to an algorithmic decoder to apply the final set of corrections resulting in a sign $S_L^{(2)}$ of the logical observables. The final sign is computed as $S_L = S_L^{(1)} \oplus S_L^{(2)}$. }
 \label{fig:PreDecOverview}
\end{figure}
 
 \subsection{Motivation for using pre-decoders}
 \label{subsec:Motivation}
 
As discussed in \cref{sec:SurfaceCodeReview}, the decoding time $T_{\text{DEC}}$ of algorithmic decoders such as minimum-weight perfect matching (MWPM) or Union Find (UF) depends strongly on the syndrome density $s$. The syndrome density itself is determined by factors such as the underlying noise model and the circuits used for syndrome extraction. This dependence becomes particularly pronounced near the error threshold, where $s$ can be large—especially for MWPM, whose runtime scales as $T_{\text{DEC}} \propto \mathcal{O}(s^3)$. Consequently, substantial reductions in decoding runtimes can be achieved by reducing the effective syndrome density prior to global decoding.

Using the definitions introduced in \cref{sec:SurfaceCodeReview}, the total time required to process $r$ syndrome measurement rounds using an algorithmic decoder alone is given by
\begin{align}
 T^{(\text{al})}_{\text{tot}}(r,s) = T_s + T_l + T^{(al)}_{\text{DEC}}(r,s),
\label{eq:TotAlgoTime}
\end{align}
where $T^{(al)}_{\text{DEC}}(r,s)$ denotes the time required to decode $r$ rounds with syndrome density $s$.

A reduction in syndrome density can be achieved by introducing an AI-based pre-decoder that performs local corrections across the space–time volume of measured syndromes \cite{Gicev2023scalablefast,ChambsLocalNN22,Australia3DConvPred}. The resulting hybrid decoding pipeline—consisting of a pre-decoder followed by a global algorithmic decoder—is illustrated in \cref{fig:PreDecOverview}. Local space–time corrections are implemented using a fully convolutional three-dimensional neural network, as described in \cref{subsec:NNArchHyperParam}.

Let $T_{l_1}$ denote the time required to transmit measured syndromes from the quantum processing unit (QPU) to the classical device implementing the pre-decoder, and let $T_{l_2}$ denote the time required to transmit the updated syndromes from the pre-decoder to the device implementing the global decoder. In this setting, the total time to process $r$ syndrome measurement rounds is
\begin{align}
T^{(\text{pra})}_{\text{tot}}(r,s) = T_s + T_{l_1} + T^{(\text{pre})}_{\text{DEC}}(r)+ T_{l_2} + T^{(al)}_{\text{DEC}}(r,s'),
\label{eq:TotPreDecPlusAlgoTime}
\end{align}
where $T^{(\text{pre})}_{\text{DEC}}(r)$ is the pre-decoder runtime and $s'$ is the reduced syndrome density obtained from $s$ after applying the pre-decoder. Crucially, due to its AI-based implementation, $T^{(\text{pre})}_{\text{DEC}}(r)$ is independent of the input syndrome density $s$.

Comparing \cref{eq:TotAlgoTime,eq:TotPreDecPlusAlgoTime}, a net speedup is achieved whenever
\begin{align}
    T^{(\text{pra})}_{\text{tot}}(r,s) < T^{(\text{al})}_{\text{tot}}(r,s).
\end{align}
In other words, the overhead introduced by pre-decoding and additional communication is offset when the reduction in global decoding time resulting from the lower syndrome density $s'$ exceeds these costs. In \cref{subsec:GPURuntimes}, we provide detailed runtime estimates of both $T^{(\text{pre})}_{\text{DEC}}(r)$ and $T^{(\text{pra})}_{\text{tot}}(r,s)$ on NVIDIA GB300 GPUs for a range of space–time volumes.
 
\subsection{Neural network architecture and hyperparameters}
\label{subsec:NNArchHyperParam}
 
 \begin{figure}
     \centering
 \subfloat[\label{fig:ConvArchV1} ]{\includegraphics[width=.5\textwidth]{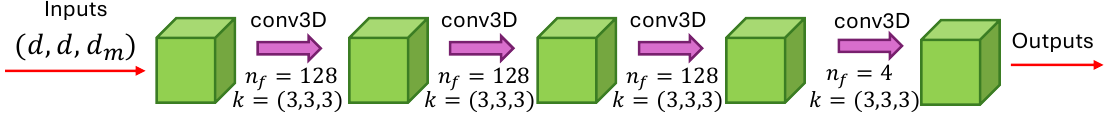}}
 \caption{ Example of a four-layer fully connected three-dimensional convolutional neural network used to train our AI-based pre-decoder. The first three layers use $n_f=128$ filters with three-dimensional kernels of size $(3,3,3)$. The final layer always uses four filters since the network has 4 output correction channels.}
 \label{fig:ConvArch}
 \end{figure}
 
In this section, we describe the neural network architecture used to construct our AI-based pre-decoders and summarize the training hyperparameters that yield optimal performance.

Our AI-based pre-decoder is implemented as a fully convolutional three-dimensional neural network, meaning that it consists exclusively of 3D convolutional layers and does not employ linear or projection layers. This fully convolutional design ensures that the network output has the same space–time dimensions as its input for each channel, enabling local corrections to be applied across the entire space–time volume of the syndrome data.

A key advantage of this architecture is its scalability: the network can be trained on input volumes of size $(d,d,d_m)$ and applied at inference time to volumes of size $(d',d',d'_m)$, with $d \neq d'$ and $d_m \neq d'_m$. An example architecture with four 3D convolutional layers is shown in \cref{fig:ConvArch}, where each layer is specified by its three-dimensional kernel size and number of filters. The final layer always uses four filters, corresponding to the four output channels described below.

Deeper architectures require skip connections to avoid vanishing gradients and were explored in Ref.~\cite{ChambsLocalNN22}. While most of the focus of the present work is on minimizing pre-decoder runtimes, we also consider them in \cref{subsec:SynDensLERCorrMatch} to enable further LER improvements.

An important architectural parameter of 3D convolutional networks is the receptive field, which quantifies the size of the local three-dimensional window of the input that influences a given output element. The receptive field plays a central role in determining the maximum effective decoding distance of the pre-decoder, since error chains with spatial or temporal extent larger than the receptive field cannot, in general, be fully corrected by local operations alone.

Consider a network with $l$ convolutional layers, where the kernel size in the $j$-th layer is $(k_j, k_j, k_j)$. Assuming unit strides and dilation coefficients $D=1$ in all layers, the receptive field is given by
\begin{align}
R_l = 1 + \sum_{i=1}^l (k_i - 1).
\label{eq:ReceptiveFieldFormula}
\end{align}
Increasing the receptive field can therefore be achieved either by increasing the number of layers or by using larger convolutional kernels. However, as shown in \cref{subsec:GPURuntimes}, increasing kernel size leads to a significantly larger increase in $T^{(\text{pre})}_{\text{DEC}}(r)$ than increasing depth, motivating the architectural choices adopted in this work.
 
\subsubsection{Input training data}
 \label{subsec:InputTrain}
 
\begin{figure*}
     \centering
 \subfloat[\label{fig:XstabMap} ]{\includegraphics[width=.4\textwidth]{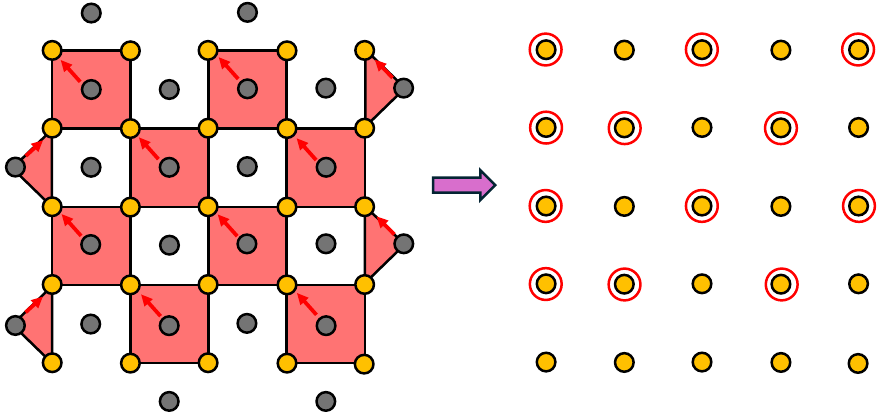}}
 \subfloat[\label{fig:ZstabMap} ]{\includegraphics[width=.4\textwidth]{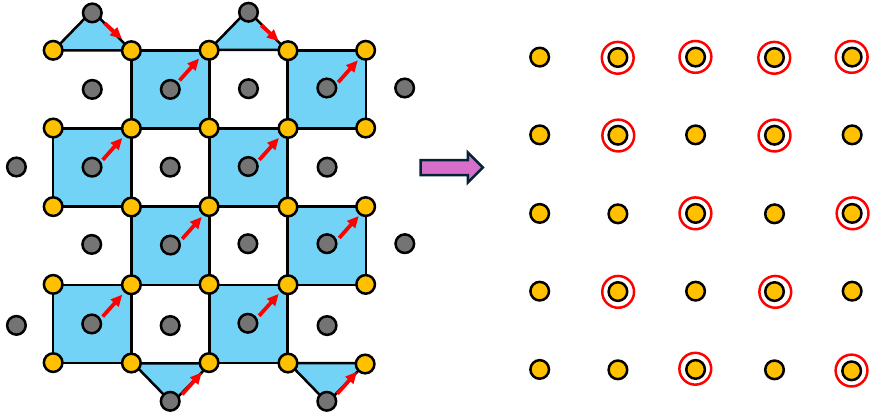}}
     \caption{(a) Example mapping of $X$-type stabilizers to a $D \times D$ grid (with $D=5$). For any $D$, measurement outcomes of weight-four $X$-type stabilizers are mapped to the top-left data qubit in its support. Weight-two stabilizers on the left or right boundary are mapped to the top data in its support. (b) Similar mapping as in (a) but for $Z$-type stabilizers.}
     \label{fig:StabMap}
\end{figure*}
 
In this subsection, we describe the structure of the input data used to train our neural networks. Throughout, tensors representing input and output training data are denoted by \texttt{trainX} and \texttt{trainY}, respectively.
 
To enable the neural network to identify both spacelike and timelike errors arising from repeated stabilizer measurements, the measured syndromes must be encoded efficiently on a two-dimensional grid for each measurement round. In addition, stabilizer statistics near the boundaries of the lattice differ from those in the bulk. To account for this, we provide the network with explicit geometric information that encodes stabilizer locations and their corresponding weights (two or four for a standard surface-code patch), as described below.

Consider a surface-code patch embedded on a $D \times D$ grid, where $D$ denotes the maximum number of data qubits (yellow vertices in \cref{fig:SurfaceCodeExamp}) along any row or column. Suppose that $N_{\text{train}}$ training samples are generated. For each sample $1 \le j \le N_{\text{train}}$, stabilizers are measured for $d_m$ syndrome measurement rounds. For each fault location in the circuit, errors are sampled according to the underlying noise model and propagated through the circuit.

After error propagation, we store (i) differences between data-qubit errors in consecutive rounds (as well as timelike failures, more on this in \cref{subsec:OutputTrain}) and (ii) differences between stabilizer measurement outcomes in consecutive rounds, commonly referred to as detector events. Let $s_{i,k}$ denote the measurement outcome of the $i$th stabilizer in round $k$. The corresponding detector event is defined as
\begin{align}
d_{i,k} = s_{i,k} \oplus s_{i,k-1}
\end{align}

Detector events for all $X$-type stabilizers in round $k$ and training sample $j$ are collected as
\begin{align}
D^{(j)}_k(X) \equiv (d_{1,k}(X), \ldots, d_{K_x,k}(X)),
\label{eq:DetectorX}
\end{align}
where for a surface code with $d_x = d_z = D$, the number of $X$ stabilizers is $K_x = (D^2 - 1)/2$. Similarly, detector events for $Z$-type stabilizers are given by
\begin{align}
D^{(j)}_k(Z) \equiv (d_{1,k}(Z), \ldots, d_{K_z,k}(Z)).
\label{eq:DetectorZ}
\end{align}

Let $E^{(j)}(X)_{(i,k)} \in \{I,X \}$ denote the $X$-error affecting the $i$-th data qubit in round $k$ for training sample $j$. We define the error difference between consecutive rounds as
\begin{align}
     \tilde{X}^{(j)}_{i,k} = E^{(j)}(X)_{i,k} \oplus E^{(j)}(X)_{i,k-1}
\end{align}
Collecting these differences over all data qubits yields
\begin{align}
\tilde{X}^{(j)}_k \equiv (\tilde{X}^{(j)}_{(1,k)}, \ldots, \tilde{X}^{(j)}_{(D^2,k)}).
\label{eq:XerrorDiff}
\end{align}
An analogous definition applies to $Z$ errors,
\begin{align}
\tilde{Z}^{(j)}_k \equiv (\tilde{Z}^{(j)}_{(1,k)}, \ldots, \tilde{Z}^{(j)}_{(D^2,k)}),
\label{eq:ZerrorDiff}
\end{align}
which together form the target labels used during training.

The input tensor \texttt{trainX} has shape $(N_{\text{train}}, D, D, d_m, N_s)$, where $N_s$ denotes the number of input channels. For the quantum-memory setting considered in this work, $N_s = 4$, as described below. In more general settings—such as lattice surgery—additional channels are required, leading to $N_s > 4$; these extensions are left for future work.

We first describe the two detector-event channels of \texttt{trainX}, following the encoding scheme introduced in Ref.~\cite{ChambsLocalNN22}. For the $k$-th syndrome measurement round and training sample $j$, we define
\begin{align}
     \texttt{trainX}(j,1{:}D,1{:}D,k,1) &= \texttt{x\_type}(k,j), \label{eq:encTrainX1} \\
     \texttt{trainX}(j,1{:}D,1{:}D,k,2) &= \texttt{z\_type}(k,j),
     \label{eq:encTrainX2}
 \end{align}
where $\texttt{x\_type}(k,j)$ and $\texttt{z\_type}(k,j)$ correspond to the detector events $D^{(j)}_k(X)$ and $D^{(j)}_k(Z)$ mapped onto the $D \times D$ grid.
 
An example of this mapping procedure is shown in \cref{fig:StabMap}. Detection events from weight-four $X$ ($Z$)-type stabilizers are mapped to the top-left (top-right) data qubit in the stabilizer’s support. For weight-two stabilizers, $X$-type detection events are mapped to the top data qubit, while $Z$-type detection events are mapped to the right data qubit. A detection event is assigned the value 1 if the stabilizer outcome changes between consecutive rounds and 0 otherwise. Grid locations receiving no detection event are always set to 0.

\begin{figure*}
     \centering
\subfloat[\label{fig:trainYvisualAidex} ]{\includegraphics[width=.75\textwidth]{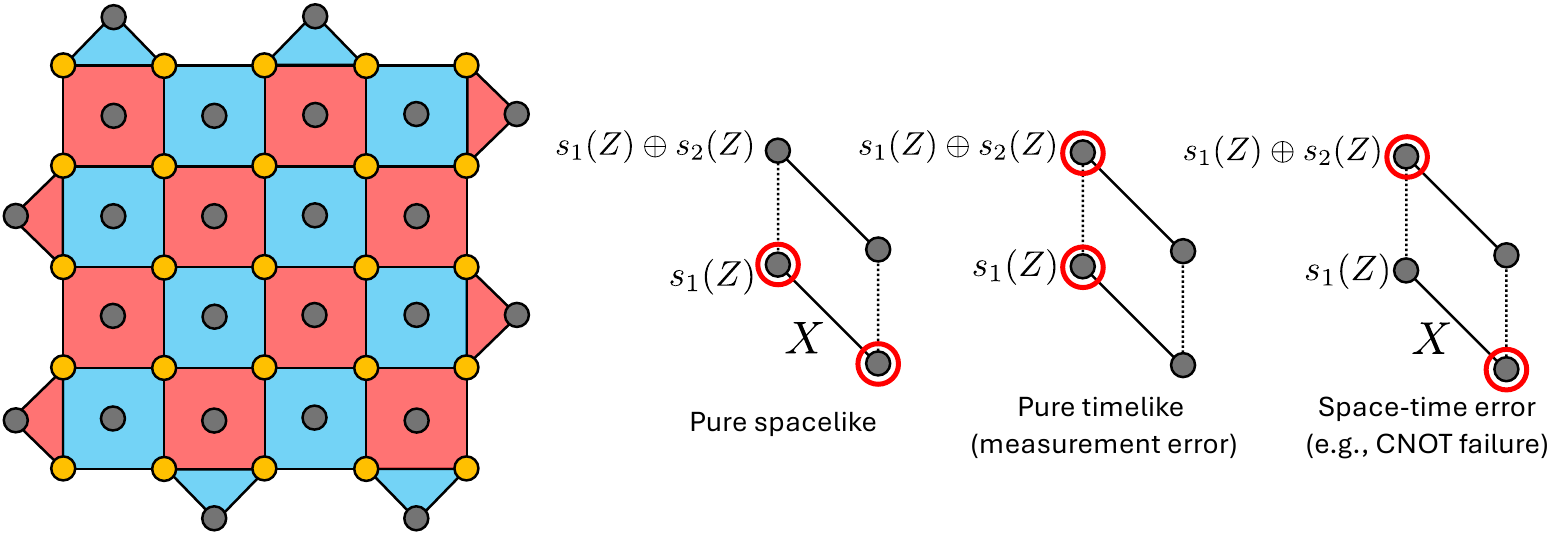}}
     \caption{Example illustrations of the computation of $s_1(Z) \oplus s_2(Z)$ used in \cref{Algo:TimelikeOutputGen}. Only pure timelike and space-time failures result in a non-trivial value for $s_1(Z) \oplus s_2(Z)$. Red circles illustrate stabilizers that are measured as $-1$ instead of $+1$ (vertices without a red circle) in a given round. }
     \label{fig:trainYvisualAid}
\end{figure*}

In addition to detector events, we encode local geometric information using the same stabilizer-to-qubit mapping. Rather than mapping detection events, these channels encode the normalized stabilizer weights at the corresponding grid locations. For each round $k$, these channels are denoted by $\texttt{x\_present}(k)$ and $\texttt{z\_present}(k)$.

During logical-qubit initialization, all entries of $\texttt{x\_present}(1)$ ($\texttt{z\_present}(1)$) are set to zero if the logical qubit is initialized in $|0\rangle$ ($|+\rangle$). Similarly, in the final measurement round $k=d_m$, all entries of $\texttt{x\_present}(d_m)$ ($\texttt{z\_present}(d_m)$) are set to zero when measuring in the $Z$ ($X$) basis.

For the $D=5$ surface-code patch shown in \cref{fig:StabMap}, the geometric channels take the form
\begin{align}
     \texttt{x\_present}(k) &= 
 \begin{bmatrix}
 1 & 0 & 1 & 0 & 0.5 \\
 0.5 & 1 & 0 & 1 & 0 \\
 1 & 0 & 1 & 0 & 0.5 \\
 0.5 & 1 & 0 & 1 & 0 \\
 0 & 0 & 0 & 0 & 0
 \end{bmatrix}, \\
     \texttt{z\_present}(k) &= 
 \begin{bmatrix}
 0 & 0.5 & 1 & 0.5 & 1 \\
 0 & 1 & 0 & 1 & 0 \\
 0 & 0 & 1 & 0 & 1 \\
 0 & 1 & 0 & 1 & 0 \\
 0 & 0 & 0.5 & 0 & 0.5
 \end{bmatrix},
 \end{align}
for $1 < k < d_m$. These channels are then incorporated into \texttt{trainX} as
\begin{align}
 \texttt{trainX}(j,1{:}D,1{:}D,k,3) &= \texttt{x\_present}(k), \label{eq:encTrainX3} \\
 \texttt{trainX}(j,1{:}D,1{:}D,k,4) &= \texttt{z\_present}(k).
\label{eq:encTrainX4}
\end{align}

\subsubsection{Output training data}
\label{subsec:OutputTrain}

We now describe the output labels used to train the pre-decoders. To reduce the syndrome density passed to a global decoder, the pre-decoder must perform both spacelike (data-qubit) and timelike (stabilizer-measurement) corrections. Accordingly, the training targets encode both types of corrections.

The output tensor \texttt{trainY} consists of four channels: two channels corresponding to $Z$- and $X$-type Pauli corrections on data qubits, and two channels corresponding to timelike corrections for $X$- and $Z$-type stabilizers.

We first describe the spacelike output channels, which occupy the first two channels of \texttt{trainY}. Using the definitions of error differences introduced in \cref{eq:XerrorDiff,eq:ZerrorDiff}, we set
\begin{align}
     \texttt{trainY}(j,1{:}D,1{:}D,k,1) &= \tilde{Z}^{(j)}_{k}, \label{eq:TrainY1} \\
     \texttt{trainY}(j,1{:}D,1{:}D,k,2) &= \tilde{X}^{(j)}_{k},
     \label{eq:TrainY2}
\end{align}
for the $j$-th training sample and the $k$-th syndrome measurement round. These channels track changes in $Z$- and $X$-type Pauli errors on data qubits between consecutive rounds, obtained by sampling faults from the noise model at each circuit location and propagating them through the syndrome-extraction circuit.
 
The remaining two output channels encode purely timelike corrections, corresponding to changes in stabilizer measurement outcomes induced by faults within a single syndrome measurement round. Because data qubits are measured in the final round, timelike corrections are defined only for rounds $k = 1, \ldots, d_m - 1$.

To construct these labels, we isolate the timelike component of each fault mechanism by comparing stabilizer syndromes obtained before and after propagating the same error configuration through an additional round of the circuit, as described in \cref{Algo:TimelikeOutputGen}.
\begin{algorithm}[H]
\caption{Timelike output channel generation}
\begin{algorithmic}
\For{$k = 1$ to $d_m - 1$}
     \State Let $E_k$ be the errors generated by the noise model at each fault location in syndrome measurement round $k$.
         \State Propagate $E_k$ and compute:
         \State \quad $X$ and $Z$ stabilizer syndromes $s_1(X)$, $s_1(Z)$
         \State Let $E^{(k)}_{\text{out}}$ be the output data qubit errors from propagating $E_k$.
         \State Propagate $E^{(k)}_{\text{out}}$ and compute:
         \State \quad $X$ and $Z$ stabilizer syndromes $s_2(X)$, $s_2(Z)$
         \State $\texttt{trainY}(j,1{:}D,1{:}D,k,3) \gets s_1(X) \oplus s_2(X)$
         \State $\texttt{trainY}(j,1{:}D,1{:}D,k,4) \gets s_1(Z) \oplus s_2(Z)$
\EndFor
\end{algorithmic}
\label{Algo:TimelikeOutputGen}
\end{algorithm}
An illustration of the computation of $s_1(Z) \oplus s_2(Z)$ used in \cref{Algo:TimelikeOutputGen} is shown in \cref{fig:trainYvisualAid}. Intuitively, the two-stage propagation procedure isolates the pure timelike contribution of faults occurring in a given syndrome measurement round by canceling spacelike effects that persist across rounds. These timelike labels enable the pre-decoder to learn local corrections that suppress time-correlated detection events, thereby further reducing the syndrome density passed to the global decoder.

\begin{figure*}
     \centering
\subfloat[\label{fig:Circuit_D5_Main} ]{\includegraphics[width=.35\textwidth]{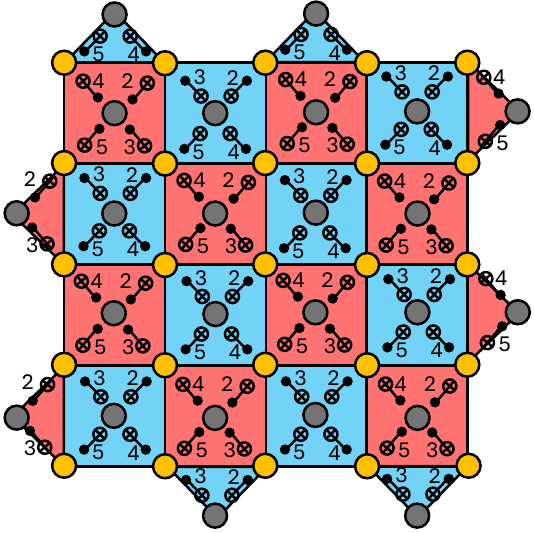}}
     \caption{ Circuit for a $d=5$ surface code showing the CNOT gates and corresponding time steps used to generate our data. The time step $t=1$ is used for preparing the ancillas (grey vertices) in the $|+\rangle$ and $|0\rangle$ basis. The time step $t=6$ is for measuring the ancillas in the $X$ or $Z$ basis.}
     \label{fig:CircuitMainCNOT}
\end{figure*}

\subsubsection{Data processing}
 \label{sec:DataProcess}

 In this subsection, we describe data-processing techniques applied during the generation of the output labels \texttt{trainY} to avoid the introduction of artificial timelike detection events. Such artifacts can arise from the temporal ordering of faults and stabilizer measurements in the syndrome-extraction circuit.

To illustrate this effect, consider the stabilizer measurement circuit shown in \cref{fig:CircuitMainCNOT}, where CNOT gates are labeled by their execution time steps. Focus on the $k$-th syndrome measurement round with $k>1$. Suppose a $Z$ error occurs at time step 6 during the ancilla measurement. The stabilizers affected by this error are not measured until round $k+1$. However, because the fault occurred during round $k$, the resulting data-qubit error could incorrectly be assigned to the spacelike output channel of \texttt{trainY} in round $k$, while the corresponding syndrome appears in \texttt{trainX} in round $k+1$.

More generally, there exist many leading-order fault processes in which a data-qubit error is generated in round $k$ but produces detectable syndrome information only in round $k+1$. If not handled carefully, such processes lead to spurious vertical pairs in space–time, artificially inflating the number of timelike events seen by the network.

To prevent the introduction of these artifacts, we apply the data-generation protocol described in \cref{Algo:DataGenOptimize}. The key idea is to update the training labels only when a fault produces a non-trivial stabilizer syndrome in the same round; otherwise, the resulting data-qubit error is deferred and treated as an input error in the subsequent round.
\begin{algorithm}[H]
\caption{Data generation protocol}
\begin{algorithmic}
 \For{$k = 1$ to $d_m - 1$}
     \State Let $E_k$ be the full set of faults generated by the noise model at each fault location in syndrome measurement round $k$.
     \State Let $N_{E_k}$ be the number of faults in $E_k$, and let $e^{(k)}_j$ denote the $j$th fault ($1 \le j \le N_{E_k}$).
     \For{$j = 1$ to $N_{E_k}$}
         \State Propagate $e^{(k)}_j$ through the surface-code stabilizer measurement circuit.
         \State Let $s_{e^{(k)}_j}$ be the resulting stabilizer syndrome.
         \State Let $|s_{e^{(k)}_j}|$ denote the Hamming weight of $s_{e^{(k)}_j}$.
         \If{$|s_{e^{(k)}_j}| > 0$}
             \State Update \texttt{trainX} and \texttt{trainY} as described in \cref{subsec:InputTrain,subsec:OutputTrain}.
         \Else
             \If{$e^{(k)}_j$ results in a non-trivial data-qubit error $e^{(k)}_{d_j}$}
                 \State Append $e^{(k)}_{d_j}$ to $E_{k+1}$ at time step 1 and ignore updates to \texttt{trainY}.
             \EndIf
         \EndIf
     \EndFor
 \EndFor
\end{algorithmic}
\label{Algo:DataGenOptimize}
\end{algorithm}

Additional care is required when processing faults containing $Y$ errors. For instance, a single $Y$ error on a data qubit can produce an $X$-type detection event in round $k$ and a $Z$-type detection event in round $k+1$, leading to mixed spacelike–timelike signatures. To avoid introducing artificial correlations of this form, all faults containing $Y$ errors are decomposed into equivalent combinations of $X$- and $Z$-only errors prior to applying \cref{Algo:DataGenOptimize}.

For single-qubit faults, this decomposition is straightforward, since $Y = X \oplus Z$ and the two components can be propagated independently. For two-qubit faults containing at least one $Y$ error, the situation is more subtle but remains systematic. Such faults arise only after CNOT gates and therefore always involve one data qubit and one ancilla qubit.

The decomposition is chosen to correlate the $X/Z$ content of the data-qubit error with the type of error detectable by the ancilla. For example, ancillas used in $X$ stabilizer measurements detect $Z$ errors. Consequently, a fault of the form $Y$(data)$Z$(ancilla) is decomposed as
\begin{align}
    YZ \to ZZ \oplus XI,
\end{align}
where each term is propagated independently. This ensures that the resulting detection events are correctly localized in time.
 
The complete set of decomposition rules used in this work is summarized in \cref{tab:Ydecomp}. After decomposition, each resulting fault is treated independently and propagated according to \cref{Algo:DataGenOptimize}.
\begin{table}[h]
 \centering
 \begin{tabular}{|c|c|c|}
 \hline
 Error & X-ancilla & Z-ancilla \\
 \hline
 YX & $XI \oplus ZI \oplus IX$ & $XX \oplus ZI$ \\
 YZ & $ZZ \oplus XI$ & $XI \oplus ZI \oplus IZ$ \\
 YY & $ZZ \oplus XI \oplus IX$ & $XX \oplus ZI \oplus IZ$ \\
 XY & $XI \oplus IX \oplus IZ$ & $XX \oplus IZ$ \\
 ZY & $ZZ \oplus IX$ & $ZI \oplus IX \oplus IZ$ \\
 \hline
 \end{tabular}
 \caption{Decomposition rules for two-qubit faults containing $Y$ errors. The first qubit is always a data qubit and the second is an ancilla qubit. Columns distinguish the ancilla type.}
 \label{tab:Ydecomp}
\end{table}
 
\subsubsection{Homological equivalence function}
\label{subsec:HomologicalEquivalence}
 
 \begin{figure*}
     \centering
 \subfloat[\label{fig:Homological_equivV1} ]{\includegraphics[width=.8\textwidth]{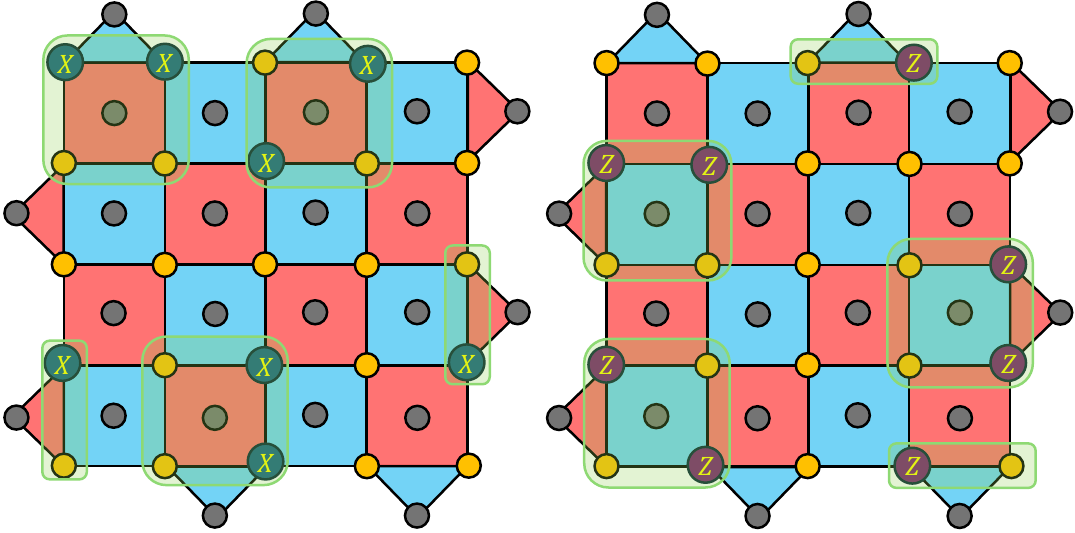}}
     \caption{ Spacelike homological equivalence convention as shown in a $d=5$ surface code lattice. On the left part of the figure, we show $X$ error configurations which are invariant under the transformations of the functions \texttt{weightReductionX} and \texttt{fixEquivalenceX}. On the right part of the figure, we show $Z$ error configurations which are invariant under the transformations of the functions \texttt{weightReductionZ} and \texttt{fixEquivalnceZ}. }
     \label{fig:Homological_equiv}
 \end{figure*}
 
 \begin{figure*}
     \centering
 \subfloat[\label{fig:Timelike_Homological_V1} ]{\includegraphics[width=.5\textwidth]{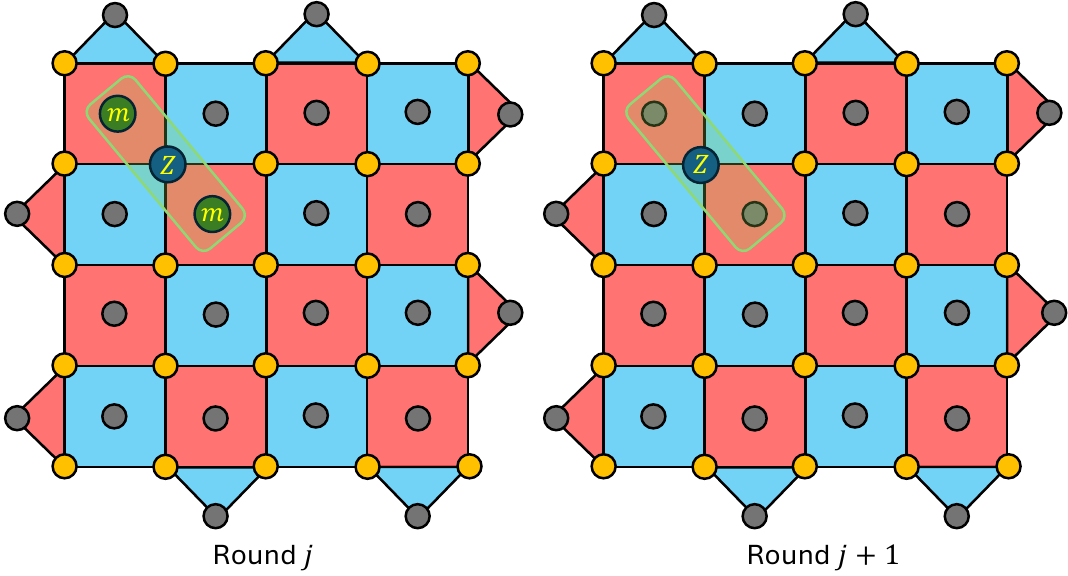}}
 \vfill
 \subfloat[\label{fig:Timelike_Homological_V2} ]{\includegraphics[width=.5\textwidth]{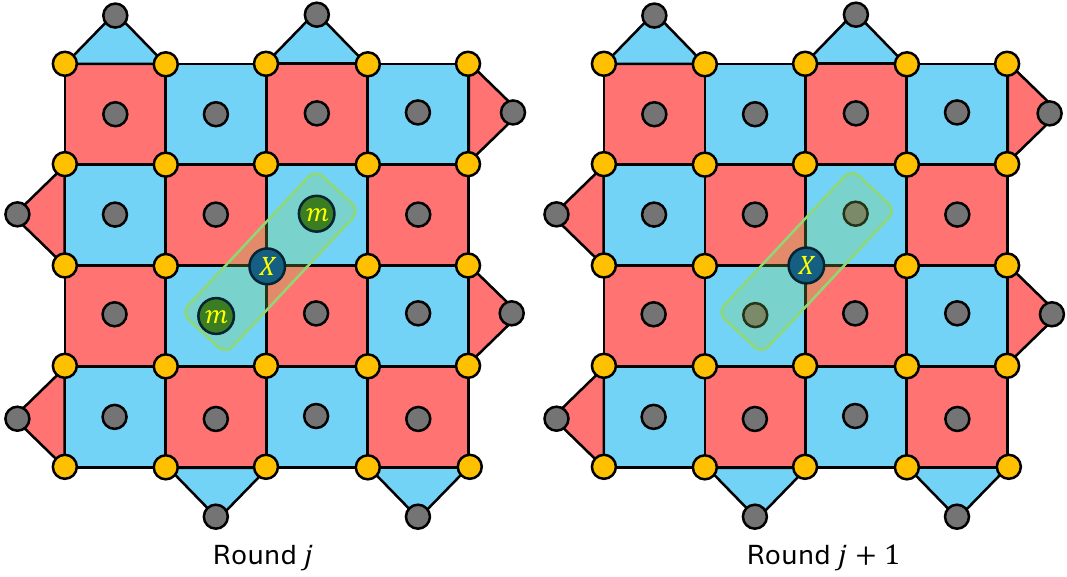}}
     \caption{ Timelike homological equivalence convention for a $d=5$ surface code. (a) For each data qubit in two consecutive syndrome measurement rounds, we apply a $Z$ correction. Measurement errors that anti-commute with the $Z$ error are added in the first round that a $Z$ data qubit error is added. If the number of 1's in \texttt{trainY} is reduced, we accept the trivial correction. (b) Same as (a) but with $X$ corrections. }
     \label{fig:Timelike_Homological}
 \end{figure*}
 
 \begin{figure*}
     \centering
 \subfloat[\label{fig:Timelike_Homological_V1_High} ]{\includegraphics[width=.5\textwidth]{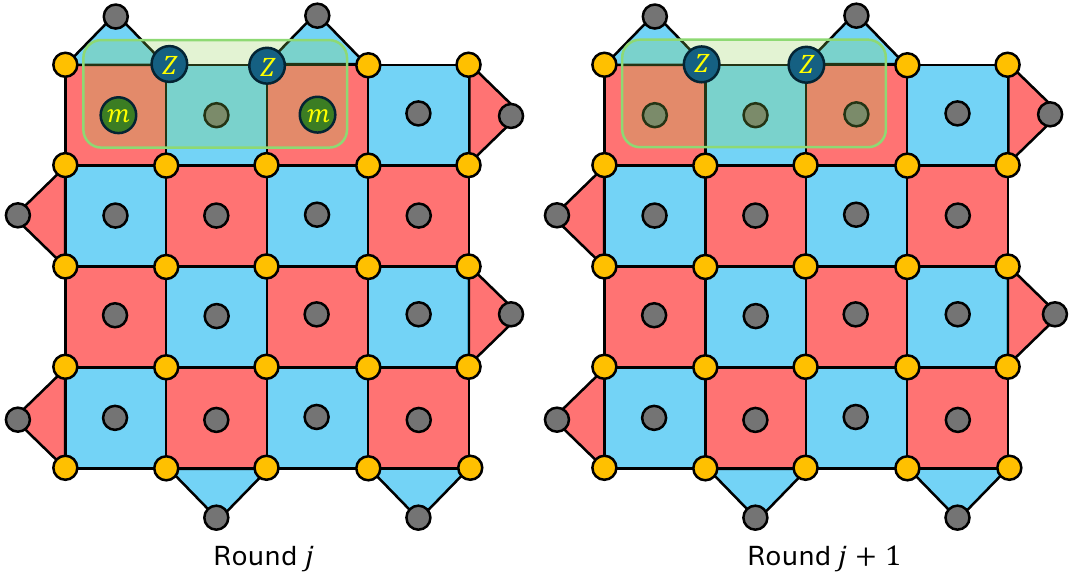}}
 \vfill
 \subfloat[\label{fig:Timelike_Homological_V2_High} ]{\includegraphics[width=.5\textwidth]{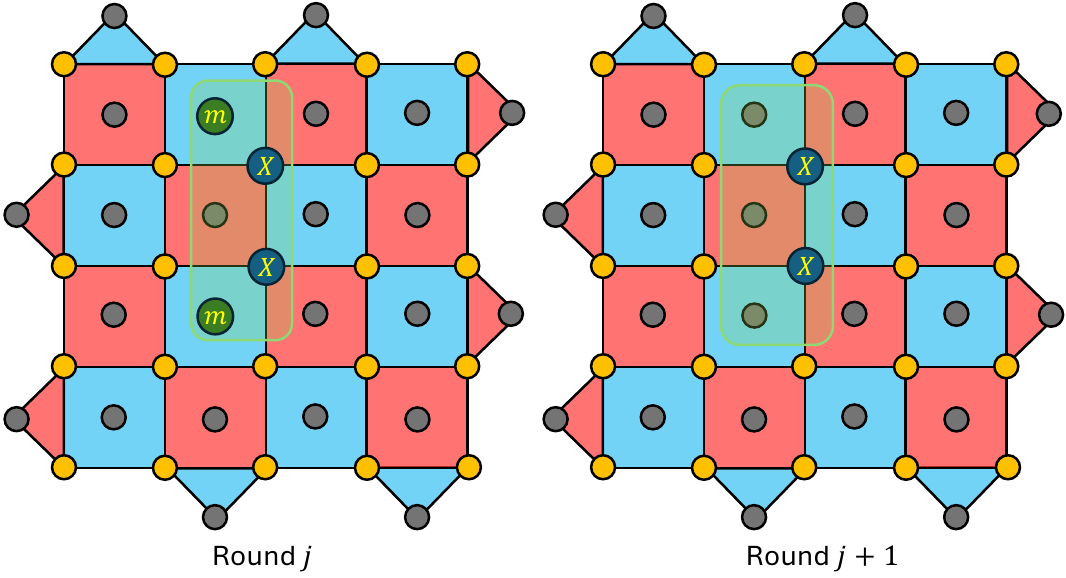}}
     \caption{ Timelike homological equivalence convention for a $d=5$ surface code for weight-two errors arising from a single fault. (a) For each weight-four $Z$-type stabilizer, after applying the \texttt{fixEquivalenceZ} function in two consecutive rounds, add a horizontal weight-two $Z$ error in the direction set by \texttt{fixEquivalenceZ} in two consecutive syndrome measurement rounds, along with measurement errors on $X$-type stabilizers that anticommute with the $Z$ errors in the first round the $Z$ errors are introduced. Apply such corrections to \texttt{trainY}. If the number of 1's in \texttt{trainY} is reduced, we accept the trivial correction. (b) Same as (a) but with $X$ corrections, and where the weight-two $X$ errors are added in the vertical direction. }
     \label{fig:Timelike_Homological_High}
 \end{figure*}

Many error configurations acting on data qubits are physically equivalent. We say that two Pauli errors $E_1$ and $E_2$ are homologically equivalent if there exists a stabilizer $g \in \mathcal{S}$ such that
\begin{align}
    E_1 = gE_2,
\end{align}
where $\mathcal{S}$ denotes the stabilizer group of the surface code. In order to reduce the complexity of the labeled training data and thereby improve training performance, we fix a canonical choice of representative within each homological equivalence class. In what follows, all transformations are chosen to preserve the induced syndrome history and the logical equivalence class of the error.

We first describe a \textbf{spacelike homological equivalence protocol}, closely following Ref.~\cite{ChambsLocalNN22}. We then introduce a complementary \textbf{timelike homological equivalence protocol} that simplifies label structure across consecutive syndrome measurement rounds.

For the spacelike protocol, consider a weight-four $X$-type stabilizer $g_k(X)$, represented by a red plaquette in \cref{fig:Homological_equiv}. Any weight-three $X$ error $E_3$ supported on $g_k(X)$ can be reduced to a weight-one error by multiplying by the stabilizer, i.e., by forming $g_k(X)E_3$. Similarly, a weight-four $X$ error supported on $g_k(X)$ is equivalent to $g_k(X)$ itself and can therefore be removed entirely. We define the function \texttt{weightReductionX} to apply these weight-reduction transformations across all relevant stabilizers. In addition, \texttt{weightReductionX} removes weight-two $X$ errors supported on weight-two $X$ stabilizers along the left and right boundaries of the surface-code patch.

Next, let $E_x$ be a weight-two $X$ error supported on a weight-four stabilizer $g_k(X)$ whose top-left data qubit has coordinates $(\alpha,\beta)$ on the $D \times D$ grid (with $\alpha$ denoting the row index and $\beta$ the column index). We define \texttt{fixEquivalenceX} via the following canonicalization rules:
\begin{itemize}
\item \textbf{Vertical $X$ chain: }If $E_x$ has support on $(\alpha,\beta)$ and $(\alpha+1,\beta)$, then \texttt{fixEquivalenceX} maps $E_x$ to support on $(\alpha,\beta+1)$ and $(\alpha+1,\beta+1)$.
\item \textbf{Horizontal $X$ chain: }If $E_x$ has support on $(\alpha+1,\beta)$ and $(\alpha+1,\beta+1)$, then \texttt{fixEquivalenceX} maps $E_x$ to support on $(\alpha,\beta)$ and $(\alpha,\beta+1)$.
\item \textbf{Diagonal $X$ chain: }If $E_x$ has support on $(\alpha,\beta)$ and $(\alpha+1,\beta+1)$, then \texttt{fixEquivalenceX} maps $E_x$ to support on $(\alpha,\beta+1)$ and $(\alpha+1,\beta)$.
\end{itemize}

Boundary stabilizers require special handling. Let $g_k(X)$ be a weight-two $X$ stabilizer along the left boundary, with the top-most qubit in its support at coordinates $(\alpha,\beta)$. If $E_x$ is a weight-one error at $(\alpha+1,\beta)$, then \texttt{fixEquivalenceX} maps it to $(\alpha,\beta)$. Conversely, if $g_k(X)$ is a weight-two $X$ stabilizer along the right boundary with top-most qubit at $(\alpha,\beta)$, then a weight-one error at $(\alpha,\beta)$ is mapped to $(\alpha+1,\beta)$. These mappings are illustrated on the left side of \cref{fig:Homological_equiv}.

We now define \texttt{simplifyX} to apply \texttt{weightReductionX} followed by \texttt{fixEquivalenceX} across all $X$-type stabilizers. The function \texttt{simplifyX} is applied iteratively until convergence. Specifically, let $M_e^{(X_{\alpha,\beta})}(j)$ be the binary matrix representing $X$ errors in syndrome measurement round $j$, where $M_e^{(X_{\alpha,\beta})}(j)=1$ indicates an $X$ error on the data qubit at $(\alpha,\beta)$ and $0$ otherwise. We apply \texttt{simplifyX} until
\begin{align}
    \texttt{simplifyX}(M_e^{(X_{\alpha,\beta})}(j)) = M_e^{(X_{\alpha,\beta})}(j),
\end{align}
for all $1 \le j \le d_m$ and all coordinates $(\alpha,\beta)$ on the $D \times D$ grid.

For $Z$-type data-qubit errors, we define \texttt{weightReductionZ} analogously. Let $E_z$ be a weight-two $Z$ error supported on a weight-four $Z$ stabilizer $g_k(Z)$ whose top-left data qubit has coordinates $(\alpha,\beta)$. The function \texttt{fixEquivalenceZ} implements the transformations:
\begin{itemize}
\item \textbf{Vertical chain: }If $E_z$ has support on $(\alpha,\beta)$ and $(\alpha+1,\beta)$, then \texttt{fixEquivalenceZ} maps it to $(\alpha,\beta+1)$ and $(\alpha+1,\beta+1)$.
\item \textbf{Horizontal chain: }If $E_z$ has support on $(\alpha+1,\beta)$ and $(\alpha+1,\beta+1)$, then \texttt{fixEquivalenceZ} maps it to $(\alpha,\beta)$ and $(\alpha,\beta+1)$.
\item \textbf{Diagonal chain: }If $E_z$ has support on $(\alpha,\beta+1)$ and $(\alpha+1,\beta)$, then \texttt{fixEquivalenceZ} maps it to $(\alpha,\beta)$ and $(\alpha+1,\beta+1)$.
\end{itemize}
For boundary weight-two $Z$ stabilizers, if $g_k(Z)$ lies along the top boundary with left-most qubit at $(\alpha,\beta)$, then a weight-one error at $(\alpha,\beta)$ is mapped to $(\alpha,\beta+1)$. If $g_k(Z)$ lies along the bottom boundary with left-most qubit at $(\alpha,\beta)$, then a weight-one error at $(\alpha,\beta+1)$ is mapped to $(\alpha,\beta)$. These mappings are shown on the right side of \cref{fig:Homological_equiv}.

We then define \texttt{simplifyZ} to apply \texttt{weightReductionZ} followed by \texttt{fixEquivalenceZ}, iterating until a $Z$-error steady state is reached. 

After applying the spacelike homological equivalence protocol independently to all syndrome measurement rounds, we apply a timelike homological equivalence protocol that simplifies label structure across consecutive rounds. Suppose there are $d_m$ syndrome measurement rounds and $d^2$ data qubits. Let $t$ index the training sample, with $1 \le t \le N_{\text{train}}$. For consecutive rounds $k$ and $k+1$, we define
\begin{align}
t^{(1)}_{Y_1}(k) &= \texttt{trainY}(t,j_1^{(1)},j_1^{(2)},k,1), \label{eq:tY1k1} \\
     t^{(3)}_{Y_1}(k) &= \texttt{trainY}(t,s^{(j_1)}_x,s^{(j_1)}_y,k,3), \label{eq:tY1k3} \\
     t^{(3)}_{Y_2}(k) &= \texttt{trainY}(t,s^{(j_2)}_x,s^{(j_2)}_y,k,3), \label{eq:tY2k3} \\    
     t^{(1)}_{p_{Y_1}}(k) &= \texttt{trainY}(t,j_1^{(1)},j_1^{(2)},k,1) \oplus 1, \label{eq:tpY1k1}\\
     t^{(3)}_{p_{Y_1}}(k) &= \texttt{trainY}(t,s^{(j_1)}_x,s^{(j_1)}_y,k,3) \oplus 1, \label{eq:tpY1k3} \\
     t^{(3)}_{p_{Y_2}}(k) &= \texttt{trainY}(t,s^{(j_2)}_x,s^{(j_2)}_y,k,3) \oplus 1.
     \label{eq:tpY2k3}
\end{align}
where $(j_1^{(1)},j_1^{(2)})$ are the coordinates of a data qubit $q_j^{(1)}$, and the coordinates of stabilizers that anticommute with $q_j^{(1)}$ are $(s^{(j_1)}_x,s^{(j_1)}_y)$ and $(s^{(j_2)}_x,s^{(j_2)}_y)$. If only a single stabilizer anticommutes with $q_j^{(1)}$, we set $t^{(3)}_{Y_2}(k)=0$ and $t^{(3)}_{p_{Y_2}}(k)=0$.
 
We further define
\begin{align}
     s_Y(k) &= t^{(1)}_{Y_1}(k) + t^{(3)}_{Y_1}(k) + t^{(3)}_{Y_2}(k), \\
     s_Y(k+1) &= t^{(1)}_{Y_1}(k+1) + t^{(3)}_{Y_1}(k+1) \nonumber \\
     &+ t^{(3)}_{Y_2}(k+1), \\
     s_{p_Y}(k) &= t^{(1)}_{p_{Y_1}}(k) + t^{(3)}_{p_{Y_1}}(k) + t^{(3)}_{p_{Y_2}}(k), \\
     s_{p_Y}(k+1) &= t^{(1)}_{p_{Y_1}}(k+1) + t^{(3)}_{Y_1}(k+1) \nonumber \\
     &+ t^{(3)}_{Y_2}(k+1), \label{eq:spykp1} 
\end{align}
as well as
\begin{align}
     s_X(k) &= \texttt{trainX}(t,s^{(j_1)}_x,s^{(j_1)}_y,k,1) \nonumber \\
     &+ \texttt{trainX}(t,s^{(j_2)}_x,s^{(j_2)}_y,k,1).
\end{align}
Note that in \cref{eq:spykp1}, the last two terms involve $t^{(3)}_{Y_1}(k+1)$ and $t^{(3)}_{Y_2}(k+1)$ rather than $t^{(3)}_{p_{Y_1}}(k+1)$ and $t^{(3)}_{p_{Y_2}}(k+1)$; see \cref{fig:Timelike_Homological} for intuition. This is because the candidate correction adds a data-qubit error to rounds $k$ and $k+1$ together with associated stabilizer measurement errors only in round $k$---the round where the error is first introduced. Since no additional measurement errors are appended at round $k+1$, the timelike labels $t^{(3)}_{Y_1}(k+1)$ and $t^{(3)}_{Y_2}(k+1)$ enter the cost sum unflipped.
 
Finally, we define
\begin{align}
     s_{\text{max}} &= \text{max}(s_Y(k) + s_X(k), s_Y(k+1) \nonumber \\
     &+ s_X(k+1)), \label{eq:smax} \\
     s^{(\text{HE})}_{\text{max}} &=\text{max}(s_{p_Y}(k) + s_X(k), s_{p_Y}(k+1) \nonumber \\
     &+ s_X(k+1)), \label{eq:smaxHE} \\
     s(k,k+1) &= s_Y(k) + s_X(k) + s_Y(k+1) \nonumber \\
     &+ s_X(k+1), \label{eq:skcalc} \\
     s^{(\text{HE})}(k,k+1) &= s_{p_Y}(k) + s_X(k) + s_{p_Y}(k+1) \nonumber \\
     &+ s_X(k+1).
     \label{eq:skcalcHE}
\end{align}

The timelike homological equivalence protocol for single data-qubit $Z$ corrections is given in \cref{Algo:TimelikeHomologicalEquivZ}. The corresponding protocol for $X$ corrections is obtained by replacing channels $(1,3)$ of \texttt{trainY} with channels $(2,4)$ in \cref{eq:tY1k1,eq:tY1k3,eq:tY2k3,eq:tpY1k1,eq:tpY1k3,eq:tpY2k3}.
\begin{algorithm}[H]
\caption{Timelike homological equivalence $Z$}
\begin{algorithmic}
\For{$k = 1$ to $d_m - 1$}
     \For{$j = 1$ to $d^2$}
         \State Let $q_j$ be a data qubit on the $d \times d$ grid with coordinates $(j_x,j_y)$.
         \State Determine the set $\mathcal{S}_j$ of stabilizers that anticommute with a $Z$ error on $q_j$.
         \If{$|\mathcal{S}_j| = 1$}
             \State Let the stabilizer coordinates be $(s^{(j_1)}_x,s^{(j_1)}_y)$.
             \State Set $t^{(3)}_{Y_2}(k) = 0$ and $t^{(3)}_{p_{Y_2}}(k) = 0$.
         \ElsIf{$|\mathcal{S}_j| = 2$}
             \State Let the stabilizer coordinates be $(s^{(j_1)}_x,s^{(j_1)}_y)$ and $(s^{(j_2)}_x,s^{(j_2)}_y)$.
         \EndIf
          \State Compute $s_{\text{max}}$ and $s^{(\text{HE})}_{\text{max}}$.
          \Statex \quad using Eqs.~\protect\ref{eq:smax} and \protect\ref{eq:smaxHE}.
 
          \State Compute $s(k,k+1)$ and $s^{(\text{HE})}(k,k+1)$.
          \Statex \quad using Eqs.~\protect\ref{eq:skcalc} and \protect\ref{eq:skcalcHE}.
 
         \If{$s^{(\text{HE})}(k,k+1) < s(k,k+1)$}
             \State Set $\texttt{trainY}(t,j_1^{(1)},j_1^{(2)},k,1) = t^{(1)}_{p_{Y_1}}(k)$.
             \State Set $\texttt{trainY}(t,s^{(j_1)}_x,s^{(j_1)}_y,k,3) = t^{(3)}_{p_{Y_1}}(k)$.
             \State Set $\texttt{trainY}(t,s^{(j_2)}_x,s^{(j_2)}_y,k,3) = t^{(3)}_{p_{Y_2}}(k)$.
             \State Set $\texttt{trainY}(t,j_1^{(1)},j_1^{(2)},k+1,1) = t^{(1)}_{p_{Y_1}}(k+1)$.
         \ElsIf{$s^{(\text{HE})}(k,k+1) = s(k,k+1)$}
             \If{$s^{(\text{HE})}_{\text{max}} > s_{\text{max}}$}
                 \State Set $\texttt{trainY}(t,j_1^{(1)},j_1^{(2)},k,1) = t^{(1)}_{p_{Y_1}}(k)$.
                 \State Set $\texttt{trainY}(t,s^{(j_1)}_x,s^{(j_1)}_y,k,3) = t^{(3)}_{p_{Y_1}}(k)$.
                 \State Set $\texttt{trainY}(t,s^{(j_2)}_x,s^{(j_2)}_y,k,3) = t^{(3)}_{p_{Y_2}}(k)$.
                 \State Set $\texttt{trainY}(t,j_1^{(1)},j_1^{(2)},k+1,1) = t^{(1)}_{p_{Y_1}}(k+1)$.
             \Else
                 \State Leave $\texttt{trainY}$ unchanged.
             \EndIf
         \Else
             \State Leave $\texttt{trainY}$ unchanged.
         \EndIf
     \EndFor
 \EndFor
 \State Repeat the above until the number of 1's in $\texttt{trainY}$ is no longer reduced.
 \end{algorithmic}
 \label{Algo:TimelikeHomologicalEquivZ}
 \end{algorithm}
An illustration of \cref{Algo:TimelikeHomologicalEquivZ} is shown in \cref{fig:Timelike_Homological}. Intuitively, applying an $X$ or $Z$ error to the same data qubit in two consecutive rounds—together with measurement errors on stabilizers that anticommute with the added error in the first of the two rounds—can correspond to a trivial operation, since no net syndrome change is registered. Exploiting this freedom can simplify \texttt{trainY} by introducing additional structure that is easier for CNNs to learn.

Without this simplification, an error that is introduced in round $k$ but masked by measurement errors (and therefore detected only in round $k+1$) would still appear as a label in \texttt{trainY} at round $k$. This can encourage the network to apply corrections in an incorrect round, leading to residual timelike failures that are then passed to the global decoder.

\cref{Algo:TimelikeHomologicalEquivZ} focuses on single data-qubit errors across two consecutive rounds. Since weight-two data-qubit errors can also arise from a single fault, we additionally consider a weight-two extension of the protocol, in which all weight-two $Z$ (or $X$) errors arising from a single fault are included. An illustration of this extension is shown in \cref{fig:Timelike_Homological_High}.

\begin{figure}
     \centering
 \subfloat[\label{fig:HomologicalSequenceEX} ]{\includegraphics[width=.5\textwidth]{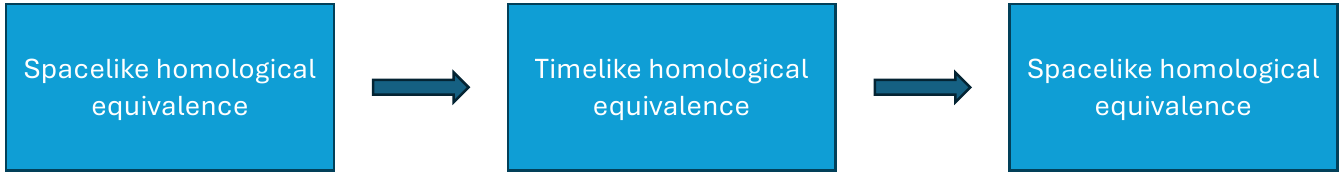}}
     \caption{Sequence of operations for the complete homological equivalence protocol. We first apply the spacelike homological equivalence protocol, followed by the timelike homological equivalence protocol (for weight-one errors), and finally reapply the spacelike protocol as a cleanup step.}
     \label{fig:HomologicalSequence}
\end{figure}

The complete homological equivalence protocol therefore combines the spacelike and timelike transformations in an iterative scheme. We first apply spacelike homological equivalence to all rounds, then apply timelike homological equivalence for weight-one data-qubit errors. Since timelike transformations can create new opportunities for spacelike simplification, we perform a final spacelike pass as a cleanup step. This sequence is illustrated in \cref{fig:HomologicalSequence}.
 
Finally, we note that many alternative choices of homological equivalence functions are possible; see, for example, the discussion of simplifier operations in Ref.~\cite{Australia3DConvPred}.

\subsubsection{Loss function}
\label{subsec:LossFunction}

To train the pre-decoder networks, we use a binary cross-entropy (BCE) objective, since the model predicts independent per-voxel probabilities for spacelike Pauli corrections and timelike syndrome flips. Concretely, the network produces four output channels and we apply a sigmoid nonlinearity to each channel to obtain probabilities in $[0,1]$.

For a surface-code patch on a $D \times D$ grid with $d_m$ syndrome measurement rounds, let the ground-truth labels $Y$ and model outputs $\hat{Y}$ be
\begin{align}
    Y &\in \{0,1\}^{4 \times D \times D \times d_m}, \\
    \hat{Y} &\in [0,1]^{4 \times D \times D \times d_m}.
\end{align}

The loss is computed as a sum of BCE terms over all channels and voxels,
\begin{align}
\mathcal{L}_{\mathrm{BCE}}(Y,\hat{Y})
&= \sum_{c=1}^{4} \sum_{\alpha=1}^{D} \sum_{\beta=1}^{D} \sum_{k=1}^{d_m}
\Big[-Y_{c,\alpha,\beta,k}\log(\hat{Y}_{c,\alpha,\beta,k})
\nonumber \\
&-(1-Y_{c,\alpha,\beta,k})\log(1-\hat{Y}_{c,\alpha,\beta,k})\Big],
\label{eq:BCE_loss}
\end{align}
which corresponds to one BCE loss per voxel per channel, for a total of $4D^2 d_m$ terms.
 
\subsubsection{Inference step}
\label{subsec:Inference}

We now describe the inference procedure for a trained pre-decoder obtained using the methods of \cref{subsec:NNArchHyperParam}. Given syndrome data formatted as \texttt{trainX}, the pre-decoder predicts local spacelike and timelike corrections, which are then used to modify the syndrome history before passing it to a global decoder.

Let \texttt{out} denote the output tensor of the trained pre-decoder. For the $j$th shot and $k$th syndrome measurement round, the predicted spacelike corrections on the $D \times D$ grid are
\begin{align}
     Z_{\text{corr}}^{(j,k)} &= \texttt{out}(j,1{:}D,1{:}D,k,1), \label{eq:ZcorreJK} \\
     X_{\text{corr}}^{(j,k)} &= \texttt{out}(j,1{:}D,1{:}D,k,2).
     \label{eq:XcorreJK}
\end{align}
and the predicted timelike stabilizer corrections are
\begin{align}
     \text{SynX}_{\text{corr}}^{(j,k)} &= \texttt{out}(j,1{:}D,1{:}D,k,3), \label{SynXCorrJK} \\
     \text{SynZ}_{\text{corr}}^{(j,k)} &= \texttt{out}(j,1{:}D,1{:}D,k,4).
     \label{SynZCorrJK}
\end{align}

Let $\text{SynX}^{(j,k)}$ and $\text{SynZ}^{(j,k)}$ denote the measured detector events for $X$- and $Z$-type stabilizers in round $k$ during inference. The syndromes induced by the predicted spacelike corrections are
\begin{align}
     S^{(j,k)}_X &= M_X\Big( Z_{\text{corr}}^{(j,k)} \Big), \label{eq:SXJK} \\
     S^{(j,k)}_Z &= M_Z\Big( X_{\text{corr}}^{(j,k)} \Big),
     \label{eq:SZJK}
\end{align}
where $M_X$ and $M_Z$ map data-qubit Pauli errors to the corresponding $X$- and $Z$-stabilizer syndromes.

If $\text{SynX}_{\text{corr}}^{(j,k)}(l)=1$, the measurement outcome of the $l$-th $X$ stabilizer is flipped in both rounds $k$ and $k+1$. Similarly, if $\text{SynZ}_{\text{corr}}^{(j,k)}(l)=1$, the outcome of the $l$-th $Z$ stabilizer is flipped in rounds $k$ and $k+1$. This implements the timelike correction predicted by the pre-decoder.

After applying both spacelike and timelike corrections, the residual syndromes passed to the global decoder are
\begin{align}
     R^{(j,1)}(X) &= \text{SynX}^{(j,1)} \oplus \text{SynX}_{\text{corr}}^{(j,1)} \oplus S^{(j,1)}_X, \label{eq:Resk1X} \\
     R^{(j,k>1)}(X) &= \text{SynX}^{(j,k)} \oplus \text{SynX}_{\text{corr}}^{(j,k)} \nonumber \\
     &\oplus \text{SynX}_{\text{corr}}^{(j,k-1)} \oplus S^{(j,k)}_X, \label{eq:ReskG1X} \\
     R^{(j,1)}(Z) &= \text{SynZ}^{(j,1)} \oplus \text{SynZ}_{\text{corr}}^{(j,1)} \oplus S^{(j,1)}_Z, \label{eq:Resk1Z} \\
     R^{(j,k>1)}(Z) &= \text{SynZ}^{(j,k)} \oplus \text{SynZ}_{\text{corr}}^{(j,k)} \nonumber \\
     &\oplus \text{SynZ}_{\text{corr}}^{(j,k-1)} \oplus S^{(j,k)}_Z.
     \label{eq:ReskG1Z}
\end{align}

Let $E^{(j,k)}(X)$ and $E^{(j,k)}(Z)$ denote the $X$- and $Z$-type data-qubit errors introduced during round $k$ (excluding accumulated errors from earlier rounds). The residual spacelike errors after applying the pre-decoder corrections are
\begin{align}
 R^{(j,k)}_e(Z) &= Z_{\text{corr}}^{(j,k)} \oplus E^{(j,k)}(Z), \\
 R^{(j,k)}_e(X) &= X_{\text{corr}}^{(j,k)} \oplus E^{(j,k)}(X).
\end{align}

Let $C^{(j,k)}(X)$ and $C^{(j,k)}(Z)$ denote the $X$- and $Z$-type corrections applied by the global algorithmic decoder in round $k$, computed from the residual syndromes in \cref{eq:Resk1X,eq:ReskG1X,eq:Resk1Z,eq:ReskG1Z}. The total accumulated corrections are
\begin{align}
L^{(j)}(X) &= \bigoplus_{k=1}^{d_m} \left[C^{(j,k)}(X) \oplus R^{(j,k)}_e(X)\right], \\
L^{(j)}(Z) &= \bigoplus_{k=1}^{d_m} \left[C^{(j,k)}(Z) \oplus R^{(j,k)}_e(Z)\right].
\end{align}

A logical $X$ ($Z$) error is said to have occurred if $L^{(j)}(X)$ ($L^{(j)}(Z)$) anticommutes with the logical operator $Z_L$ ($X_L$) of the $D \times D$ surface-code patch.

\section{Noise learning architecture from syndrome statistics}
\label{sec:EffectivePreDecNoiseModel}

\begin{figure*}
     \centering
 \subfloat[\label{fig:NoiseLearnArch} ]{\includegraphics[width=.8\textwidth]{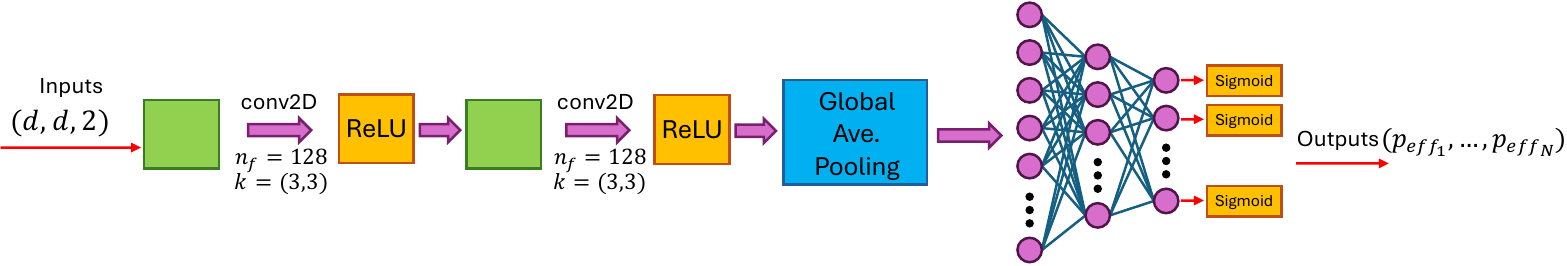}}
 \caption{ Architecture for learning the circuit-level noise parameters of the gates used to implement the surface code. Two-dimensional convolutional layers extract local spatial features from two consecutive syndrome-measurement rounds mapped to a 2D grid following the procedure in \cref{fig:StabMap}. A global average pooling layer aggregates these features into global statistics that capture syndrome-motif frequencies. The final MLP head maps these global features to the estimated noise parameters for each circuit-level component. }
 \label{fig:NoiseLearnArch}
\end{figure*}

When operating a quantum device, it is not always possible to fully characterize the underlying circuit-level noise model with sufficient accuracy to compute optimal decoding weights. In practice, noise processes may be partially unknown, drift over time, or deviate from simplified assumptions used in simulations. However, syndrome measurement data from repeated stabilizer rounds is experimentally accessible and contains statistical information about the effective error processes affecting the code. This motivates approaches that infer decoding parameters directly from syndrome statistics rather than relying on an explicit circuit-level noise model.

When a pre-decoder is applied to measured syndrome data, the resulting residual syndromes passed to the global decoder are modified according to \cref{eq:Resk1X,eq:ReskG1X,eq:Resk1Z,eq:ReskG1Z}. As a result, the statistics of the residual syndromes are governed by an effective noise model that generally differs from the original circuit-level model used to generate the physical errors. Global decoders such as PyMatching compute matching-graph edge weights using probabilities derived from an assumed noise model \cite{HiggottPyMatch}. If the effective noise statistics differ from those assumed by the decoder, the resulting edge weights may be suboptimal.

In this section, we introduce a neural network architecture that learns the effective noise parameters required to compute near-optimal edge weights and correlation structure for PyMatching directly from syndrome statistics of two consecutive bulk measurement rounds. The learned parameters support both standard (uncorrelated) matching and correlated matching, which incorporates hyperedge information through two-pass reweighting. During training, the network is provided with syndrome data generated from a known circuit-level noise model. At inference time, the trained network can be applied to experimentally obtained syndrome statistics—or to the residual syndromes produced by a pre-decoder—to estimate the corresponding effective noise model. These learned probabilities can then be used to construct the detector error model supplied to PyMatching.

A key observation enabling this approach is that the probability formulas for both edges and hyperedges in the surface code matching graph are independent of code distance. For both the $X$- and $Z$-stabilizer matching graphs, there are 18 distinct edge types and 43 distinct hyperedge type compositions whose probability expressions are identical for all code distances $d \ge 5$ (see \cref{app:EdgeWeights}). While the number of instances of each type scales with the code distance, their functional dependence on the underlying noise parameters does not.

This distance-independence, combined with the use of global average pooling in our neural network architecture, allows the noise-learning model to be trained at a single code distance and to generalize to arbitrary distances during inference.

\subsection{Architecture}
\label{subsec:NoiseLearnArchitecture}

An overview of the noise-learning architecture is shown in \cref{fig:NoiseLearnArch}. The input to the network consists of syndrome data from two consecutive bulk syndrome measurement rounds, mapped onto a two-dimensional grid using the same conventions described in \cref{subsec:InputTrain} and illustrated in \cref{fig:StabMap}. The input tensor has shape $(B, 4, 2, D, D)$, where $B$ is the number of syndrome samples, the 4 channels correspond to the encodings defined in \cref{eq:encTrainX1,eq:encTrainX2,eq:encTrainX3,eq:encTrainX4}, and the two rounds are extracted from the bulk (middle) portion of a $d_m$-round experiment to avoid temporal boundary effects from initialization and final measurement.

The architecture consists of three stages:

\textbf{Convolutional feature extractor.} A 4-layer 2D CNN processes each syndrome pair independently. The input channels ($4 \times 2 = 8$ after reshaping) are processed through layers with filter counts $[128, 256, 256, 128]$, each using $3 \times 3$ kernels with padding to preserve spatial dimensions. We use GroupNorm (32 groups) for normalization and GeLU activations. Dropout ($p=0.1$) is applied only to the final CNN layer.

\textbf{Global average pooling.} The output of the final CNN layer, $H \in \mathbb{R}^{128 \times D \times D}$, is reduced to a 128-dimensional feature vector by averaging over all spatial positions:
\begin{align}
g_c = \frac{1}{D^2} \sum_{x,y} H_{c,x,y}.
\end{align}
This operation is distance-preserving: the pooled features have the same dimensionality regardless of code distance $d$, enabling a single trained model to generalize across distances.

\textbf{MLP prediction head with post-MLP averaging.} We apply a 3-layer MLP independently to each sample's pooled features, producing per-sample logits:
\begin{align}
\boldsymbol{z}_k = \text{MLP}(\boldsymbol{g}_k) \in \mathbb{R}^{25},
\end{align}
where the MLP has hidden dimensions $[256, 128]$ with GeLU activations and dropout ($p=0.2$). The logits are then averaged across the batch:
\begin{align}
\bar{\boldsymbol{z}} = \frac{1}{B} \sum_{k=1}^{B} \boldsymbol{z}_k.
\label{eq:PostMLPAvg}
\end{align}

Finally, the averaged logits are mapped to noise parameters using a bounded log-space transformation:
\begin{align}
\hat{p}_i = \exp\!\Big(\log p_{\min}' + \big(\log p_{\max}' - \log p_{\min}'\big) \cdot \sigma(\bar{z}_i)\Big),
\label{eq:BoundedLogSpace}
\end{align}
where $\sigma$ is the sigmoid function, $p_{\min}' = p_{\min}/100$, and $p_{\max}' = 3 \, p_{\max}$, with $p_{\min} = 10^{-3}$ and $p_{\max} = 10^{-2}$. The extended bounds account for the fact that individual noise parameters (e.g., individual CNOT Pauli channels) can be significantly smaller or larger than the base error rate. This log-space parameterization enables the network to naturally span multiple orders of magnitude in probability values while ensuring all predictions lie in a valid range.

The post-MLP averaging in \cref{eq:PostMLPAvg} allows each syndrome sample to contribute its own parameter estimate in logit space before aggregation. During training, $B$ is the batch size; during inference, $B = N_{\text{test}}$ where $N_{\text{test}} \gg 1$ syndrome pairs are used for reliable estimation. The network is trained using the same aggregation procedure used during inference, eliminating train--test mismatch.

\subsection{Edge and hyperedge probability formulas}
\label{subsec:EdgeHyperedgeFormulas}

The matching graph used by PyMatching contains edges connecting pairs of detectors that could arise from the same error, as well as hyperedges representing correlated multi-detector events that decompose into pairs of edges. To compute both edge weights (for standard matching) and conditional probabilities (for correlated matching), we derive closed-form probability formulas for all edge and hyperedge types as functions of the 25 noise parameters.

\textbf{Edge formulas.} By systematically activating each single-Pauli error in the circuit and tracing which detector pairs it flips, we identify all error mechanisms contributing to each edge. When multiple independent mechanisms flip the same detector pair, their probabilities combine via XOR:
\begin{align}
P_1 \oplus P_2 = P_1 + P_2 - 2P_1 P_2.
\label{eq:XORcombine}
\end{align}
Each edge probability is thus expressed as an XOR combination of sums of noise parameters. For both the $X$- and $Z$-stabilizer matching graphs, this analysis yields \textbf{18 distinct edge types}: 3 spacelike, 4 timelike, 5 diagonal, and 6 boundary types. These formulas are distance-independent: the same expressions apply for all $d \ge 5$, with only the instance count of each type scaling with distance.

\textbf{Hyperedge formulas.} When Stim generates a detector error model with \texttt{decompose\_errors=True}, correlated multi-detector events are decomposed into pairs of edges separated by the \texttt{\^{}} operator. PyMatching uses these decomposed hyperedges for correlated two-pass matching, where conditional probabilities $P(E_2 \mid E_1) = P_{\text{joint}} / P(E_1)$ are used to reweight edges in a second pass after an initial matching solution.

Using the same single-error tracing methodology as for edges, we identify all error mechanisms that produce each decomposed hyperedge pattern. The joint probability of each hyperedge is computed as the XOR combination of contributing error probabilities. Classifying hyperedges by their component edge types yields \textbf{43 distinct type compositions}. These formulas are distance-independent: all 86 types derived at $d=5$ cover all hyperedge types observed at $d = 5, 7, 9, 11, 21$, and $31$. The formulas are verified against Stim's detector error model.

\subsection{Loss function}
\label{subsec:LossFuncCompEffectiveNoise}

The noise-learning network predicts parameters $\hat{\boldsymbol{p}}$ from which we compute predicted edge and hyperedge probabilities. The loss function combines contributions from both edge and hyperedge loss functions as
\begin{align}
\mathcal{L} = \mathcal{L}_{\text{edge}} + \mathcal{L}_{\text{hyper}}.
\label{eq:CombinedLoss}
\end{align}
The edge loss is a count-weighted MSE over the $N_e = 18$ edge types for the relevant basis, and the hyperedge loss is a count-weighted MSE over the $N_h = 43$ hyperedge type compositions:
\begin{align}
\mathcal{L}_{\text{edge}} = \sum_{j=1}^{N_e} c_j \big(\hat{P}_{e_j} - P_{e_j}\big)^2,
\label{eq:EdgeLoss}
\end{align}
\begin{align}
\mathcal{L}_{\text{hyper}} = \sum_{k=1}^{N_h} d_k \big(\hat{H}_k - H_k\big)^2,
\label{eq:HyperedgeLoss}
\end{align}
where $c_j$ and $d_k$ denote instance counts for edges and hyperedges, and $P_{e_j} = \mathcal{E}_j(\boldsymbol{p})$ and $H_k = \mathcal{H}_k(\boldsymbol{p})$ are the ground-truth probabilities computed from the known noise parameters (see \cref{app:EdgeWeights}). Because all XOR formulas involve only additions and multiplications, both $\mathcal{E}_j$ and $\mathcal{H}_k$ are fully differentiable, enabling end-to-end gradient-based training.

During training, the base error rate is sampled from a log-uniform distribution over $[p_{\text{min}},p_{\text{max}}]$. With this sampling, terms in the loss functions can be biased towards sampled values closer to $p_{\text{max}}$. To correct for this, we introduce a variance-stabilizing weight
\begin{align}
    w(p) = \Big( \frac{p_0}{p} \Big)^2,
\end{align}
with $p_0 = \sqrt{p_{\text{min}} \cdot p_{\text{max}}}$ the geometric mean, yielding the unbiased edge and hyperedge losses:
\begin{align}
    \mathcal{L}_{\text{edge}} &= w(p)\sum_{j=1}^{N_e} c_j \cdot \left( \hat{P}_{e_j} - P_{e_j} \right)^2, \label{eq:EdgeLoss_v1} 
\end{align}
\begin{align}
    \mathcal{L}_{\text{hyper}} &= w(p)\sum_{k=1}^{N_h} d_k \cdot \left( \hat{H}_k - H_k \right)^2. \label{eq:HyperedgeLossv1} 
\end{align}

The inclusion of hyperedge terms serves two purposes: it provides the conditional probability information needed for correlated matching, and it acts as a beneficial regularizer by breaking the parameter degeneracy inherent in edge-only optimization. Empirically, the edge and hyperedge losses are naturally comparable in magnitude without any relative scaling, and no additional regularization is needed.

\subsection{Training procedure}
\label{subsec:TrainingProcedure}

The training data is generated on-the-fly using a GPU-accelerated Pauli frame simulator. Let $d$ be the surface code distance used to train the noise learning model. For each training step we do the following:
\begin{enumerate}[leftmargin=*]
\item Sample a base error rate $p_{\text{base}}$ from a log-uniform distribution over $[p_{\min}, p_{\max}]$, then derive the 25 noise parameters with location-specific random multipliers and random Pauli-type distributions (see \cref{subsec:NotationMethodology}).
\item Generate $B$ independent syndrome samples at the training distance $d$ using the sampled noise model.
\item For each sample $k$, compute $\boldsymbol{z}_k = \text{MLP}(\text{GAP}(\text{CNN}(\boldsymbol{x}_k)))$.
\item Average logits: $\bar{\boldsymbol{z}} = \frac{1}{B}\sum_k \boldsymbol{z}_k$, then $\hat{\boldsymbol{p}} = \text{BoundedLogSpace}(\bar{\boldsymbol{z}})$ via \cref{eq:BoundedLogSpace}.
\item Compute $\hat{P}_{e_j} = \mathcal{E}_j(\hat{\boldsymbol{p}})$ and $\hat{H}_k = \mathcal{H}_k(\hat{\boldsymbol{p}})$.
\item Minimize $\mathcal{L} = \mathcal{L}_{\text{edge}} + \mathcal{L}_{\text{hyper}}$ and backpropagate through the differentiable formulas.
\end{enumerate}

The hierarchical noise sampling ensures diverse training data spanning multiple orders of magnitude while maintaining physically reasonable correlations between parameters.

\subsection{Inference strategy}
\label{subsec:InferenceStrat}

At inference time, the trained network is applied to syndrome data produced by the pre-decoder. From any surface code experiment with $d_m \ge 3$ syndrome measurement rounds, we extract a pair of consecutive bulk rounds (avoiding the first and last rounds to exclude temporal boundary effects). These two rounds are formatted as the input tensor and fed through the network along with $N_{\text{test}} \gg 1$ shots, producing per-sample logits that are averaged and converted to noise parameters via \cref{eq:PostMLPAvg,eq:BoundedLogSpace}.

The learned parameters $\hat{\boldsymbol{p}}$ are used to construct a complete Stim circuit with the corresponding noise model, from which a detector error model is generated with \texttt{decompose\_errors=True} and \texttt{approximate\_disjoint\_errors=True}. This detector error model is then loaded into PyMatching, supporting both uncorrelated matching (using edge weights only) and correlated matching (using edge weights and hyperedge conditional probabilities).
 
\section{Numerical results and performance benchmarks}
\label{sec:Numerics}

\begin{table*}[t]
\centering
\footnotesize
\setlength{\tabcolsep}{4pt}
\renewcommand{\arraystretch}{0.95}
\begin{tabular}{|c|c|c|c|c|}
\hline
 & \texttt{num\_filters} & \texttt{kernel\_size} & RF size & \texttt{num\_params} \\
\hline
Model 1 & [128,128,128,4] & [3,3,3,3] & 9  & 912,272 \\
Model 2 & [256,256,256,4] & [3,3,3,3] & 9  & 3,595,012 \\
Model 3 & [128,128,128,4] & [5,5,5,5] & 17 & 4,224,388 \\
Model 4 & [128,128,128,128,128,4] & [3,3,3,3,3,3] & 13 & 1,797,764 \\
Model 5 & [256,256,256,256,256,4] & [3,3,3,3,3,3] & 13 & 7,134,468 \\
\hline
\end{tabular}
\caption{Pre-decoder models considered in this work. The size of the vectors used for \texttt{num\_filters} and \texttt{kernel\_size} indicate how many 3DConv layers are used. The entries in \texttt{num\_filters} and \texttt{kernel\_size} indicate the number of filters and kernel size used in that given layer. Note that if an entry in the $j$-th column of \texttt{kernel\_size} is $K$, a kernel size of $K \times K \times K$ is used in that layer. We use \cref{eq:ReceptiveFieldFormula} to compute the receptive field size. All models use stride 1 and no dilation.  }
\label{tab:models}
\end{table*}

In this section we present numerical results for the family of pre-decoder models summarized in \cref{tab:models}. All models are based on fully convolutional three-dimensional CNN architectures (see \cref{subsec:NNArchHyperParam}), in which successive layers extract increasingly higher-order features from the spatiotemporal syndrome volume. Early layers specialize in local, low-order patterns such as single-fault detection-event pairs or short timelike chains, while deeper layers hierarchically combine these primitives to represent more complex correlations arising from hook errors, bursts of measurement faults, and multi-fault space–time structures.

The number of filters in each convolutional layer controls the expressiveness of the local feature basis: wider layers allow multiple distinct syndrome motifs to be represented in parallel, increasing the network’s capacity to model diverse physical error mechanisms. The kernel size determines the spatial and temporal neighborhood over which features are computed. Small kernels enforce locality consistent with the fault-propagation structure of the surface code, while increased depth allows longer-range correlations to be assembled hierarchically.

The five models in \cref{tab:models} are designed to explore architectural tradeoffs between expressive power and pre-decoding runtimes. Increasing the number of filters (model width) generally improves representational capacity but increases the number of floating-point operations per convolution, leading to higher runtimes during inference. For example, Model 1 uses three hidden layers with 128 filters and $3\times3\times3$ kernels, yielding a relatively lightweight architecture with low runtimes but limited capacity. Model 2 increases the filter count to 256 per layer, resulting in roughly a four-fold increase in parameter count and GPU runtime, but with improved modeling capability.

Model 3 keeps the network width fixed while increasing the kernel size to $5\times5\times5$, expanding the receptive field from 9 to 17 lattice units. This allows longer-range space–time correlations to be captured earlier in the network, at the cost of substantially more parameters and slower convolutions. Models 4 and 5 instead increase network depth while retaining small kernels, thereby expanding the receptive field hierarchically while keeping each convolution computationally cheaper than a large-kernel alternative. These models therefore probe the tradeoff between deeper hierarchical feature extraction and inference speed.

Collectively, this suite of models spans multiple orthogonal architectural axes—width, depth, and kernel size—enabling a systematic assessment of how design choices affect logical error rate performance and GPU runtimes. Runtime results for each model are presented in \cref{subsec:GPURuntimes}.

\begin{table*}
\centering
\resizebox{\textwidth}{!}{%
\begin{tabular}{|c|c|}
\hline
\textbf{Hyperparameters} &  \textbf{Values} \\ \hline
Shots per epoch   &  67,108,864 \\ \hline
Number of epochs   & 100 \\ \hline
Batch size per GPU   & Epoch 1: 512, $\text{Epoch} \ge 2$: 2048  \\ \hline
Number of GPUs   & 8 \\ \hline
Optimizer   &  Lion: $\text{Weight decay} = 10^{-7}$, $\text{beta2} = 0.95$ \\ \hline
Learning rate schedule   & Warmup then decay (100 warmup steps). Apply $\gamma=0.7$ at milestones $[0.25,0.5,1.0]$  \\ \hline
Learning rates   & $\text{Model 1} = 3\times10^{-4}$, $\text{Model 2} = 2\times10^{-4}$, $\text{Model 3} = 1\times10^{-4}$, $\text{Model 4} = 2\times10^{-4}$, $\text{Model 5} = 1\times10^{-4}$  \\ \hline
Activation function   & GeLU (tanh approximation) \\ \hline
Dropout   & 0.05  \\ \hline
Exponential moving average (ema)   & $\text{decay} = 0.0001$  \\ \hline
 \end{tabular}
} 
\caption{ Hyperparameters used to train models 1 to 5 from \cref{tab:models}. The $\gamma=0.7$ is applied to the learning rate at milestones $[0.25,0.5,1.0]$. For instance, the first milestone 0.25 indicates that at $25\%$ of training steps, the learning rate becomes $0.7 \times \text{base}$. The tanh approximation of GeLU uses the function $\text{GeLU}(x) \approx 0.5 x (1 +\tanh{(\sqrt{2 / \pi}(x + 0.044715 x^3))})$.  }
\label{tab:hyperparameters}
\end{table*}

All pre-decoder models are trained using the hyperparameters listed in \cref{tab:hyperparameters}. Unless otherwise stated, simulations throughout this section employ the following depolarizing circuit-level noise model:

\begin{itemize}
\item A $|0\rangle$ ($|+\rangle$) state preparation is followed by an $X$ ($Z$) error with probability $2p/3$.
\item Prior to each $Z$ ($X$) basis measurement, an $X$ ($Z$) error occurs with probability $2p/3$.
\item With probability $p$, each two-qubit gate is followed by a two-qubit Pauli error drawn uniformly from $\{I,X,Y,Z\}^{\otimes 2} \setminus \{I\otimes I\}$.
\item During idle locations associated with either CNOT gates or state-preparation and measurement, a Pauli error is drawn uniformly from $\{X,Y,Z\}$ with probability $p$.
\end{itemize}

When applying the homological equivalence scheme in \cref{fig:HomologicalSequence} during training, the timelike homological equivalence protocol is constrained to include only weight-one corrections (i.e., we apply corrections like in \cref{fig:Timelike_Homological} but not those of \cref{fig:Timelike_Homological_High}) as this was found to produce the best results. 

This section is structured as follows. In \cref{subsec:SynDensLER}, we quantify the reduction in syndrome density produced by each pre-decoder and the resulting improvements in logical error rates when the processed syndromes are passed to uncorrelated PyMatching for the global decoder. In \cref{subsec:SynDensLERCorrMatch} we perform the same analysis but for correlated PyMatching used as the global decoder. In \cref{subsec:GPURuntimes}, we report both the standalone pre-decoder inference runtimes and the end-to-end decoding runtimes of the combined pre-decoder + PyMatching pipeline, demonstrating substantial speedups relative to PyMatching alone. In \cref{subsec:TimeLikeParallel}, we show how the pre-decoder per-round runtimes can be substantially reduced to numbers well below $1 \mu s$ when implemented in a parallel-window decoding fashion with multiple GPUs. Finally, in \cref{subsec:NoiseLearnImprove}, we demonstrate numerically that the noise learning model of \cref{sec:EffectivePreDecNoiseModel} is able to recover the correct circuit-level noise probabilities that produce near optimal edge weights in the matching graphs used for uncorrelated and correlated PyMatching. We also show that applying the noise learning model to pre-decoder outputs, and using the predicted probabilities in the global decoder did not result in lower logical failure rates. This is due to structure of residual errors after the pre-decoder is applied.
 
\subsection{Logical error rates and syndrome densities for uncorrelated PyMatching}
\label{subsec:SynDensLER}
 
\begin{figure*}
    \centering
\subfloat[\label{fig:ler_vs_p_Model_1} ]{\includegraphics[width=.8\textwidth]{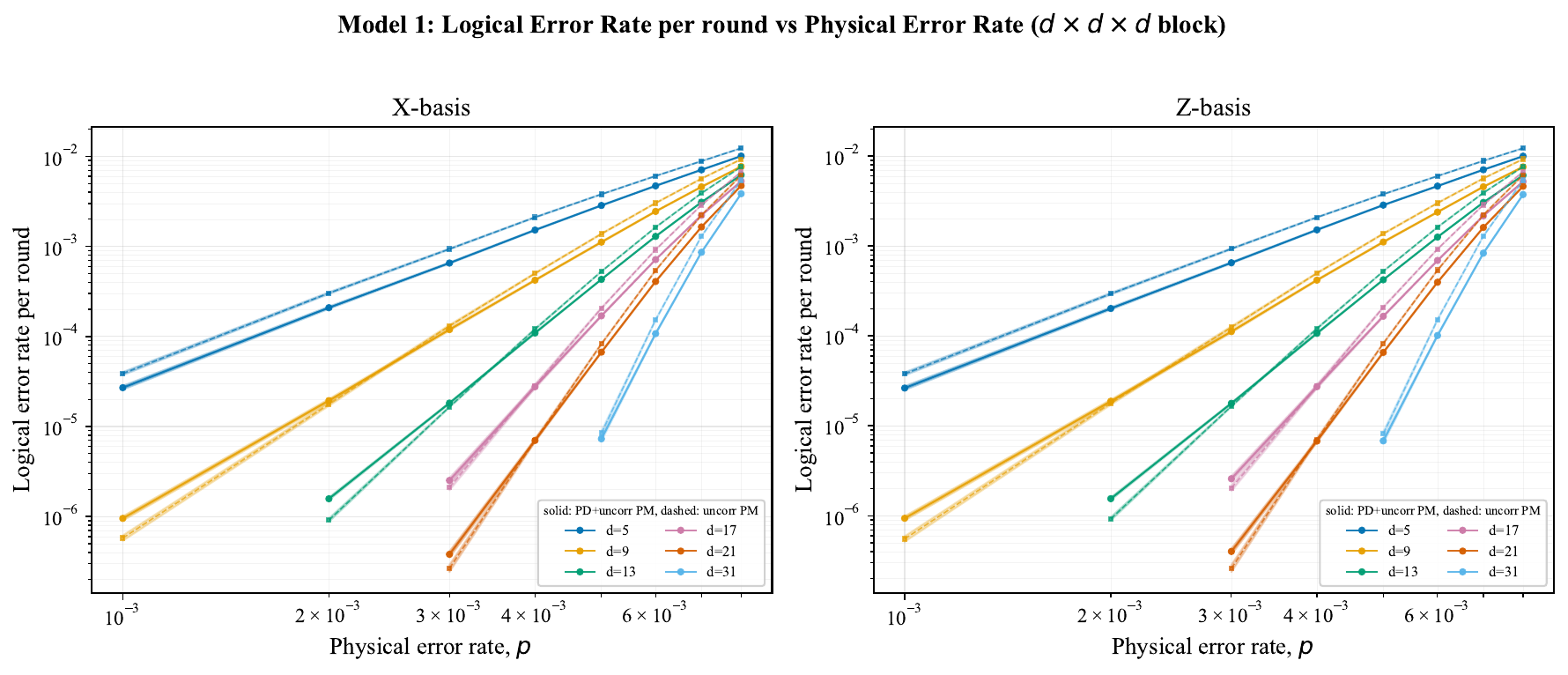}}
\vfill
\subfloat[\label{fig:ler_vs_p_Model_5} ]{\includegraphics[width=.8\textwidth]{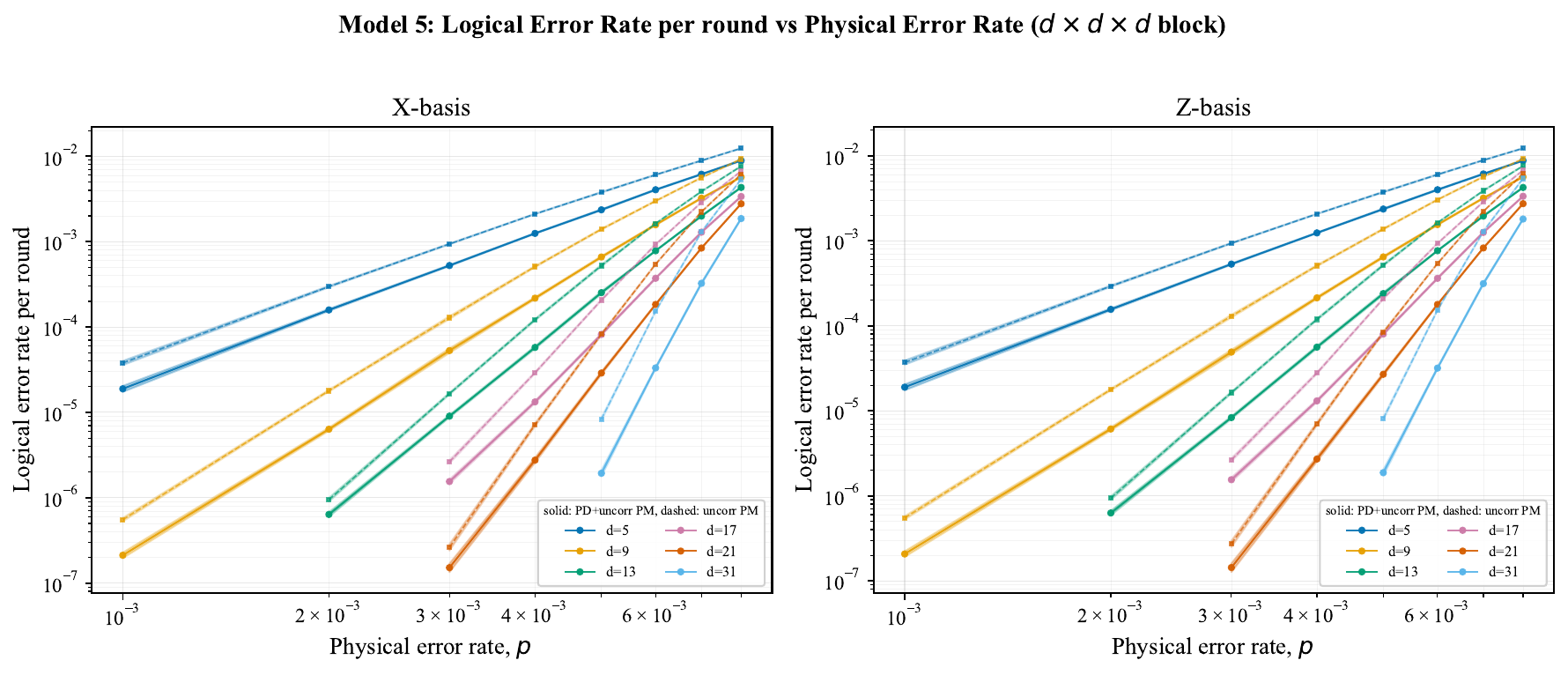}}
\caption{Plots of per-round LER for uncorrelated PyMatching (dashed lines) vs per-round LER of a pre-decoder model followed by uncorrelated PyMatching (solid lines). Due to the low LER's at $(31,31,31)$, we only provide data near threshold. In (a) we use model 1 from \cref{tab:models} (which corresponds to the fastest model, see \cref{subsec:GPURuntimes}) whereas in (b) we use model 5. }
\label{fig:LERPlots}
\end{figure*}

\begin{table*}[htbp]
\centering
 \resizebox{\textwidth}{!}{%
\begin{tabular}{|l|c|c|c|c|c|c|}
\hline
\textbf{Model} & LER improvement $d=5$ & LER improvement $d=9$ & LER improvement $d=13$ & LER improvement $d=17$ & LER improvement $d=21$ & LER improvement $d=31$ \\
\hline
Model 1 & $1.29 \text{x}$ & $1.24 \text{x}$ & $1.27 \text{x}$ & $1.29 \text{x}$ & $1.33 \text{x}$ & $1.44 \text{x}$ \\
Model 4 & $1.44 \text{x}$ & $1.66 \text{x}$ & $1.76 \text{x}$ & $1.98 \text{x}$ & $2.28 \text{x}$ & $3.21 \text{x}$ \\
Model 5 & $1.50 \text{x}$ & $1.90 \text{x}$ & $2.08 \text{x}$ & $2.48 \text{x}$ & $2.96 \text{x}$ & $4.66 \text{x}$ \\
\hline
\end{tabular}
}
\caption{LER improvement factor ($X$-basis) for models 1, 4 and 5 of \cref{tab:models} followed by uncorrelated PyMatching compared to uncorrelated PyMatching alone. All data is obtained at $p=0.006$.}
\label{tab:LER_Improvement}
\end{table*}

\jan{
\begin{table*}[htbp]
\centering
 \resizebox{\textwidth}{!}{%
\begin{tabular}{|l|c|c|c|c|c|c|}
\hline
\textbf{Model} & LER improvement $d=5$ & LER improvement $d=9$ & LER improvement $d=13$ & LER improvement $d=17$ & LER improvement $d=21$ & LER improvement $d=31$ \\
\hline
Model 1 & $1.43 \text{x}$ & $1.10 \text{x}$ & $0.91 \text{x}$ & $0.84 \text{x}$ & $0.70 \text{x}$ & $1.37 \text{x}$(*) \\
Model 4 & $1.71 \text{x}$ & $1.90 \text{x}$ & $1.32 \text{x}$ & $1.17 \text{x}$ & $1.31 \text{x}$ & $3.02 \text{x}$(*) \\
Model 5 & $1.79 \text{x}$ & $2.43 \text{x}$ & $1.83 \text{x}$ & $1.70 \text{x}$ & $1.73 \text{x}$ & $3.89 \text{x}$(*) \\
\hline
\end{tabular}
}
\caption{LER improvement factor ($X$-basis) for models 1, 4 and 5 of \cref{tab:models} followed by uncorrelated PyMatching compared to uncorrelated PyMatching alone. All data is obtained at $p=0.003$. (*) Extrapolated}
\label{tab:LER_Improvement_p003}
\end{table*}
}
\begin{table*}[htbp]
\centering
 \resizebox{\textwidth}{!}{%
\begin{tabular}{|l|c|c|c|c|c|c|}
\hline
\textbf{Model} & LER improvement $d=5$ & LER improvement $d=9$ & LER improvement $d=13$ & LER improvement $d=17$ & LER improvement $d=21$ & LER improvement $d=31$ \\
\hline
Model 1 & $1.16 \text{x}$ & $1.05 \text{x}$ & $1.01 \text{x}$ & $0.971 \text{x}$ & $0.942 \text{x}$ & $0.846 \text{x}$ \\
\hline
\end{tabular}
}
\caption{LER improvement factor ($X$-basis) for model 1 followed by PyMatching compared to PyMatching alone. In this table, model 1 is trained using ReLU activation functions rather than GeLU. ReLU activations result in faster inference times as shown in \cref{subsec:GPURuntimes}. All data is obtained at $p=0.006$. }
\label{tab:LER_ImprovementReLU}
\end{table*}

\begin{figure*}
    \centering
\subfloat[\label{fig:sdr_vs_p_Model_1} ]{\includegraphics[width=.8\textwidth]{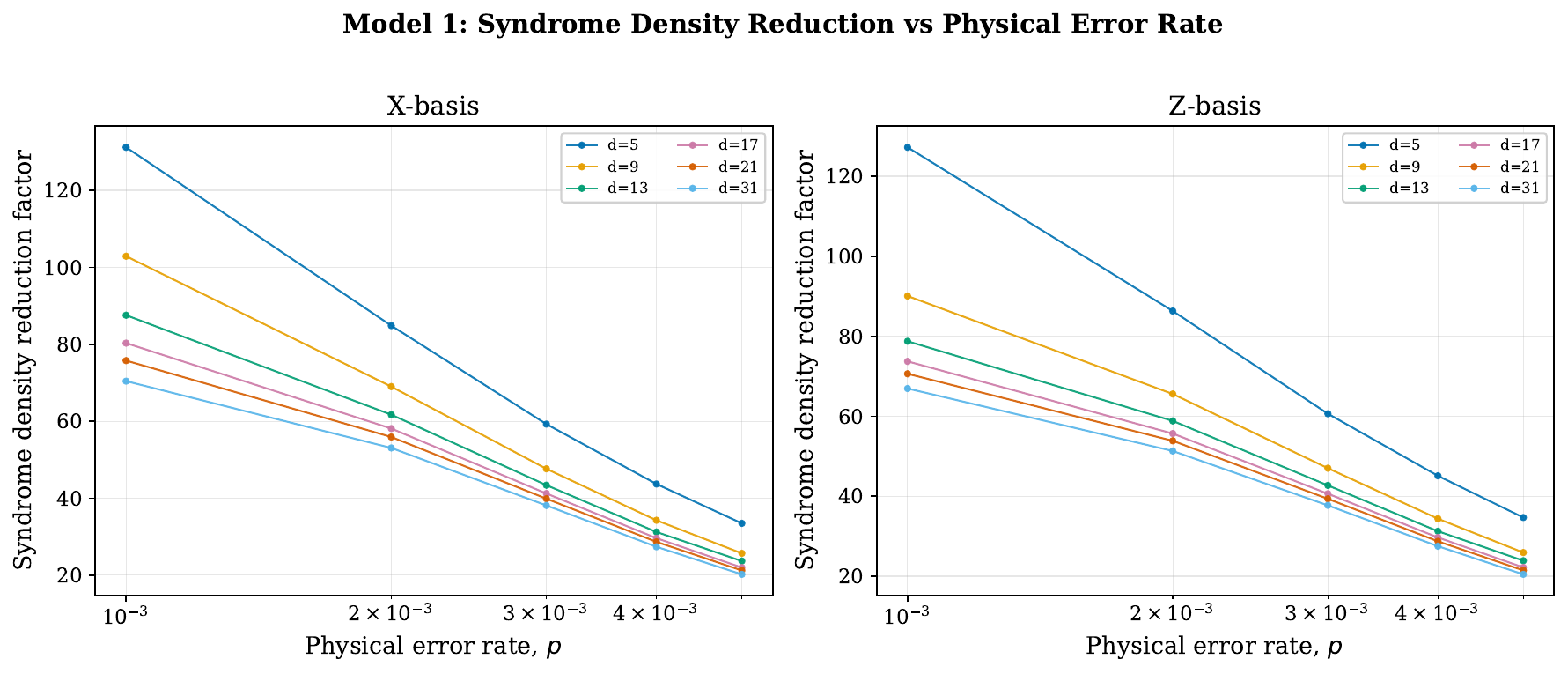}}
\vfill
\subfloat[\label{fig:sdr_vs_p_Model_5} ]{\includegraphics[width=.8\textwidth]{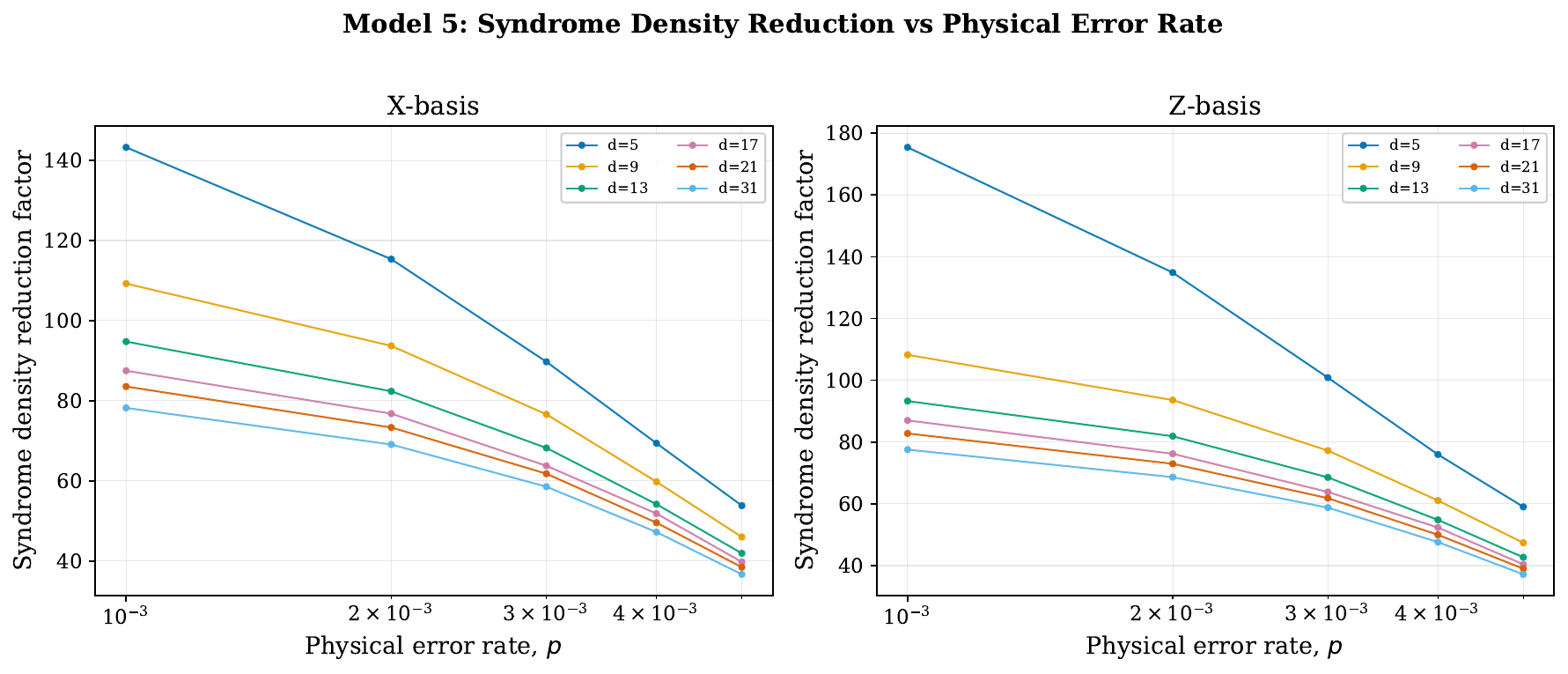}}
\caption{Plots of the syndrome density reduction factor for models 1 and 5 as a function of the physical error rate $p$ at various code distances. In (a) we show results for model 1 and in (b) for model 5.  }
\label{fig:SDRPlots}
\end{figure*}

In this subsection, we compare the logical error rates (LERs) obtained using uncorrelated PyMatching alone with those obtained using a pre-decoder followed by uncorrelated PyMatching. In what follows in this subsection, we will omit the word \textit{uncorrelated} and should be understood that when mentioning PyMatching we refer to the uncorrelated version. We focus on models 1 and 5 from \cref{tab:models}, which respectively represent the fastest and the highest-capacity pre-decoder architectures considered in this work. These comparisons quantify the extent to which local pre-decoding can improve logical performance by reducing the effective syndrome density passed to the global decoder. The results are shown in \cref{fig:LERPlots}.

All models were trained using the hyperparameters listed in \cref{tab:hyperparameters}. During training, each model was trained on a surface-code space–time volume of size $(d_r,d_r,d_r)$ where $d_r$ was chosen to match the receptive field of the network. For example, model~1 has a receptive field of 9 lattice units (see \cref{tab:models}), and was therefore trained with $d_r=9$. We found that using training volumes larger than the receptive field did not improve performance, while using volumes smaller than the receptive field degraded generalization when the trained model was applied to larger code distances.

During training, the shots per epoch listed in \cref{tab:hyperparameters} were generated by using the physical error rate $p=0.006$, since we saw the best performance with a $p$ close to surface-code threshold from below due to the larger syndrome density producing more non-trivial events. We do not consider larger values of $p$, since the surface-code threshold is near $p\approx 0.007$.

As shown in \cref{subsec:GPURuntimes}, model 1 achieves the lowest inference runtimes among all pre-decoders considered, but also exhibits the smallest LER improvements due to its limited depth and channel width. For $p \gtrsim 0.004$, the LER obtained using model 1 followed by PyMatching is lower than that of PyMatching alone for all considered code distances. At lower values of $p$, however, there exist regimes in which model 1 + PyMatching slightly underperforms PyMatching alone. This behavior is expected, since during training most contributions to the loss originate from higher-$p$ samples. Fine-tuning the training distribution toward lower $p$ values would likely improve performance in this regime. We also note that LERs can be further reduced when using the noise learning architecture described in \cref{sec:EffectivePreDecNoiseModel}. Numerical results are provided in \cref{subsec:NoiseLearnImprove}.

\begin{table*}[htbp]
\centering
 \resizebox{\textwidth}{!}{%
\begin{tabular}{|l|c|c|c|c|c|c|}
\hline
\textbf{Model} & \textbf{d=13, $p=0.003$ ($\mu$s/\text{round})} & \textbf{d=13, $p=0.006$ ($\mu$s/\text{round})} & \textbf{d=21, $p=0.003$ ($\mu$s/\text{round})} & \textbf{d=21, $p=0.006$ ($\mu$s/\text{round})} & \textbf{d=31, $p=0.003$ ($\mu$s/\text{round})} & \textbf{d=31, $p=0.006$ ($\mu$s/\text{round})} \\
\hline
Uncorrelated PyMatching & 3.38 & 9.97 & 13.41 & 29.95 & 28.78 & 91.06 \\
Uncorrelated PyMatching after model 1 (GeLU) & 1.32 & 3.05 & 5.26 & 11.30 & 11.92 & 30.45 \\
Uncorrelated PyMatching after model 4 (GeLU) & 1.22 & 2.55 & 4.92 & 9.26 & 10.81 & 22.86 \\
Uncorrelated PyMatching after model 5 (GeLU) & 1.20 & 2.38 & 4.80 & 8.43 & 10.70 & 20.50 \\
Pre-decoder model 1 (GeLU) & 2.397 & 2.397 & 1.872 & 1.872 & 2.609 & 2.609 \\
Pre-decoder model 4 (GeLU) & 3.252 & 3.252 & 2.703 & 2.703 & 3.774 & 3.774 \\
Pre-decoder model 5 (GeLU) & 4.364 & 4.364 & 5.056 & 5.056 & 9.263 & 9.263 \\
Pre-decoder model 1 (ReLU) & 2.297 & 2.297 & 1.719 & 1.719 & 2.139 & 2.139 \\
Pre-decoder model 4 (ReLU) & 3.091 & 3.091 & 2.312 & 2.312 & 2.892 & 2.892 \\
Pre-decoder model 5 (ReLU) & 4.201 & 4.201 & 3.746 & 3.746 & 6.511 & 6.511 \\
\hline
\end{tabular}
}
\caption{Comparison of runtimes for uncorrelated PyMatching (both with and without syndromes processed by pre-decoder models) and  pre-decoder models. All results correspond to the task of decoding a single (batch size $=1$) $d\times d\times d$ block, and we report averaged runtimes per syndrome measurement round. PyMatching runtimes are computed using a Grace Neoverse-V2 CPU. The label ``PyMatching after model $X$'' refers to PyMatching runtimes after processing syndromes by the pre-decoder model $X$ (i.e. one of the 5 models in \cref{tab:models}). GPU runtimes for all five pre-decoder models are computed using an NVIDIA GB300 GPU using TensorRT with FP8 precision. }
\label{tab:runtimes_mwpm_bs1}
\end{table*}

\begin{table}[htbp]
\centering
\begin{tabular}{|l|c|c|c|c|}
\hline
$d$ & $p$ & M1 speedup & M4 speedup & M5 speedup \\
\hline
13 & 0.003 & \textbf{0.91x} & 0.76x & 0.61x \\
13 & 0.006 & \textbf{1.83x} & 1.72x & 1.48x \\
21 & 0.003 & \textbf{1.88x} & 1.76x & 1.36x \\
21 & 0.006 & 2.27x & \textbf{2.50x} & 2.22x \\
31 & 0.003 & \textbf{1.98x} & 1.97x & 1.44x \\
31 & 0.006 & 2.75x & \textbf{3.42x} & 3.06x \\
\hline
\end{tabular}
\caption{Total speedup factors when using a pre-decoder (model MX with GeLU activation) + uncorrelated PyMatching compared to uncorrelated PyMatching alone. Speedup is defined as the ratio between raw uncorrelated PyMatching runtimes and the sum of pre-decoder inference runtimes plus uncorrelated PyMatching runtimes after pre-decoding (see \cref{tab:runtimes_mwpm_bs1}). The largest speedup factor for each input setting is shown in bold.}
\label{tab:Summary_Speedup}
\end{table}

In contrast, model 5, which uses additional layers and a larger number of filters per layer, consistently outperforms PyMatching alone across nearly all distances and physical error rates considered, as shown in \cref{fig:ler_vs_p_Model_5}. This improved performance comes at the cost of increased inference runtimes (see \cref{subsec:GPURuntimes}), reflecting a tradeoff between decoding accuracy and runtime. For $p=0.006$, the LER improvement factors obtained using models 1 and 5 are summarized in \cref{tab:LER_Improvement}.

We note that the results in \cref{fig:LERPlots} and \cref{tab:LER_Improvement} correspond to models trained using GeLU activation functions (see \cref{tab:hyperparameters}). As shown in \cref{subsec:GPURuntimes}, replacing GeLU with ReLU results in faster inference on NVIDIA GB300 GPUs (see also \cref{tab:runtimes_mwpm_bs1}). The corresponding LER results for model 1 trained with ReLU activations are shown in \cref{tab:LER_ImprovementReLU}. While ReLU yields a modest LER improvement for most code distances, a slight degradation is observed at $d=31$, illustrating a tradeoff between inference speed and logical performance.

Finally, we examine the syndrome density reduction (SDR) achieved by the pre-decoders. The SDR factors for models~1 and~5 are shown in \cref{fig:SDRPlots}. Larger syndrome density reductions directly translate into faster global decoding, explaining the runtimes improvements observed for the combined pre-decoder + PyMatching pipeline. As seen in \cref{fig:SDRPlots}, the largest SDR gains occur at lower values of $p$, which is consistent with the local nature of the pre-decoder and the fact that the probability of an error chain of length $k$ scales as $cp^k$ for some constant $c$.

\subsection{Logical error rates and syndrome densities for a correlated matching global decoder}
\label{subsec:SynDensLERCorrMatch}

\begin{figure*}
    \centering
\subfloat[\label{fig:Model8RepresentationEX} ]{\includegraphics[width=.8\textwidth]{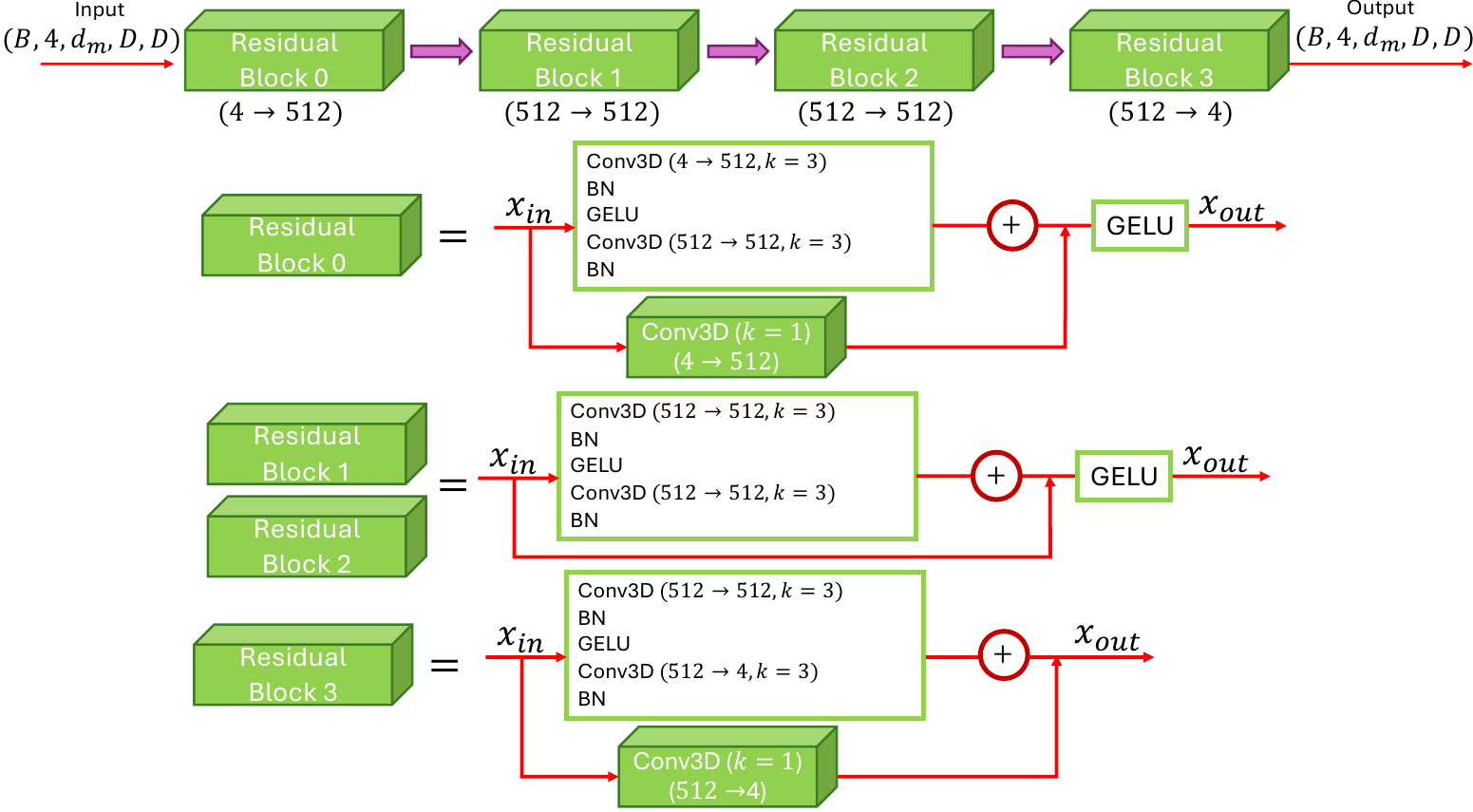}}
\caption{ Pre-decoder neural network architecture used when the global decoder employs correlated matching. The model is a fully convolutional 3D residual network composed of four residual blocks, each containing two 3×3×3 convolutions with BatchNorm. The first block expands channel dimension from 4 to 512, and the final block compresses from 512 to 4 via 1×1×1 projection shortcuts; intermediate blocks use identity skip connections. Residual connections are employed to improve gradient flow and stabilize deep optimization. The network has a receptive field of size 17 and the total number of parameters for this network is 42,593,296. }
\label{fig:Model8Representation}
\end{figure*}

\begin{figure*}
    \centering
\subfloat[\label{fig:ler_vs_p_Model_6} ]{\includegraphics[width=.8\textwidth]{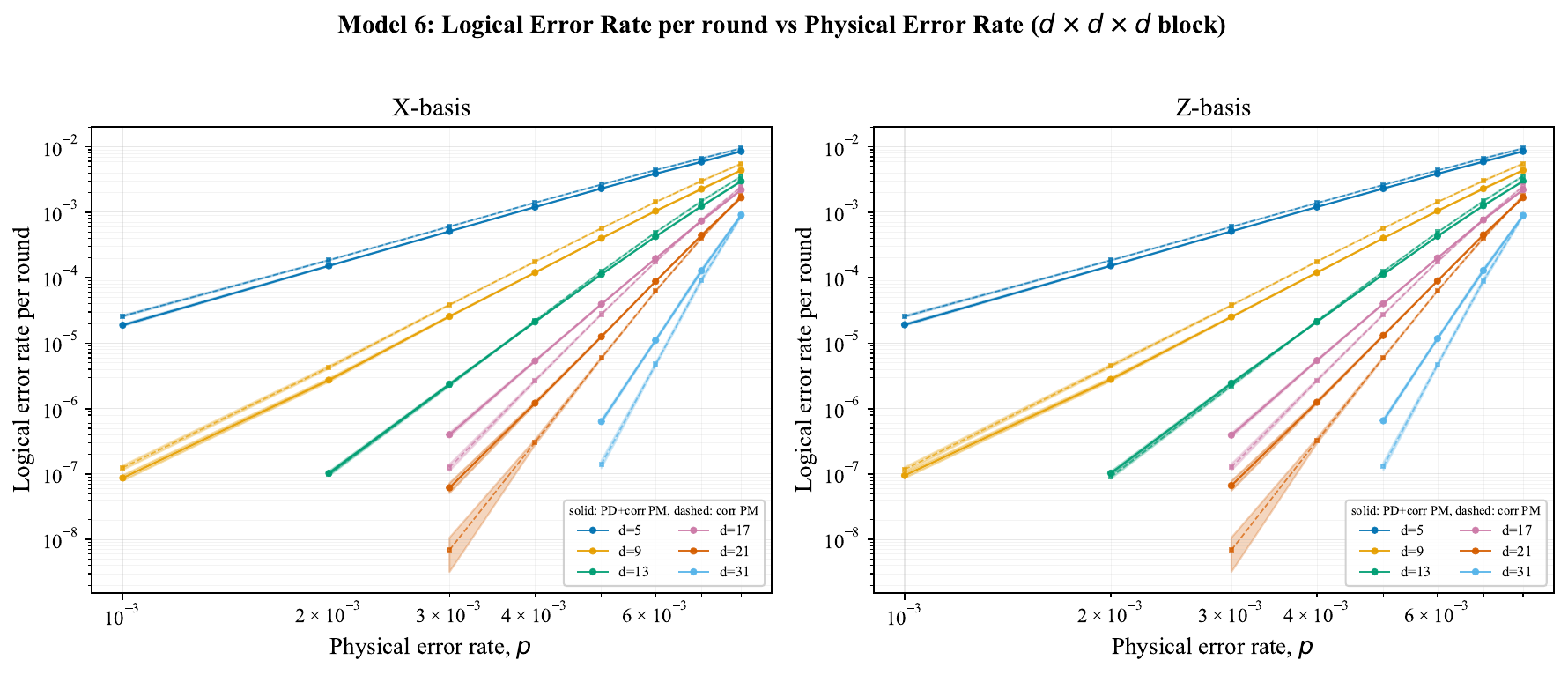}}
\vfill
\subfloat[\label{fig:sdr_vs_p_Model_6} ]{\includegraphics[width=.8\textwidth]{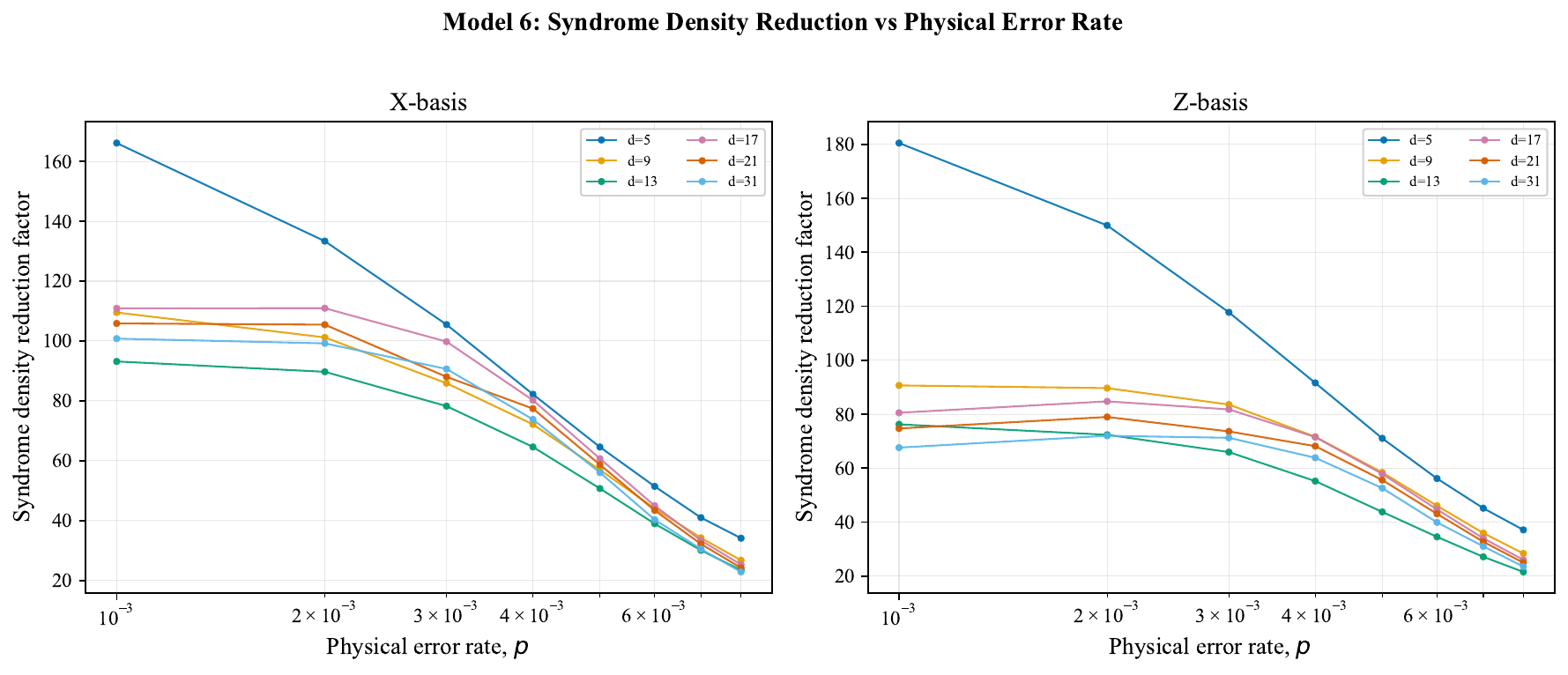}}
\caption{Per-round LERs obtained from using pre-decoder model 6 described in \cref{fig:Model8Representation} with correlated PyMatching as the global decoder. The pre-decoder is trained at $p=0.006$. The LERs are improved compared to baseline correlated matching at $d=5,9$ and 13. At $d\ge 17$, the LER is slightly worse with a growing gap as $p$ decreases. (b) Syndrome density reduction factor obtained by applying the model 6 pre-decoder to input syndromes.  }
\label{fig:Model8LERs}
\end{figure*}

In this subsection, we perform an analogous analysis to \cref{subsec:SynDensLER} but where the global decoder corresponds to a correlated matching decoder \cite{HiggottPyMatch,Higgott2025sparseblossom}. The correlated matching decoder achieves lower LERs relative to uncorrelated PyMatching by using hyperedges in the matching graph for fault mechanism that produce errors which anticommute with more than two detectors \cite{TransformerGoogle}. 

When considering correlated matching as the global decoder, we found that the pre-decoder models given in \cref{tab:models} result in a higher LER than correlated matching alone. The reason for this is that most of the residual errors from the application of a pre-decoder that produce a logical fault when applying either PyMatching or correlated matching have structure such that they form strings of size greater than $(d-1)/2$ which are parallel to a logical observable. As such, a logical fault would result from any global decoder performing a minimum-weight correction. To mitigate this problem, we use a larger CNN network shown in \cref{fig:Model8Representation}. The network uses more 3D convolutional layers (eight in this case excluding projection layers) thereby increasing its ability to learn from more complex fault mechanisms. Due to the larger number of layers, we partition the network into residual blocks, with each residual block using skip connections for improved gradient flow and to stabilize deep optimization. In what follows, we refer to the network in \cref{fig:Model8Representation} as model 6.

Since the receptive field of model 6 is 17, we train it on a $d=17$ lattice with $d_m=17$ syndrome measurement rounds. The model is trained at $p=0.006$ and applied to $p \in [0.001, 0.008]$ during inference. We also scale by 4 the resources for training with respect to the numbers shown in \cref{tab:hyperparameters} (GPUs and number of epochs) while keeping the effective batch size and number of shots per epoch constant and use a learning rate of $1 \times 10^{-4}$. In \cref{fig:ler_vs_p_Model_6}, we showcase the LERs obtained by applying model 6 to input syndrome data, followed by using correlated matching as the global decoder. As can be seen, for $d=5,9$ and 13, the LER improves from the use of the pre-decoder at all sampled $p$ values. However, at $d\ge 17$, the LER slightly increases, with a widening gap as $p$ decreases. This can be remedied by adding additional layers to the model in \cref{fig:Model8Representation} (thus increasing the size of the receptive field and model capacity), at the cost of higher pre-decoder runtimes. However, standard techniques like model distillation \cite{hinton2015distilling} can compress these larger models into smaller with almost no loss in accuracy. Such explorations are left for future work.

In \cref{fig:sdr_vs_p_Model_6}, we show the SDR achieved from using model 6. At low error rates, the syndrome density is reduced by nearly two orders of magnitude. In \cref{subsec:GPURuntimes} we show the total correlated PyMatching speedups achieved from the application of the model 6 pre-decoder.

\subsection{GPU runtimes and optimizations}
\label{subsec:GPURuntimes}

\begin{figure*}
    \centering
\subfloat[\label{fig:blackwell_13x13x13} ]{\includegraphics[width=.45\textwidth]{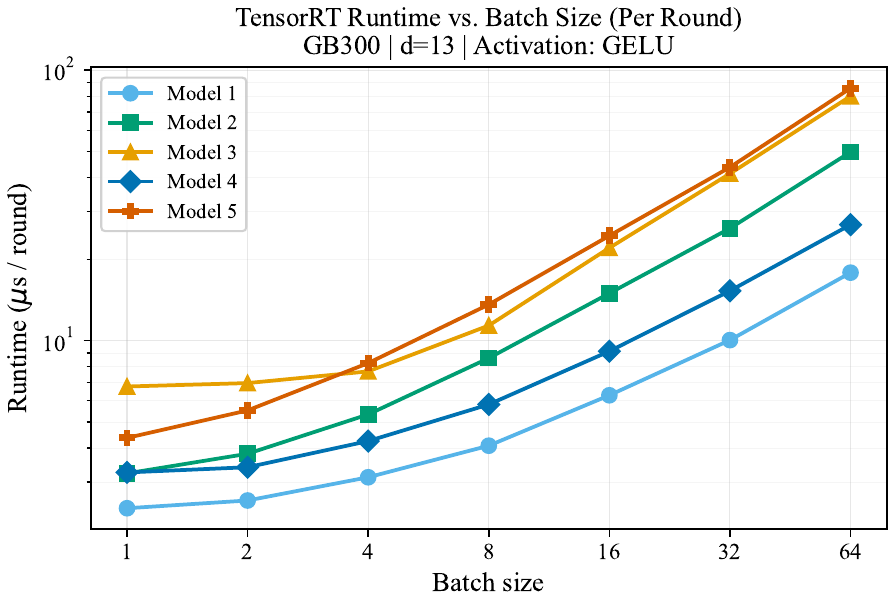}}
\subfloat[\label{fig:blackwell_21x21x21V1} ]{\includegraphics[width=.45\textwidth]{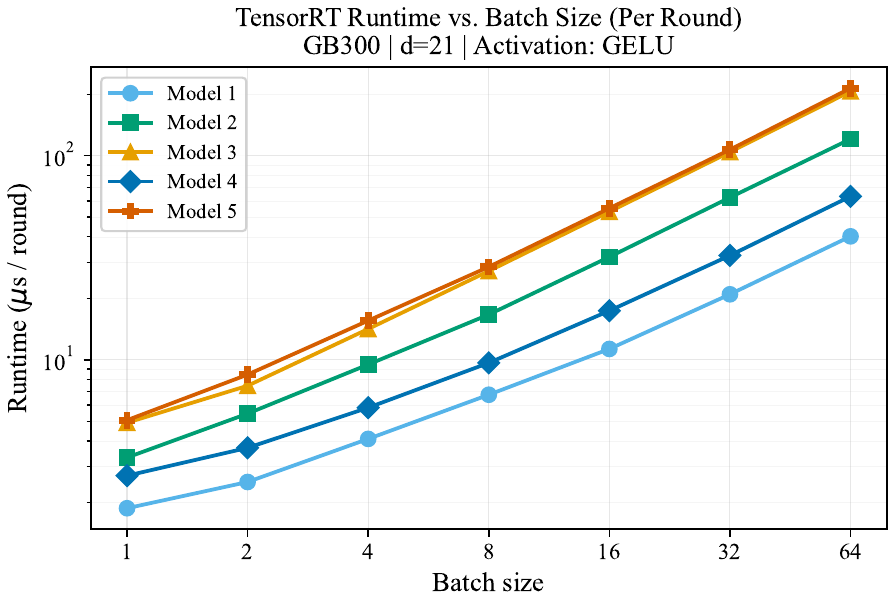}}
\vfill
\subfloat[\label{fig:blackwell_21x21x21V2} ]{\includegraphics[width=.45\textwidth]{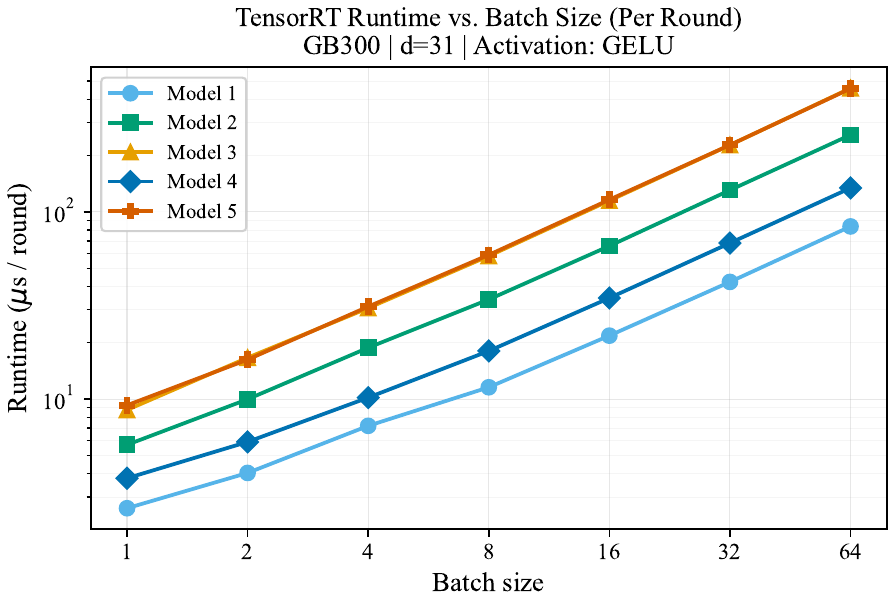}}
\subfloat[\label{fig:blackwell_21x21x21V3} ]{\includegraphics[width=.45\textwidth]{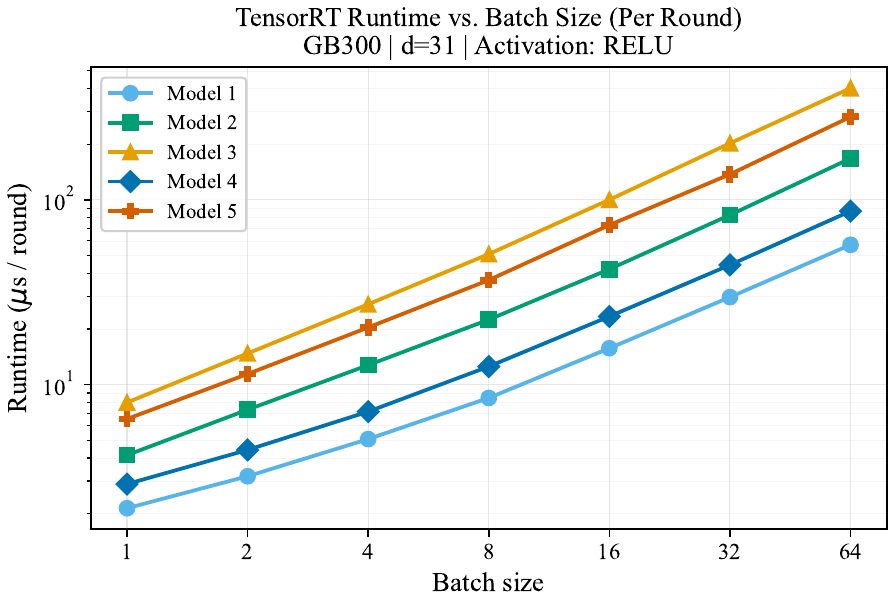}}
\caption{GPU runtime performance on an NVIDIA GB300 GPU using TensorRT with FP8 precision. (a) runtimes measurements for $13 \times 13 \times 13$ space-time volumes across the five pre-decoder models listed in \cref{tab:models}, trained with the GeLU activation function. (b) and (c) same as (a) but with $21 \times 21 \times 21$ and $31 \times 31 \times 31$ space-time volumes. (d) Same as (c) but with the GeLU activation function replaced with ReLU. As can be seen, such a replacement results in faster runtimes.}
\label{fig:blackwell_runtimes}
\end{figure*}

\begin{figure}
    \centering
\subfloat[\label{fig:runtimes_per_round_GB300_gelu_model6_D13_D21_D31} ]{\includegraphics[width=.4\textwidth]{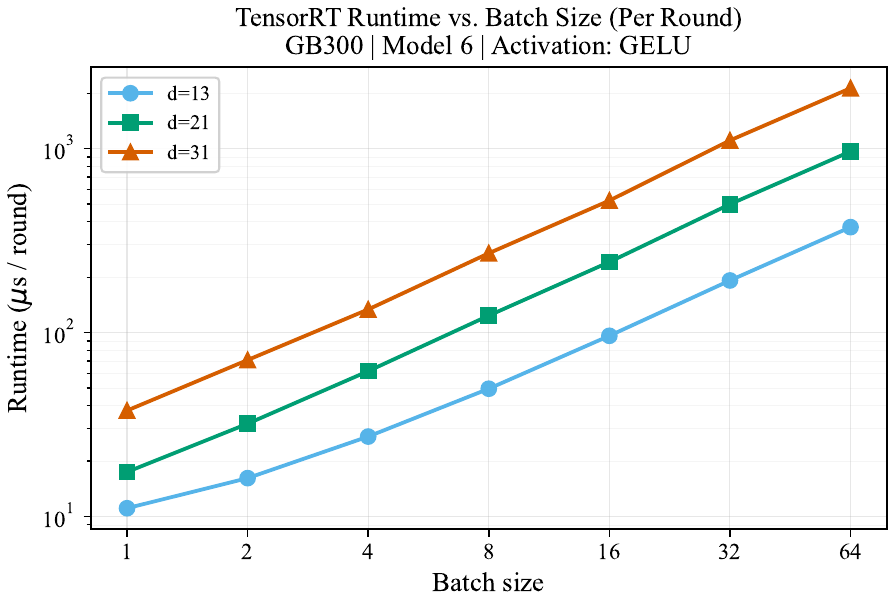}}
\caption{ Pre-decoder runtime as a function of the batch size for model 6 given in \cref{fig:Model8Representation} for various input volumes at FP8 precision. Batch sizes greater than one can be used for space and time parallelization in a parallel block-wise decoding scheme.  }
\label{fig:BatchSizeCorrelatedMatching}
\end{figure}

\begin{figure*}
    \centering
\subfloat[\label{fig:ler_vs_time_p0.003_0.006_X} ]{\includegraphics[width=.9\textwidth]{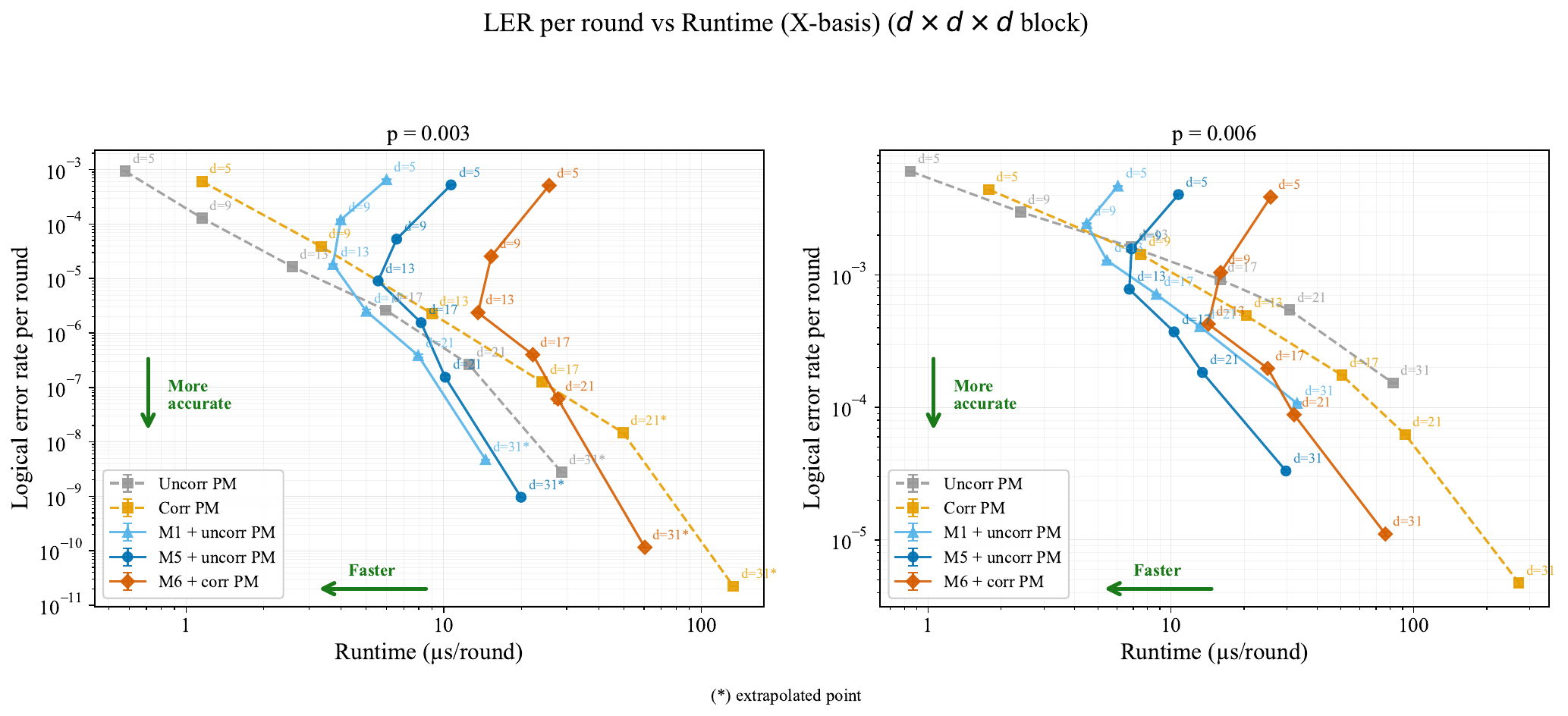}}
\caption{End-to-end per-round logical error rates (LER) and \textbf{single-shot (batch size 1) runtimes} across different decoding strategies with representative physical error rates $p=0.003$ (left) and $p=0.006$ (right). Pre-decoder models M1, M5 and M6 run at FP8 precision and were timed on a single GB300 GPU, while PyMatching (PM) was timed on a single Grace Neoverse-V2 CPU. We see how there is a tradeoff between pre-decoder model and global decoder choice. Our strategy of combining an AI pre-decoder with PyMatching offers a favorable tradeoff: at small $d$, the pre-decoder inference cost dominates and raw PyMatching is faster. However, at large $d$, a reduction in syndrome density from the pre-decoder accelerates PyMatching enough to offset the pre-decoder cost, making the full pipeline faster than raw PyMatching. Lighter models (M1, M5) with uncorrelated PyMatching offer the lowest runtimes at moderate accuracy, while M6 with correlated PyMatching targets the highest-accuracy regime. Points marked with (*) have their LER extrapolated, while their runtimes are measured directly.}
\label{fig:Global_Runtime_Compare}
\end{figure*}

\begin{table}[htbp]
\centering
\small
\begin{tabular}{|l|c|c|c|c|}
\hline
$d$ & $p$ & 
\makecell{Corr PM \\ ($\mu s$/round)} & 
\makecell{Corr PM after PD \\ ($\mu s$/round)} & 
Speedup \\
\hline
5 & 0.003 & 1.15 & 0.61 & 1.9x \\
5 & 0.006 & 1.78 & 0.69 & 2.6x \\
9 & 0.003 & 3.35 & 1.01 & 3.3x \\
9 & 0.006 & 7.51 & 1.73 & 4.3x \\
13 & 0.003 & 9.14 & 2.67 & 3.4x \\
13 & 0.006 & 21.51 & 4.53 & 4.8x \\
17 & 0.003 & 24.12 & 5.82 & 4.1x \\
17 & 0.006 & 50.63 & 8.68 & 5.8x \\
21 & 0.003 & 49.75 & 10.31 & 4.8x \\
21 & 0.006 & 92.27 & 14.72 & 6.3x \\
31 & 0.003 & 133.31 & 22.78 & 5.9x \\
31 & 0.006 & 270.83 & 38.78 & 7.0x \\
\hline
\end{tabular}
\caption{Decoding times of correlated PyMatching both with and without the use of the pre-decoder model 6 given in \cref{fig:Model8Representation}. The final column gives the speedup of correlated PyMatching alone when using model 6 to process the input syndromes.  }
\label{tab:runtimes_mwpm_bs1_correlated}
\end{table}

\begin{table}[htbp]
\centering
\begin{tabular}{|l|c|c|}
\hline
$d$ & $p$ & Total Speedup \\
\hline
13 & 0.003 & 0.66x \\
13 & 0.006 & 1.38x \\
21 & 0.003 & 1.79x \\
21 & 0.006 & 2.87x \\
31 & 0.003 & 2.21x \\
31 & 0.006 & 3.54x \\
\hline
\end{tabular}
\caption{Total speedup of using both the pre-decoder with correlated PyMatching compared to correlated PyMatching alone.}
\label{tab:runtimes_mwpm_bs1_correlated_total_speedup}
\end{table}

In this subsection, we analyze both the runtime of the pre-decoders themselves and the end-to-end decoding runtimes achieved when combining a pre-decoder with both uncorrelated and correlated PyMatching. All results are compared against baseline uncorrelated and correlated PyMatching and runtimes obtained using unprocessed syndrome data. GPU runtime measurements for the pre-decoders are performed on a single NVIDIA GB300 GPU with FP8 precision, while uncorrelated and correlated PyMatching runtimes are measured on a Grace Neoverse-V2 CPU.

We begin with runtime results for uncorrelated matching, with a summary  provided in \cref{tab:runtimes_mwpm_bs1}. Pre-decoder runtimes measurements were obtained using NVIDIA TensorRT’s \texttt{trtexec} utility with FP8 inference. To minimize measurement overhead and isolate steady-state device-side inference time, we enabled CUDA graph capture (\texttt{--useCudaGraph}), disabled host–device transfers (\texttt{--noDataTransfers}), and used spin-wait synchronization (\texttt{--useSpinWait}) for low-runtimes timing. Each configuration was executed with 200 warmup iterations followed by 100 timed iterations to mitigate cold-start effects. All benchmarks were collected using TensorRT v25.12 on an NVIDIA GB300 GPU.

We benchmarked five pre-decoder architectures across batch sizes $B \in \{1,2, 4,8,16,32,64\}$ and three input tensor shapes: $4 \times 13 \times 13 \times 13$, $4 \times 21 \times 21 \times 21$, and $4 \times 31 \times 31 \times 31$, corresponding to 13, 21, and 31 syndrome measurement rounds, respectively. runtimes results for batch size $B=1$ are reported in \cref{tab:runtimes_mwpm_bs1}, while batch-size scaling is shown in \cref{fig:blackwell_runtimes}.

Several remarks are in order regarding the runtime results in \cref{tab:runtimes_mwpm_bs1}. First, pre-decoder runtimes are independent of the physical error rate $p$, whereas both uncorrelated and correlated PyMatching runtimes depend strongly on $p$ through the syndrome density, as reviewed in \cref{sec:SurfaceCodeReview}. The first row of \cref{tab:runtimes_mwpm_bs1} reports baseline uncorrelated PyMatching runtimes for surface codes of distance $d=13,21$ and 31, using $d$ syndrome measurement rounds in each case. Results for $(13,13,13)$ and $(21,21,21)$ are shown at $p=0.003$ and $p=0.006$, while for $(31,31,31)$ we report results only at $p=0.006$ to emphasize near-threshold behavior.

Rows 2--4 of \cref{tab:runtimes_mwpm_bs1} show uncorrelated PyMatching runtimes when provided with syndromes processed by the pre-decoder. For example, for inputs of size $(21,21,21)$ at $p=0.006$, the uncorrelated PyMatching runtime is reduced from $29.95\,\mu\text{s}$ to $11.30\,\mu\text{s}$ when using syndromes produced by model 1 in \cref{tab:models}, corresponding to a $\approx 2.65\times$ speedup in the global decoder alone.

Rows 5--7 report standalone pre-decoder runtimes on the NVIDIA GB300 GPU using GeLU activation functions. For instance, model 1 achieves a runtime of $1.872\,\mu\text{s}$ per round for $(21,21,21)$ inputs. Estimates of the time required to transfer syndrome data between the pre-decoder and the global decoder using NVIDIA’s NVQLink architecture \cite{NVQLink2025} indicate that this overhead is negligible relative to both the pre-decoder and PyMatching runtimes and is therefore ignored. Consequently, the total decoding runtimes at $p=0.006$ is $13.17\,\mu\text{s}$, representing an overall $\approx 2.27\times$ speedup relative to PyMatching alone. At $p=0.003$, the total speedup is reduced to $\approx 1.88\times$, as expected since at lower error rates PyMatching becomes faster and the pre-decoder overhead becomes relatively more significant.

In the hypothetical limit of negligible pre-decoder runtimes, the speedup at $p=0.006$ for $(31,31,31)$ inputs would approach $\approx 3.0\times$ for model 1 and $\approx 4.4\times$ for model 5, illustrating the extent to which global-decoder runtime dominates near threshold. Rows 8--10 of \cref{tab:runtimes_mwpm_bs1} report pre-decoder runtimes obtained using ReLU activation functions in place of GeLU, yielding additional runtimes reductions. Total end-to-end speedups achieved for all five models at $p=0.006$ are summarized in \cref{tab:Summary_Speedup}. Interestingly, for volumes of size $(31,31,31)$, model 4 achieves the largest overall speedup.

\begin{table*}
\centering
\begin{tabular}{|l|c|c|c|c|c|c|}
\hline
Model & Precision & Batch size & $d$ & Number of rounds & Time ($\mu s$) / Round & Number of GPUs \\
\hline
1 & FP8 & 1 & 13 & 1000 & 0.11 & 13 \\
1 & FP8 & 2 & 13 & 1000 & 0.13 & 7 \\
1 & FP8 & 4 & 13 & 1000 & 0.179 & 4 \\
1 & FP8 & 1 & 21 & 1000 & 0.179 & 8 \\
1 & FP8 & 2 & 21 & 1000 & 0.244 & 4 \\
1 & FP8 & 4 & 21 & 1000 & 0.423 & 2 \\
4 & FP8 & 1 & 13 & 1000 & 0.138 & 13 \\
4 & FP8 & 2 & 13 & 1000 & 0.211 & 7 \\
4 & FP8 & 4 & 13 & 1000 & 0.282 & 4 \\
4 & FP8 & 1 & 21 & 1000 & 0.231 & 8 \\
4 & FP8 & 2 & 21 & 1000 & 0.324 & 4 \\
4 & FP8 & 4 & 21 & 1000 & 0.551 & 2 \\
\hline
\end{tabular}
\caption{Decoding time per round as a function of batch size for 1000 rounds of stabilizer measurements when using the time parallel-window decoding scheme of Ref.~\cite{CampbellParallelV1,AlibabaParallel}. We provided the number of GPUs needed to decode each block in parallel.}
\label{tab:Time_Like_Parallel_Table}
\end{table*}

The trends in \cref{tab:runtimes_mwpm_bs1} demonstrate that runtimes speedups increase with both code distance and physical error rate $p$. This behavior is consistent with the reduction in effective syndrome density produced by the pre-decoder and the resulting improvement in global-decoder runtime near threshold. Given the relatively high physical error rates expected in early fault-tolerant quantum computers, operation at large code distances $(d \ge 21)$ is anticipated, making these scaling trends particularly relevant.

Comparing pre-decoder architectures, we find that model 3— which uses $5 \times 5 \times 5$ convolutional kernels—exhibits the highest runtimes for smaller input volumes $(13,13,13)$, while model 5 becomes the slowest for larger volumes $(21,21,21)$ and $(31,31,31)$. When these runtimes results are considered alongside the logical error rate improvements reported in \cref{subsec:SynDensLER}, they indicate that deeper architectures with smaller convolutional kernels ($3 \times 3 \times 3$) offer a more favorable tradeoff between runtime and decoding performance than shallower architectures with larger kernels.

Next, we examine pre-decoder runtimes as a function of batch size in \cref{fig:blackwell_runtimes} for models 1-5. Using batch sizes greater than one enables multiple logical qubits or decoding blocks to be processed in parallel, which is particularly well suited to parallel block-wise decoding architectures \cite{CampbellParallelV1,AlibabaParallel}. Because our pre-decoders jointly predict spacelike and timelike corrections on data qubits and stabilizers, they naturally support parallel decoding windows in both space and time \cite{CampbellParallelV1,NVQLink2025}. When the number of available GPUs is insufficient to achieve the desired level of parallelism, increased batch sizes can be used to partially compensate. In \cref{sec:BatchingImprove} we provide greater details showing how increasing the batch size can reduce overall resource costs for enabling real-time decoding when using the results in \cref{fig:blackwell_runtimes}.

We now consider speedups when using the model-6 pre-decoder of \cref{fig:Model8Representation} with correlated PyMatching as the global decoder. In \cref{tab:runtimes_mwpm_bs1_correlated} we provide the decoding runtimes (in $\mu s$) of correlated matching using both raw syndrome and syndromes processed by the model-6 pre-decoder. Similarly to the results obtained for uncorrelated matching, we see that speedups improve as the code distance increases and as $p$ increases. Including the runtimes of the model-6 pre-decoder on an NVIDIA GB300 with FP8 precision, the total speedups using the pre-decoder + correlated matching pipeline compared to correlated matching alone are given in \cref{tab:runtimes_mwpm_bs1_correlated_total_speedup}. The GPU runtimes used to produce the results in \cref{tab:runtimes_mwpm_bs1_correlated_total_speedup} are shown in \cref{fig:BatchSizeCorrelatedMatching} for a batch size of one. The plot in \cref{fig:BatchSizeCorrelatedMatching} also shows the runtimes of model 6 for batch sizes which are greater than 1 with FP8 precision. Runtimes increase in a near linear fashion with increasing batch size. 

Lastly in \cref{fig:Global_Runtime_Compare}, we provide two plots (one for $p=0.003$ and another for $p=0.006$) of the logical error rates achieved with various decoding strategies considered above (both with and without the use of pre-decoders) as a function of the runtimes. Such plots highlights the tradeoffs between LER and runtimes while clearly illustrating regimes where a given decoding strategy is favorable over another. For example, when $p=0.006$, we see both a reduction in LER and runtimes of model 5 + uncorrelated PyMatching (dark blue curve) compared to correlated PyMatching (grey curve) alone for $d \ge 13$.

In future work, we will extend these methods to lattice-surgery protocols and demonstrate fully parallel block-wise decoding across spatial and temporal dimensions. In such settings, we anticipate that using large batch sizes will play a crucial role in reducing classical resource costs for real-time decoding.

\subsection{Faster pre-decoders with parallel-window decoding in time}
\label{subsec:TimeLikeParallel}

Once trained, the pre-decoder can be deployed within a temporal parallel window decoding protocol following the methods of Ref.~\cite{CampbellParallelV1,AlibabaParallel}. Specifically, the pre-decoder is applied to both commit regions (together with their associated buffer rounds) and cleanup regions. Each commit block—and likewise each cleanup block—can be decoded independently and in parallel when a dedicated GPU is assigned per block. Alternatively, a single GPU may process multiple blocks simultaneously by using a batch size greater than one, trading reduced hardware requirements for increased per-block decoding runtimes.

In \cref{tab:Time_Like_Parallel_Table}, we report the per-round decoding time for our Model 1 and Model 4 pre-decoders when processing 1000 rounds of syndrome measurements under this parallel time-window scheme. We assume that all blocks of size $d \times d \times 3d$ are decoded in parallel for both commit and cleanup regions. The factor of three comes from the buffer regions used for each commit region. We also list the number of GPUs required to achieve these runtimes. As expected, increasing the batch size reduces the number of GPUs needed, while correspondingly increasing the decoding time per round. Nevertheless, in all configurations considered, the per-round decoding time remains well below $1 \mu\text{s}$. We note that increasing the total number of rounds beyond 1000 in \cref{tab:Time_Like_Parallel_Table} would result in even smaller per-round runtimes if enough GPUs (and/or larger batch sizes) were used to ensure that all blocks of size $d \times d \times 3d$ were decoded in parallel. In particular, if a large number of syndrome measurement rounds is performed, using larger batch sizes may become more advantageous even if the per-block runtime increases. 

To obtain the results in \cref{tab:Time_Like_Parallel_Table}, we assume that the GPUs used to decode all commit regions in parallel can be reused to subsequently decode all cleanup regions in parallel. We further neglect communication latencies between the commit and cleanup stages. Since such overheads are expected to contribute primarily a constant time offset, their relative impact diminishes as the number of syndrome measurement rounds increases.

\subsection{Numerical results with noise learning}
\label{subsec:NoiseLearnImprove}

\begin{table*}
\centering
\resizebox{\textwidth}{!}{%
\begin{tabular}{|c|c|}
\hline
\textbf{Hyperparameters} &  \textbf{Values} \\ \hline
CNN filters per layer & $[128, 256, 256, 128]$ \\ \hline
CNN kernel size per layer  & $3 \times 3$ \\ \hline
CNN normalization & GroupNorm (32 groups) \\ \hline
CNN dropout & 0.1 (last layer only) \\ \hline
MLP neurons per layer & $[256, 128, 25]$ \\ \hline
MLP dropout & 0.2 \\ \hline
Activation function (CNN and MLP) & GeLU (tanh approximation) \\ \hline
Pooling function & Global average pooling (GAP) \\ \hline
Batch aggregation & Post-MLP logit averaging (\cref{eq:PostMLPAvg}) \\ \hline
Output parameterization & Bounded log-space (\cref{eq:BoundedLogSpace}) \\ \hline
Loss function & $\mathcal{L}_{\text{edge}}$ (18 edge formulas) $+$ $\mathcal{L}_{\text{hyper}}$ (43 hyperedge formulas) \\ \hline
Optimizer   &  AdamW (weight decay $3 \times 10^{-2}$) \\ \hline
Exponential moving average (EMA)   & decay $= 0.0001$  \\ \hline
Learning rate schedule   & Warmup then decay (100 warmup steps). Apply $\gamma=0.7$ at milestones $[0.25,0.5,1.0]$  \\ \hline
Learning rate   & $5 \times 10^{-4}$  \\ \hline
Samples per epoch & 250 randomly sampled $\boldsymbol{p}$ vectors $\times$ 4096 shots each  \\ \hline
Training distance & $d = 21, 31$ \\ \hline
Batch size per GPU & 4,096  \\ \hline
Number of GPUs & 32 (8 nodes $\times$ 4 GPUs)  \\ \hline
Total parameters & $\sim$1.26M \\ \hline
 \end{tabular}
} 
\caption{ Hyperparameters used to train the noise learning architecture described in \cref{sec:EffectivePreDecNoiseModel}. The model uses post-MLP logit averaging with bounded log-space output and a combined edge + hyperedge loss function.}
\label{tab:hyperparametersNoiseLearning}
\end{table*}

\begin{figure*}
    \centering
\subfloat[\label{fig:nl_effect_raw} ]{\includegraphics[width=.7\textwidth]{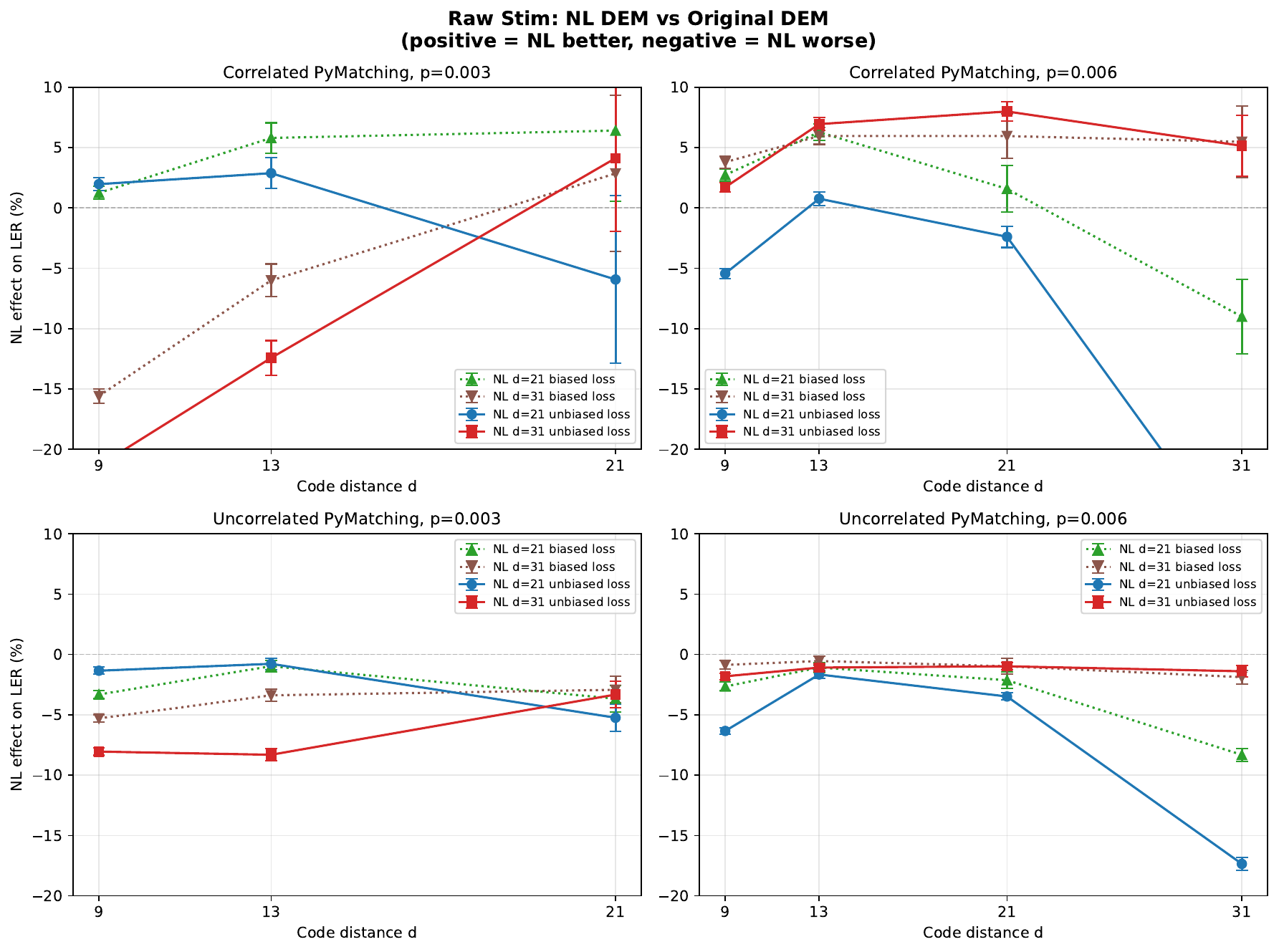}}
\vfill
\subfloat[\label{fig:nl_effect_predec} ]{\includegraphics[width=.7\textwidth]{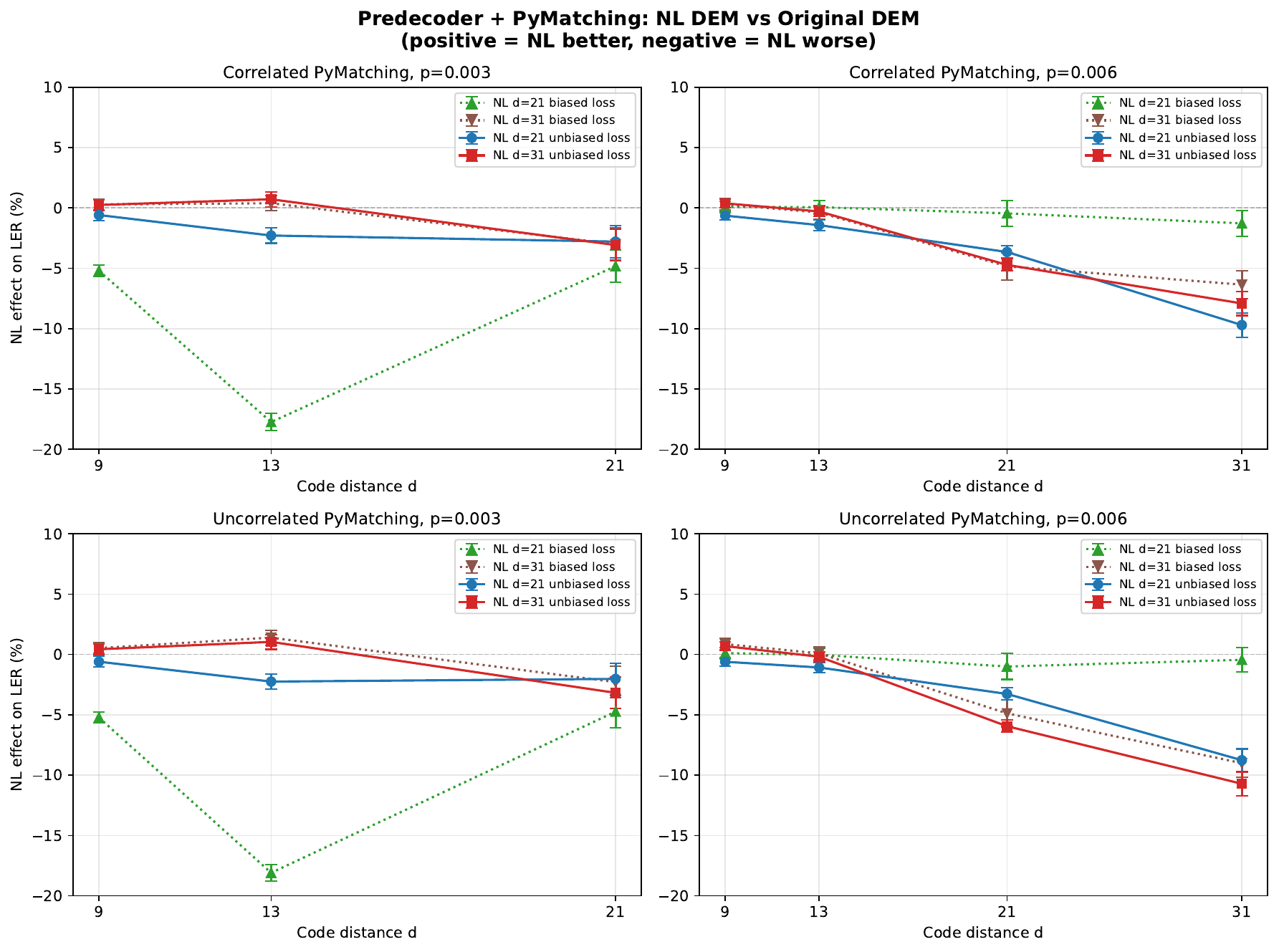}}
\caption{(a) LER for correlated and uncorrelated PyMatching when using probability vectors in a detector error model (DEM) obtained from the trained noise learning architecture. The biased losses are given in \cref{eq:EdgeLoss,eq:HyperedgeLoss} and the unbiased losses in \cref{eq:EdgeLoss_v1,eq:HyperedgeLossv1}. The noise learning models are trained at $d=21$ and $d=31$ with $p_{\text{base}} \in [0.001, 0.01]$. The learned models are then applied to syndrome data generated with stim at $d=9,13,$ and 21. At $p=0.003$, the biased model trained at $d=21$ produces the most competitve results across code distances. However at $p=0.006$, the unbiased model trained at $d=31$ produces the best overall results across correlated and uncorrelated matching. (b) Same as (a), but where the noise learning model is applied to syndrome statistics produced by the Model 5 pre-decoder. The best performance at $d=13$ comes from the unbiased noise model trained at $d=31$. However at larger distances, the $d=21$ biased loss model offers the best overall performance.}
\label{fig:NoiseLearnPlots}
\end{figure*}

In this section, we evaluate the trained noise learning model of \cref{fig:NoiseLearnArch}, using the hyperparameters listed in \cref{tab:hyperparametersNoiseLearning}, on syndrome statistics from two consecutive rounds of the surface code. The model outputs probability vectors that are then used to construct detector error models for both uncorrelated and correlated PyMatching. We compare the resulting LERs with those obtained when PyMatching is provided with probabilities derived directly from the original circuit-level noise model used to generate the syndrome data. The goal of this experiment is to demonstrate that the trained noise learning model can infer probability vectors that closely approximate the edge and hyperedge weights obtained directly from the original circuit-level noise model, yielding LERs that closely match those obtained when the true circuit-level noise parameters are known.

We next apply the trained noise learning model to syndrome statistics obtained from the outputs of the Model 5 pre-decoder described in \cref{tab:models}. The probability vectors predicted by the noise learning model are used to construct detector error models for both uncorrelated and correlated PyMatching. We then compute the resulting LERs and compare them with those obtained when PyMatching uses probabilities derived directly from the original circuit-level noise model.

In \cref{fig:nl_effect_raw}, we show the relative LERs obtained with correlated and uncorrelated PyMatching when DEMs are constructed from probability vectors predicted by the noise learning model, compared to DEMs constructed directly from the circuit-level noise model used to generate the syndrome data. Four noise learning models were trained, two at $d=21$ and two at $d=31$. For each distance, we consider both biased and unbiased loss functions given in \cref{eq:EdgeLoss,eq:HyperedgeLoss} and \cref{eq:EdgeLoss_v1,eq:HyperedgeLossv1}. As can be seen across the four plots, the model trained at $d=31$ using an unbiased loss function generally offers the best results when applied to $d=21$ and $d=31$ data, with the $d=21$ models (both with biased and unbiased losses) giving better results at $d=9$ and $d=13$. Such results are expected given that boundary effects of the surface code lattice play a bigger role at smaller distances, with bulk-like effects dominating at larger distances. We also note that both the biased and unbiased models trained at $d=31$ give very similar results when applied to $d=21$ and $d=31$ data. However the biased noise learning model gives better performance at lower distances. Lastly, we notice an \textit{improvement} in LER with correlated PyMatching compared to the baseline result where probabilities are computed directly from the circuit-level noise model. However for uncorrelated matching, the edge weights computed from the noise learned models approach the baseline result but slightly underperforms.  This can be understood by noting that correlated PyMatching is a heuristic algorithm that performs a second decoding pass using reweighted edges derived from the first-pass matching solution. As a result, the true circuit-level probabilities are not necessarily optimal inputs for this approximate pipeline. In contrast, the probabilities predicted by the noise learning model can sometimes produce a first-pass matching that triggers more effective reweighting, leading to improved second-pass corrections. For uncorrelated matching, however, there are gauge degrees of freedom in choosing the probability vector, since the edge weights depend only on sums of probabilities (e.g., \cref{eq:SpaceLikePS1X} in \cref{app:EdgeWeights}) rather than on the individual probability values. Consequently, the true DEM probability vector provides a lower bound on the achievable LER for uncorrelated matching, which explains why the noise learning model slightly under-performs in this case.

Now looking at the results in \cref{fig:nl_effect_predec}, we see that applying the noise learning model to syndrome outputs from model 5 of the pre-decoder and using the predicted probabilities in either correlated or uncorrelated PyMatching results in slightly worse performance compared to using to raw circuit-level probabilities in the DEM. At first this may seem counterintuitive since the pre-decoder results in different syndrome statistics than those that would be obtained from the original DEM. However, the majority of residual errors from corrections applied by model 5 of the pre-decoder have a very specific structure. We found numerically that nearly all residual errors that lead to a logical fault when applying a global decoder form strings of length \textit{greater than} $(d-1)/2$ and which are parallel to the logical observable of interest. Given this structure, regardless of what global decoder is applied, a minimum-weight correction will always produce a logical fault. This explains in large part why the LER is not improved in \cref{fig:nl_effect_predec} when applying the noise learning model to pre-decoder output syndrome statistics. It also explains the need for the larger model 6 given in \cref{fig:Model8Representation} of \cref{subsec:SynDensLERCorrMatch} to obtain better LERs than correlated PyMatching.

\section{Improved parallelization through batching}
\label{sec:BatchingImprove}

\begin{table}
\centering
\small   
\begin{tabular}{|c|c|c|}
\hline
Batch size & $N_{\text{par}}$ improvement & Speedup factor \\
\hline
2 & $3.2\text{x}$ & $1.993\text{x}$ \\
4 & $3.56\text{x}$ & $0.996\text{x}$ \\
64 & $12.49\text{x}$ & $0.2\text{x}$ \\
\hline
\end{tabular}
\caption{Improvements to $N_{\text{par}}$ and the corresponding speedup factor between uncorrelated PyMatching and the pre-decoder + uncorrelated PyMatching as a function of the batch size (data obtained from \cref{fig:blackwell_13x13x13}). All data is obtained with $p=0.006$ and input volumes of size $(13,13,13)$. We use model~1 for the pre-decoder implemented with a ReLU activation function.}
\label{tab:Summary_Speedup_Npar}
\end{table}

\begin{figure}
    \centering
\subfloat[\label{fig:LER_vs_pEX} ]{\includegraphics[width=.45\textwidth]{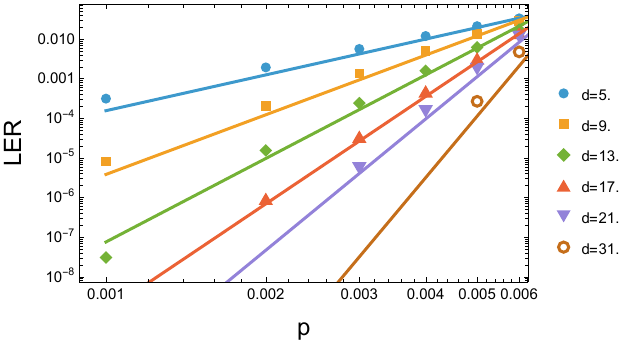}}
\caption{LER of the surface code using the uncorrelated PyMatching decoder. We use the data to obtain the constants $c_1$ and $c_2$ in \cref{eq:PLdp}. Solid lines correspond to $p_L(p,d)$ in \cref{eq:PLdp}. }
\label{fig:LER_vs_p}
\end{figure}

Recall that the number of parallel resources $N_{\text{par}}$ required to avoid the exponential backlog is given by \cref{eq:NparSats}. From \cref{tab:runtimes_mwpm_bs1}, at $p=0.006$ and input volumes of size $(13,13,13)$, a decoder using pure uncorrelated PyMatching requires $N_{\text{par}}=8$. On the other hand, our pre-decoder followed by uncorrelated PyMatching requires $N_{\text{par}}=5$ while simultaneously giving an overall speedup per block of $1.993 \text{x}$ when using model 1 in \cref{tab:models} (assuming ReLU activation functions are used). 

Using the results from \cref{fig:blackwell_13x13x13}, we can further improve $N_{\text{par}}$ when increasing the batch size used by the GPU. For instance, at a batch size of 2, the pre-decoder runtime to process an input volume of size $(13,13,13)$ is unchanged. As such, two logical qubits can be decoded in parallel without affecting $T_{\text{DEC}}$ in \cref{eq:NparSats}. Results for batch sizes of 2,4 and 64 are summarized in \cref{tab:Summary_Speedup_Npar}. As can be seen, for a batch size of 2, the pre-decoder + PyMatching requires $3.2 \text{x}$ fewer parallel resources than PyMatching alone while simultaneously resulting in a $T_{\text{DEC}}$ which is $1.993\text{x}$ faster. Using a batch size of 4 gives a slight improvement in the number of parallel resources compared to the batch size 2 case, but $T_{\text{DEC}}$ is nearly identical to using PyMatching alone. A batch size of 64 results in a large reduction in the number of parallel resources ($12.49 \text{x}$). However, $T_{\text{DEC}}$ is about $80 \%$ slower than PyMatching alone. On the surface, such a tradeoff might seem not to be worthwhile. However when running a quantum algorithm using lattice surgery with parallel block-wise decoding in both space and time, given the very large code distances that can be obtained from merged patches, such parallelization may require hundreds of thousands of GPU's. As such a reduction of $12.49 \text{x}$ could substantially reduce the cost of classical resources required to enable real-time decoding. 

Since the results in \cref{tab:Summary_Speedup_Npar} use model 1 with ReLU activation functions, the LER is slightly worse than the one obtained with GeLU (compare \cref{tab:LER_ImprovementReLU} with \cref{tab:LER_Improvement} showing a $1.01\text{x}$ LER improvement compared to $1.27\text{x}$ at $d=13$). On the surface, it may seem as though the decrease in pre-decoder runtimes when using ReLU compared to GeLU (and thus the overall $T_{\text{DEC}}$) is not a worthwhile given the increase in LER. However we conclude this section by showing that in most settings of interest, a large reducion in LER is required to implement a quantum algorithm with a smaller surface code distance, thus making the ReLU tradeoff worthwhile. 

As was shown in Refs.~\cite{Chamberland22,Chamberland22b}, we can approximate the logical failure rate of the surface code at distance $d$ and failure probability $p$ to be
\begin{align}
    p_L(p,d) \approx c_1 d (c_2 p)^{(d+1)/2},
    \label{eq:PLdp}
\end{align}
for some constants $c_1$ and $c_2$ when $p$ is below the surface code threshold. Using logical failure rates obtained from uncorrelated PyMatching, we find that $c_1 = 0.01938$ and $c_2 = 116.95$. In \cref{fig:LER_vs_p}, the polynomial $p_L(p,d)$ (solid lines) is compared to LERs obtained for PyMatching using Monte Carlo methods. As can be seen there is good agreement between the data and the approximation in \cref{eq:PLdp}.

Now suppose all the logical operations required to run a quantum algorithm must fail with probability no greater than $\delta$. For a given $p$, we can determine the distance $d$ by setting $p_L(p,d) < \delta$. For the sake of this argument, we set $\delta = 10^{-10}$ which is applicable for moderate sized algorithms \cite{Chamberland22}. At $p=0.001$ and using the constants $c_1$ and $c_2$ obtained above, we require $d=21$ to ensure $p_L(p,d) < \delta$. Suppose now we set $p_L^{(2)}(p,d) = \alpha p_L(p,d)$ where $\alpha >1$ quantifies the worsening of the LER when using a different decoder (for instance a pre-decoder + uncorrelated PyMatching rather than uncorrelated PyMatching alone). We find that alpha must be at least $\alpha \approx 4.39$ for $d$ to go from 21 to 23 to ensure that $p_L^{(2)}(p,d) < \delta$. In other words, the decoder would require the LER to be $4.39 \text{x}$ worse than the LER obtained from PyMatching to require a larger code distance ensuring that $p_L^{(2)}(p,d) < \delta$. As such, for most quantum algorithms, we believe the decrease in $T_{\text{DEC}}$ obtained by using ReLU activations for our pre-decoders compared to GeLU is worthwhile even though ReLU results in slightly worse LERs.

\section{Conclusion}
\label{sec:Conclusion}

In this work we developed a surface code pre-decoder architecture to correct local space-time failures, with residual errors corrected by a global decoder such as uncorrelated and correlated PyMatching. Architectural improvements compared to previous works (especially with how we process output labels for spacelike and timelike errors) as well as the deployment of our pre-decoders on NVIDIA GB300 GPUs resulted in substantial speedups when considering pre-decoder + PyMatching runtimes compared to PyMatching alone while also producing LER improvements relative to PyMatching (both uncorrelated and correlated). Runtimes for physical error rates of $p=0.003$ and $p=0.006$ at moderate to large code distances are summarized in \cref{tab:Summary_Speedup,tab:runtimes_mwpm_bs1_correlated_total_speedup} and are up to $3.42\text{x}$ faster than pure uncorrelated PyMatching and $3.5\text{x}$ faster than pure correlated PyMatching. To our knowledge, our work is the first to demonstrate \textit{both} LER and full end-to-end speedup improvements when using AI-based pre-decoder. We also developed a novel neural network noise learning architecture that can learn circuit-level noise rates from pure syndrome statistics. The noise learning architecture produced near-optimal edge weights when used in uncorrelated PyMatching, and performance improvements for correlated PyMatching were observed (see \cref{fig:nl_effect_raw}). 

There are several compelling directions for future work. The first one involves closing the performance gap with correlated PyMatching at smaller physical error rates and larger code distances. In this regime, failures are dominated by rare error patterns that are vastly underrepresented in the training data. To address this, future work could explore improvements in both training data and model architecture. On the data side, models could be fine-tuned on curated datasets enriched with these rare events. On the architectural side, while fully convolutional networks successfully provide fast, highly parallelizable inference on arbitrary-sized volumes, it would be very interesting to find alternative architectures with these same properties but that deliver significantly better LER performance than fully convolutional networks.

A second major avenue for improvement is model distillation. While simply scaling up the parameter count of our pre-decoders improves logical error rates, deploying massive models incurs unacceptable pre-decoder runtime penalties. If one were to take the scaling route, one should investigate training highly over-parameterized ``teacher" models that successfully learn to correct complex, rare error events, and subsequently distilling that knowledge into smaller, faster ``student'' models. This approach could decouple the capacity required to learn optimal decoding strategies from the strict runtime constraints required for real-time execution.

A third critical direction for real-time execution is further optimizing inference runtimes and throughput through extreme quantization. While in this work we successfully deployed our pre-decoders in FP8 precision on NVIDIA GB300 GPUs, pushing to the next frontier of efficiency will require adopting 4-bit floating-point (NVFP4) precision. Because of the limited dynamic range and precision at 4 bits, future efforts must therefore integrate Quantization-Aware Training (QAT) directly into the pre-decoder training pipeline to maintain logical error rate performance while unlocking the massive compute throughput of NVFP4 tensor cores. This effect will be more substantial with every new NVIDIA GPU generation.

From a broader perspective, expanding this framework to other error-correcting codes represents another key direction for future work. The immediate natural progression is to consider color codes, which works almost identically to the framework we presented here and will be the focus of a forthcoming manuscript.

Finally, an important direction for future work is to adapt our architecture to decoding logical operations performed via lattice surgery in a parallel block-wise decoding fashion (in \textbf{both} space and time). One reason we did not go beyond $d=31$ in this work is that parallelizing in both space and time limits the block size needed to decode lattice surgery operations. Further we believe our pre-decoders will adapt well to such settings, pushing us closer towards realizing real-time decoding for full universal fault-tolerant quantum computation. 
 
\newpage 
\appendix
 
\section{Edge weight calculations}
\label{app:EdgeWeights}
 
\begin{figure*}
    \centering
\subfloat[\label{fig:Zstab_Graph_2D_pymatching} ]{\includegraphics[width=.3\textwidth]{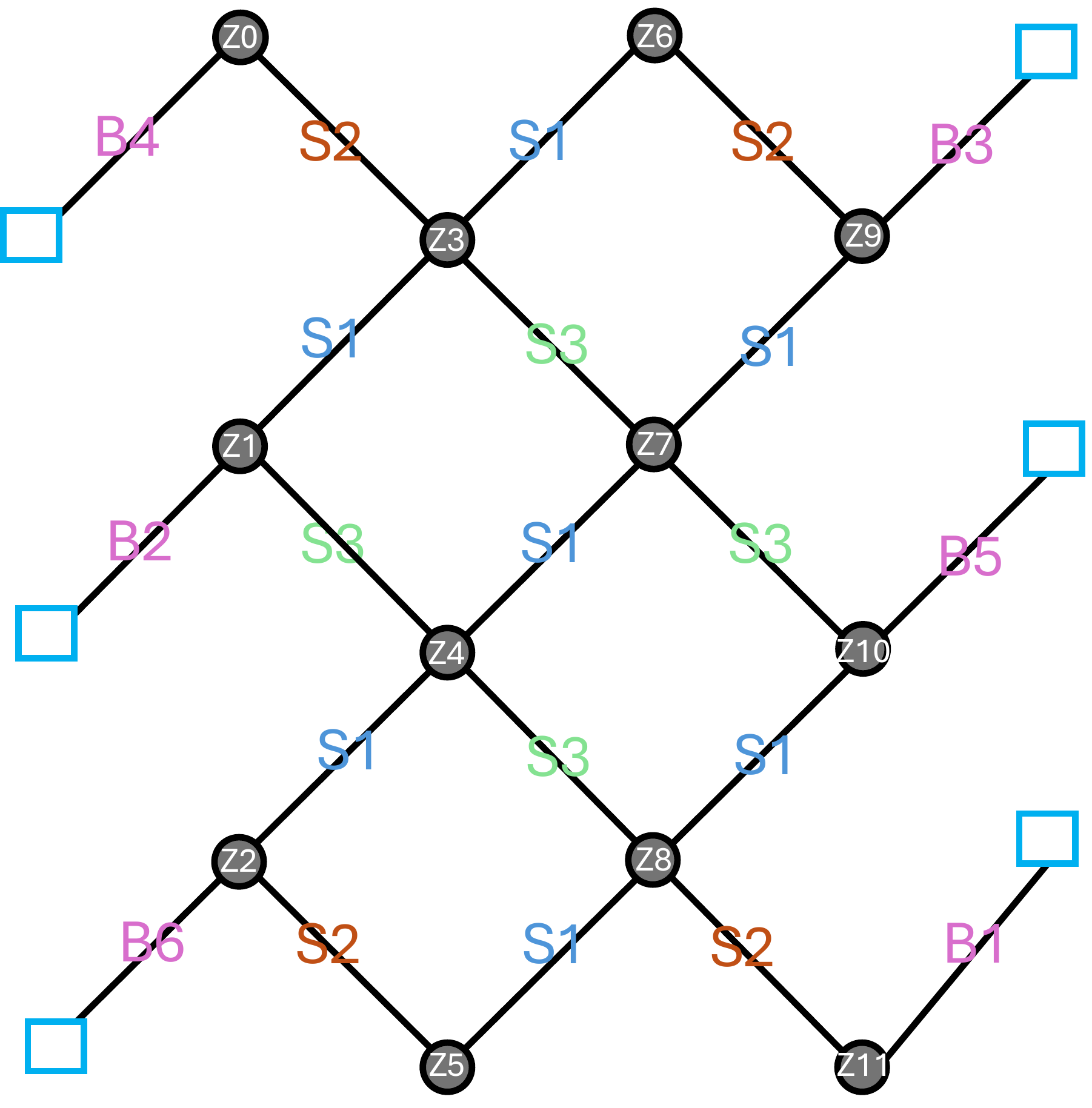}}
\hfill
\subfloat[\label{fig:Xstab_Graph_2D_pymatching} ]{\includegraphics[width=.3\textwidth]{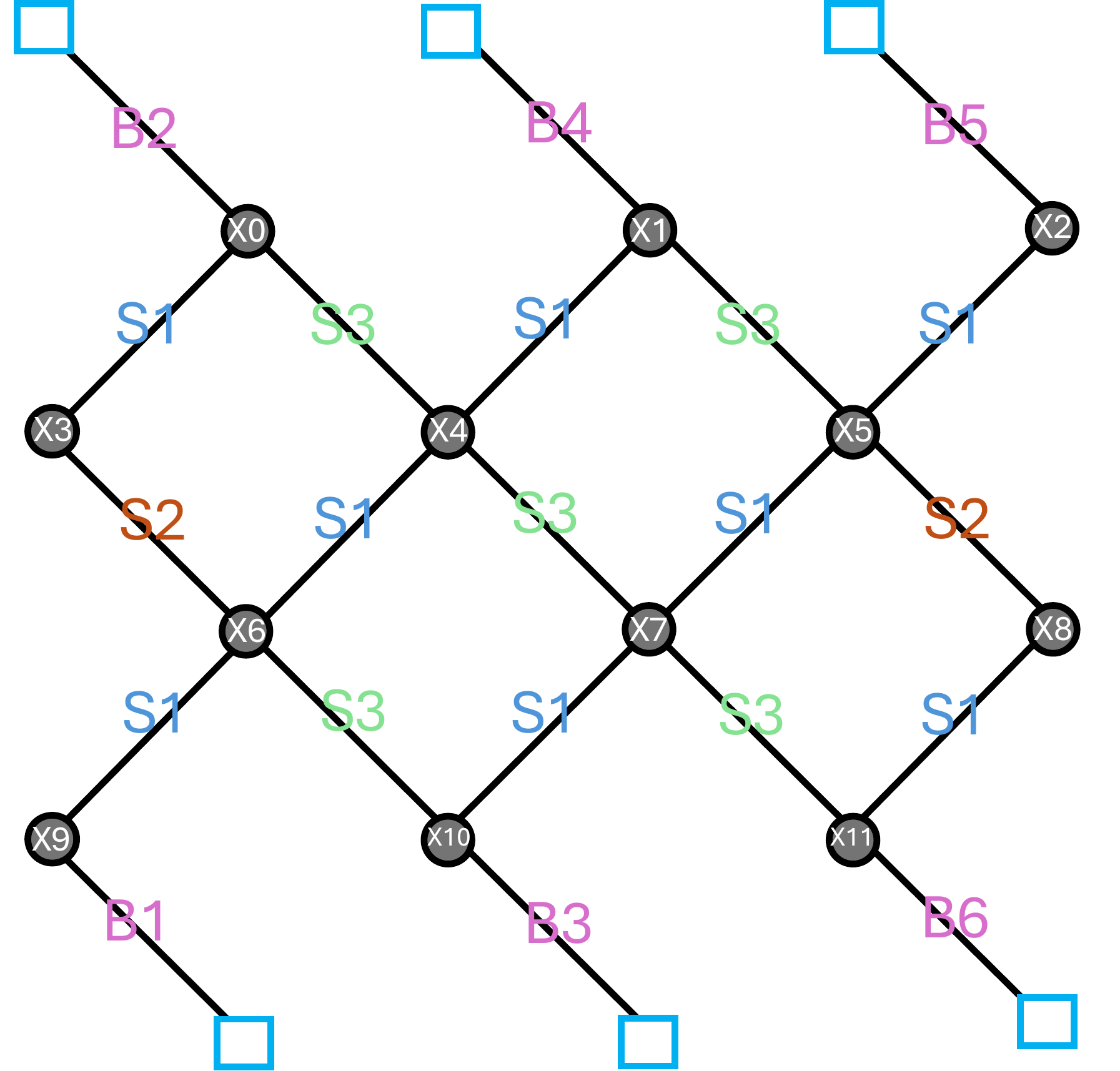}}
\hfill
\subfloat[\label{fig:Meas_Graph_Vert_Z_pymatching} ]{\includegraphics[width=.3\textwidth]{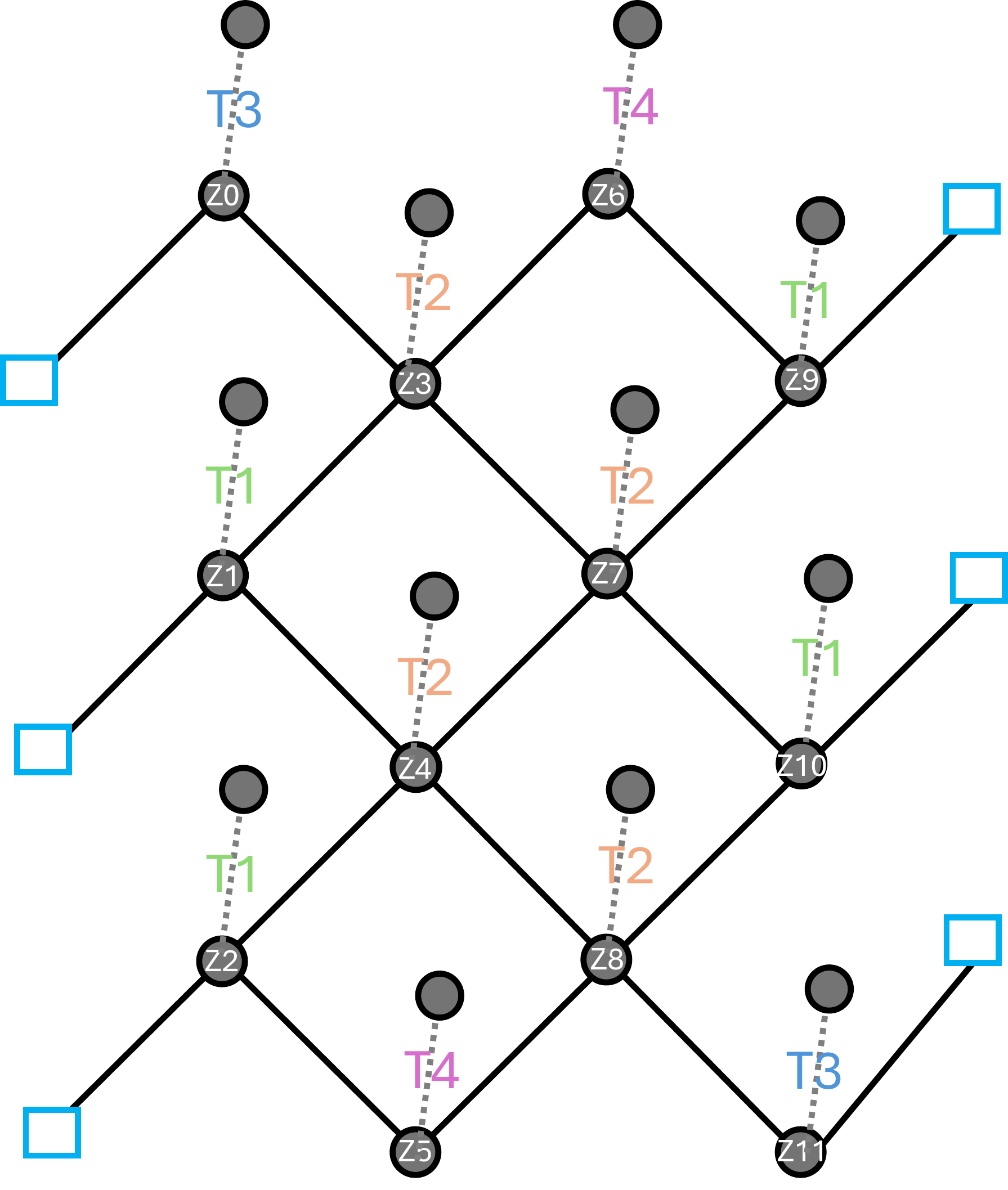}}
\vfill
\subfloat[\label{fig:Meas_Graph_Vert_X_pymatching} ]{\includegraphics[width=.3\textwidth]{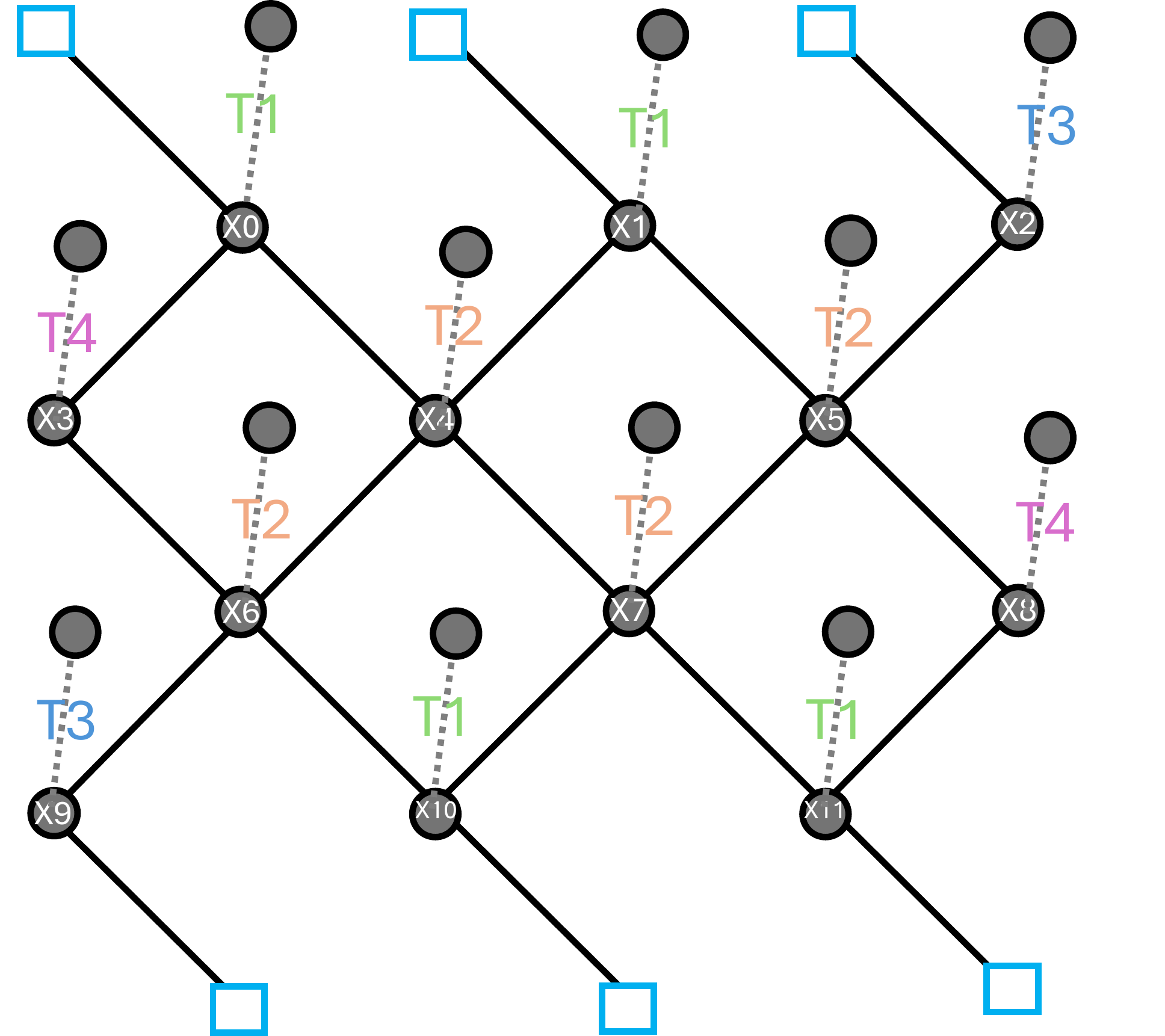}}
\hfill
\subfloat[\label{fig:Diagonal_Graph_Z} ]{\includegraphics[width=.3\textwidth]{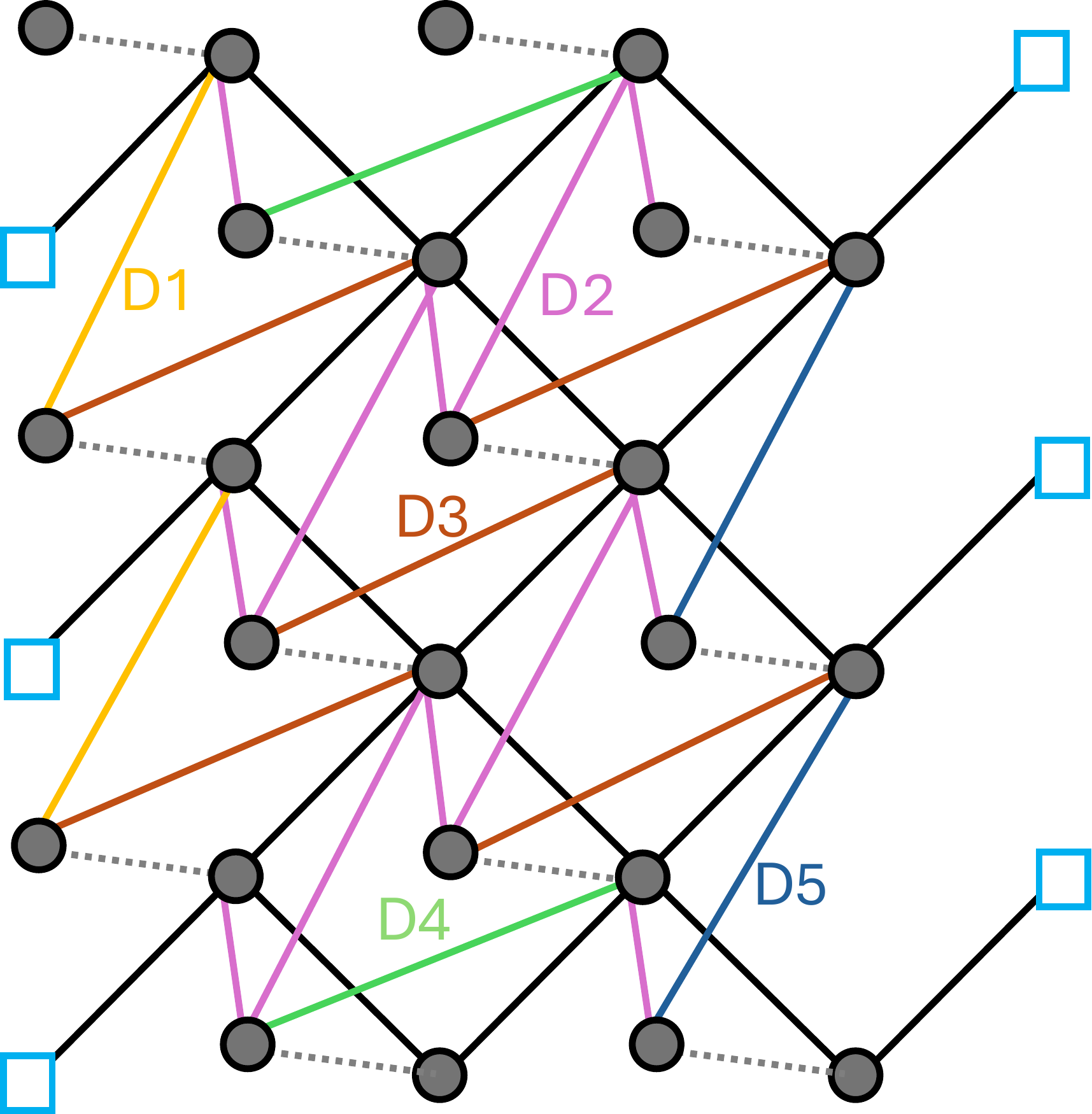}}
\hfill
\subfloat[\label{fig:Diagonal_Graph_X} ]{\includegraphics[width=.3\textwidth]{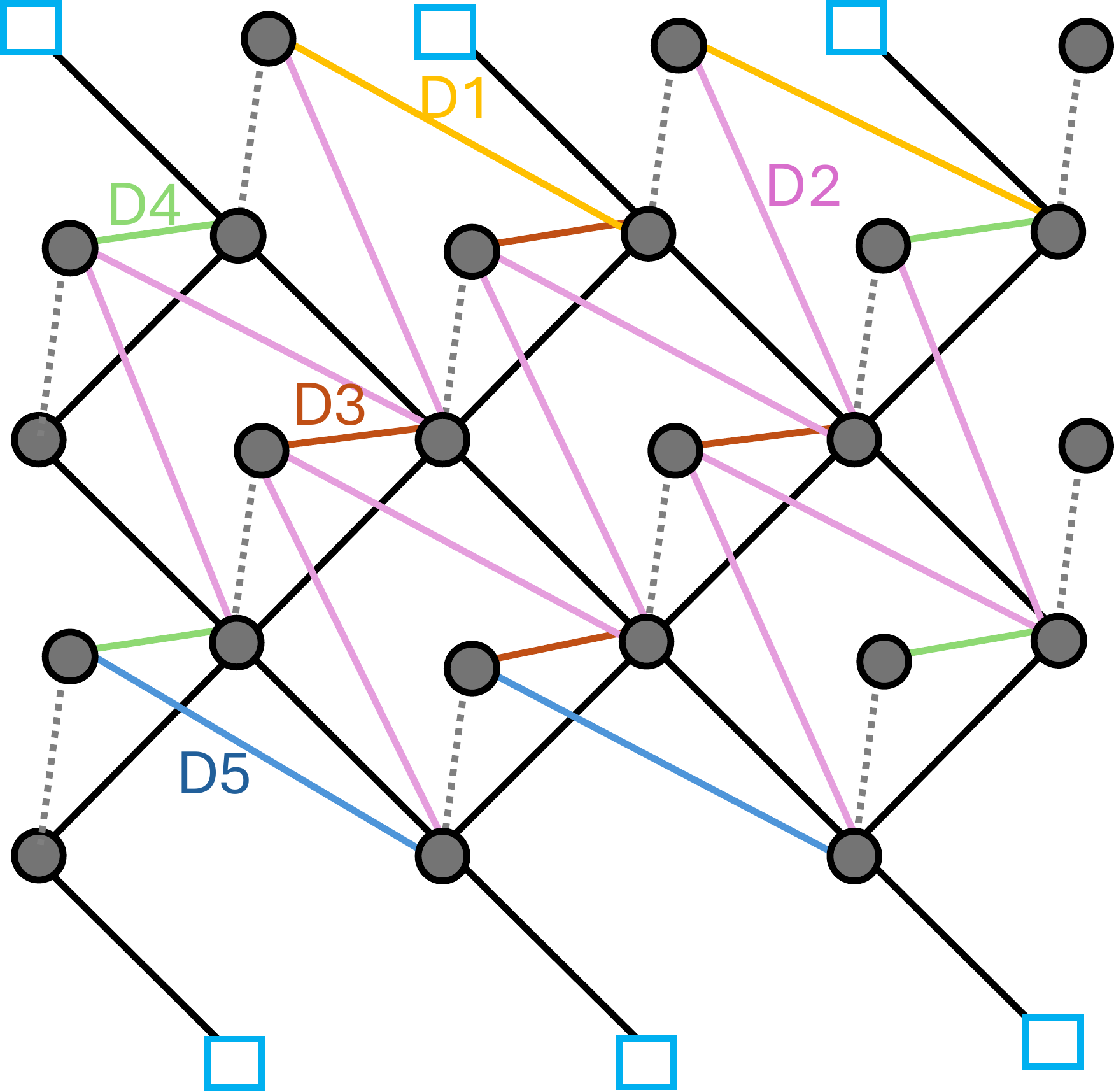}}
\caption{ (a) Two-dimensional graph for $Z$-stabilizers for the circuit in FIG. \ref{fig:CircuitMainCNOT}. We add labels for each edge type (i.e., both boundary and bulk edges). (b) Same as (a) but for $X$-stabilizers. (c) $Z$-stabilizer graph showing vertical edge labels used for measurement errors. (d) Same as (c) but for $X$-stabilizers. (e) Labels of diagonal edges for $Z$-type stabilizers. (f) Same as (e) but for $X$-type stabilizers.}
\label{fig:Graphs}
\end{figure*}
 
 In this appendix we provide the details for computing the edge weights used in the matching graphs for the surface code. The circuit used for a $d=5$ surface code is shown in \cref{fig:CircuitMainCNOT} and contains all the different types of edges that are obtained at arbitrary distances.
 
 \subsection{Notation and methodology}
 \label{subsec:NotationMethodology}
 
The circuit-level noise model is parameterized by 25 probabilities:
\begin{itemize}
    \item \textbf{State preparation errors (2):} $P_{SX}$ for $|+\rangle$ preparation, $P_{SZ}$ for $|0\rangle$ preparation.
    \item \textbf{Measurement errors (2):} $P_{mX}$ for $X$-basis measurement, $P_{mZ}$ for $Z$-basis measurement.
    \item \textbf{Idle errors during CNOT layers (3):} $P_{\text{idle,CNOT}}^{(X)}$, $P_{\text{idle,CNOT}}^{(Y)}$, $P_{\text{idle,CNOT}}^{(Z)}$ for single-qubit Pauli errors during two-qubit gate operations.
    \item \textbf{Idle errors during SPAM window (3):} $P_{\text{idle,SPAM}}^{(X)}$, $P_{\text{idle,SPAM}}^{(Y)}$, $P_{\text{idle,SPAM}}^{(Z)}$ for single-qubit Pauli errors on data qubits during ancilla preparation/reset.
    \item \textbf{CNOT errors (15):} $P_{\text{CX}}^{(P_iP_j)}$ for each two-qubit Pauli $P_i \otimes P_j$ (with $P_i$ at control, $P_j$ at target), where $P_i, P_j \in \{I, X, Y, Z\}$ excluding the identity $II$.
 \end{itemize}
 Given a probability $P$, the edge weight used by PyMatching is obtained by taking $w = -\log{P}$.
 
 When computing edge probabilities for the matching graph, errors from multiple fault locations can contribute to the same edge. When multiple independent error mechanisms flip the same pair of detectors, their probabilities are combined using the XOR operation:
 \begin{align}
     P_1 \oplus P_2 = P_1 + P_2 - 2 P_1 P_2.
     \label{eq:XORcombine}
 \end{align}
 For multiple components $\{c_1, c_2, \ldots, c_n\}$, the XOR is applied sequentially:
 \begin{align}
     \bigoplus_{i=1}^n c_i = c_1 \oplus c_2 \oplus \cdots \oplus c_n.
     \label{eq:XORmulti}
 \end{align}
 
 Each component $c_i$ may itself be a sum of Pauli probabilities that create the same detector pattern from the same fault location:
 \begin{align}
      c_i = \sum_{P \in \mathcal{P}_i} P_{\text{CX}}^{(P)} \quad \text{or} \quad c_i = P_I^{(P)},
     \label{eq:ComponentSum}
 \end{align}
 where $\mathcal{P}_i$ is the set of Paulis that create the same detector pattern from a given CNOT location.
 
 \subsection{Edge classification}
 \label{subsec:EdgeClassification}
 
 The matching graph contains four categories of edges:
 \begin{itemize}
     \item \textbf{Spacelike edges}: Connect different stabilizers within the same measurement round. Arise from data qubit errors.
     \item \textbf{Timelike edges}: Connect the same stabilizer across adjacent measurement rounds. Arise from ancilla/measurement errors.
     \item \textbf{Diagonal edges}: Connect different stabilizers across adjacent measurement rounds. Arise from combined data and measurement errors.
     \item \textbf{Boundary edges}: Connect a single stabilizer to the logical boundary. Arise from measurement errors near the code boundary.
 \end{itemize}
 
For a $d=5$ surface code, there are 12 $X$-stabilizers and 12 $Z$-stabilizers. Both matching graphs contain \textbf{18 distinct edge types} each, which are \textbf{distance-independent}---the same formulas apply for any $d \ge 5$. The edge types are:
\begin{itemize}
    \item \textbf{Spacelike}: 3 types (S1, S2, S3)
    \item \textbf{Timelike}: 4 types (T1, T2, T3, T4)
    \item \textbf{Diagonal}: 5 types (D1, D2, D3, D4, D5)
    \item \textbf{Boundary}: 6 types (B1, B2, B3, B4, B5, B6)
\end{itemize}
While the $X$-graph and $Z$-graph have the same number of edge types, the distribution of edges among types differs due to the different lattice orientations. Note that under symmetric (uniform) noise, some edge types have identical probabilities (e.g., D1/D5, boundary pairs B1/B5, B2/B6, B3/B4), but differ under asymmetric noise and must be treated separately.
 
 \subsection{X-stabilizer graph edge formulas}
 \label{subsec:XstabFormulas}
 
 We provide the verified edge probability formulas for the $X$-stabilizer matching graph. These formulas detect $Z$ and $Y$ errors on data qubits.
 
\subsubsection{Spacelike edges}

\textbf{Type $P_{S1}^{(X)}$:}
\begin{align}
    P_{S1}^{(X)} &= \bigoplus \Big[ P_{\text{CX}}^{(YY)} + P_{\text{CX}}^{(ZZ)}, \; P_{\text{CX}}^{(IZ)} + P_{\text{CX}}^{(XZ)}, \nonumber \\
    &\quad P_I^{(Z)}, \; P_I^{(Z)}, \; P_{\text{CX}}^{(YZ)} + P_{\text{CX}}^{(ZY)}, \nonumber \\
    &\quad P_{\text{CX}}^{(IY)} + P_{\text{CX}}^{(XY)}, \; P_I^{(Y)}, \; P_I^{(Y)} \Big].
    \label{eq:SpaceLikePS1X}
\end{align}

\textbf{Type $P_{S2}^{(X)}$:}
\begin{align}
    P_{S2}^{(X)} &= \bigoplus \Big[ P_{\text{CX}}^{(IY)}, \; P_{\text{CX}}^{(XY)}, \; P_{\text{CX}}^{(YZ)} + P_{\text{CX}}^{(ZZ)}, \nonumber \\
    &\quad P_{\text{CX}}^{(IZ)}, \; P_{\text{CX}}^{(IZ)}, \; P_{\text{CX}}^{(ZI)} + P_{\text{CX}}^{(ZZ)}, \nonumber \\
    &\quad P_I^{(Z)}, \; P_I^{(Z)}, \; P_I^{(Z)}, \; P_{\text{CX}}^{(IY)}, \nonumber \\
    &\quad P_{\text{CX}}^{(YX)} + P_{\text{CX}}^{(YY)}, \; P_{\text{CX}}^{(XY)}, \; P_{\text{CX}}^{(YY)} + P_{\text{CX}}^{(ZY)}, \nonumber \\
    &\quad P_{\text{CX}}^{(YI)} + P_{\text{CX}}^{(YZ)}, \; P_I^{(Y)}, \; P_I^{(Y)}, \; P_I^{(Y)}, \nonumber \\
    &\quad P_{\text{CX}}^{(XZ)}, \; P_{\text{CX}}^{(ZX)} + P_{\text{CX}}^{(ZY)}, \; P_{\text{CX}}^{(XZ)} \Big].
\end{align}

\textbf{Type $P_{S3}^{(X)}$:}
\begin{align}
    P_{S3}^{(X)} &= \bigoplus \Big[ P_{\text{CX}}^{(IY)}, \; P_{\text{CX}}^{(YX)} + P_{\text{CX}}^{(YY)}, \; P_{\text{CX}}^{(IY)}, \; P_{\text{CX}}^{(ZX)} + P_{\text{CX}}^{(ZY)}, \nonumber \\
    &\quad P_{\text{CX}}^{(XY)}, \; P_{\text{CX}}^{(XY)}, \; P_{\text{CX}}^{(IZ)} + P_{\text{CX}}^{(ZI)}, \; P_{\text{CX}}^{(ZZ)}, \; P_{\text{CX}}^{(ZZ)}, \nonumber \\
    &\quad P_{\text{CX}}^{(IZ)}, \; P_{\text{CX}}^{(IZ)}, \; P_{\text{CX}}^{(ZI)} + P_{\text{CX}}^{(ZZ)}, \; P_I^{(Z)}, \; P_I^{(Z)}, \nonumber \\
    &\quad P_{\text{CX}}^{(YY)}, \; P_{\text{CX}}^{(YZ)}, \; P_{\text{CX}}^{(YY)}, \; P_{\text{CX}}^{(XY)} + P_{\text{CX}}^{(YX)}, \nonumber \\
    &\quad P_{\text{CX}}^{(YI)} + P_{\text{CX}}^{(YZ)}, \; P_I^{(Y)}, \; P_I^{(Y)}, \; P_{\text{CX}}^{(XZ)}, \nonumber \\
    &\quad P_{\text{CX}}^{(IY)} + P_{\text{CX}}^{(ZX)}, \; P_{\text{CX}}^{(XZ)}, \; P_{\text{CX}}^{(XZ)} + P_{\text{CX}}^{(YI)}, \nonumber \\
    &\quad P_{\text{CX}}^{(ZY)}, \; P_{\text{CX}}^{(YZ)}, \; P_{\text{CX}}^{(ZY)} \Big].
\end{align}
 
\subsubsection{Timelike edges}

\textbf{Type $P_{T1}^{(X)}$:}
\begin{align}
    P_{T1}^{(X)} &= \bigoplus \Big[ P_{\text{CX}}^{(ZI)}, \; P_{\text{CX}}^{(YI)} + P_{\text{CX}}^{(ZI)}, \; P_{SX}, \; P_{SX}, \nonumber \\
    &\quad P_{\text{CX}}^{(YX)}, \; P_{\text{CX}}^{(YI)}, \; P_{\text{CX}}^{(ZX)}, \; P_{\text{CX}}^{(YX)} + P_{\text{CX}}^{(ZX)} \Big].
\end{align}

\textbf{Type $P_{T2}^{(X)}$:}
\begin{align}
    P_{T2}^{(X)} &= \bigoplus \Big[ P_{\text{CX}}^{(YX)} + P_{\text{CX}}^{(ZI)}, \; P_{\text{CX}}^{(ZI)}, \; P_{\text{CX}}^{(ZI)}, \nonumber \\
    &\quad P_{\text{CX}}^{(YI)} + P_{\text{CX}}^{(ZI)}, \; P_{SX}, \; P_{SX}, \nonumber \\
    &\quad P_{\text{CX}}^{(YI)}, \; P_{\text{CX}}^{(YX)}, \; P_{\text{CX}}^{(YI)} + P_{\text{CX}}^{(ZX)}, \; P_{\text{CX}}^{(YX)}, \nonumber \\
    &\quad P_{\text{CX}}^{(ZX)}, \; P_{\text{CX}}^{(ZX)}, \; P_{\text{CX}}^{(YI)}, \; P_{\text{CX}}^{(YX)} + P_{\text{CX}}^{(ZX)} \Big].
\end{align}

\textbf{Type $P_{T3}^{(X)}$:}
\begin{align}
    P_{T3}^{(X)} &= \bigoplus \Big[ P_{\text{CX}}^{(YI)} + P_{\text{CX}}^{(ZI)}, \; P_I^{(Y)} + P_I^{(Z)}, \nonumber \\
    &\quad P_I^{(Y)} + P_I^{(Z)}, \; P_{SX}, \; P_{SX}, \; P_{\text{CX}}^{(YX)} + P_{\text{CX}}^{(ZX)} \Big].
\end{align}

\textbf{Type $P_{T4}^{(X)}$:}
\begin{align}
    P_{T4}^{(X)} &= \bigoplus \Big[ P_{\text{CX}}^{(YX)} + P_{\text{CX}}^{(ZI)}, \; P_{\text{CX}}^{(YI)} + P_{\text{CX}}^{(ZI)}, \nonumber \\
    &\quad P_I^{(Y)} + P_I^{(Z)}, \; P_I^{(Y)} + P_I^{(Z)}, \; P_{SX}, \; P_{SX}, \nonumber \\
    &\quad P_{\text{CX}}^{(YI)} + P_{\text{CX}}^{(ZX)}, \; P_{\text{CX}}^{(YX)} + P_{\text{CX}}^{(ZX)} \Big].
\end{align}
 
\subsubsection{Diagonal edges}

\textbf{Type $P_{D1}^{(X)}$:}
\begin{align}
    P_{D1}^{(X)} &= \bigoplus \Big[ P_{\text{CX}}^{(ZZ)}, \; P_{\text{CX}}^{(YY)}, \; P_{\text{CX}}^{(ZY)}, \; P_{\text{CX}}^{(YZ)} \Big].
\end{align}

\textbf{Type $P_{D2}^{(X)}$:}
\begin{align}
    P_{D2}^{(X)} &= \bigoplus \Big[ P_{\text{CX}}^{(IZ)}, \; P_{\text{CX}}^{(ZZ)}, \; P_{\text{CX}}^{(XY)}, \; P_{\text{CX}}^{(XZ)}, \nonumber \\
    &\quad P_{\text{CX}}^{(YY)}, \; P_{\text{CX}}^{(IY)}, \; P_{\text{CX}}^{(ZY)}, \; P_{\text{CX}}^{(YZ)} \Big].
\end{align}

\textbf{Type $P_{D3}^{(X)}$:}
\begin{align}
    P_{D3}^{(X)} &= \bigoplus \Big[ P_{\text{CX}}^{(IZ)} + P_{\text{CX}}^{(XY)}, \; P_{\text{CX}}^{(ZI)}, \; P_{\text{CX}}^{(ZI)}, \nonumber \\
    &\quad P_{\text{CX}}^{(YZ)} + P_{\text{CX}}^{(ZZ)}, \; P_{\text{CX}}^{(YI)}, \; P_{\text{CX}}^{(YX)}, \nonumber \\
    &\quad P_{\text{CX}}^{(IY)} + P_{\text{CX}}^{(XZ)}, \; P_{\text{CX}}^{(YX)}, \; P_{\text{CX}}^{(ZX)}, \nonumber \\
    &\quad P_{\text{CX}}^{(ZX)}, \; P_{\text{CX}}^{(YI)}, \; P_{\text{CX}}^{(YY)} + P_{\text{CX}}^{(ZY)} \Big].
\end{align}

\textbf{Type $P_{D4}^{(X)}$:}
\begin{align}
    P_{D4}^{(X)} &= \bigoplus \Big[ P_{\text{CX}}^{(IZ)} + P_{\text{CX}}^{(XY)}, \; P_{\text{CX}}^{(ZI)}, \; P_{\text{CX}}^{(YZ)} + P_{\text{CX}}^{(ZZ)}, \nonumber \\
    &\quad P_I^{(Z)}, \; P_{\text{CX}}^{(IY)} + P_{\text{CX}}^{(XZ)}, \; P_{\text{CX}}^{(YX)}, \nonumber \\
    &\quad P_{\text{CX}}^{(YI)}, \; P_{\text{CX}}^{(YY)} + P_{\text{CX}}^{(ZY)}, \; P_I^{(Y)}, \; P_{\text{CX}}^{(ZX)} \Big].
\end{align}

\textbf{Type $P_{D5}^{(X)}$:}
\begin{align}
    P_{D5}^{(X)} &= \bigoplus \Big[ P_{\text{CX}}^{(IZ)}, \; P_{\text{CX}}^{(XY)}, \; P_{\text{CX}}^{(XZ)}, \; P_{\text{CX}}^{(IY)} \Big].
\end{align}
 
\subsubsection{Boundary edges}

\textbf{Type $P_{B1}^{(X)}$:}
\begin{align}
    P_{B1}^{(X)} &= \bigoplus \Big[ P_{\text{CX}}^{(IY)}, \; P_{\text{CX}}^{(ZY)}, \; P_{\text{CX}}^{(XY)}, \; P_{\text{CX}}^{(YY)}, \nonumber \\
    &\quad P_{\text{CX}}^{(IY)} + P_{\text{CX}}^{(XY)}, \; P_{\text{CX}}^{(YX)} + P_{\text{CX}}^{(YY)}, \; P_I^{(Y)}, \nonumber \\
    &\quad P_{\text{CX}}^{(YX)} + P_{\text{CX}}^{(ZX)}, \; P_{\text{CX}}^{(IZ)} + P_{\text{CX}}^{(XZ)} + P_{\text{CX}}^{(YZ)} + P_{\text{CX}}^{(ZZ)}, \nonumber \\
    &\quad P_{\text{CX}}^{(IZ)}, \; P_{\text{CX}}^{(ZI)} + P_{\text{CX}}^{(ZZ)}, \; P_I^{(Z)}, \; P_I^{(Z)}, \; P_I^{(Z)}, \; P_I^{(Z)}, \nonumber \\
    &\quad P_{\text{CX}}^{(YI)} + P_{\text{CX}}^{(ZI)}, \; P_{\text{CX}}^{(ZZ)}, \; P_{\text{CX}}^{(XZ)}, \; P_{\text{CX}}^{(YZ)}, \nonumber \\
    &\quad P_{\text{CX}}^{(ZX)} + P_{\text{CX}}^{(ZY)}, \; P_{\text{CX}}^{(YY)} + P_{\text{CX}}^{(ZY)}, \; P_{\text{CX}}^{(YI)} + P_{\text{CX}}^{(YZ)}, \nonumber \\
    &\quad P_I^{(Y)}, \; P_I^{(Y)}, \; P_I^{(Y)}, \; P_{\text{CX}}^{(YX)} + P_{\text{CX}}^{(ZX)}, \; P_{\text{CX}}^{(YI)} + P_{\text{CX}}^{(ZI)} \Big].
\end{align}

\textbf{Type $P_{B2}^{(X)}$:} This formula has 52 XOR components. A representative subset:
\begin{align}
    P_{B2}^{(X)} &= \bigoplus \Big[ P_{\text{CX}}^{(IY)} + P_{\text{CX}}^{(XZ)}, \; P_{\text{CX}}^{(YX)}, \; P_{\text{CX}}^{(YI)} + P_{\text{CX}}^{(ZX)}, \nonumber \\
    &\quad P_{\text{CX}}^{(XY)}, \; P_{\text{CX}}^{(IY)} + P_{\text{CX}}^{(ZX)}, \; P_{\text{CX}}^{(YI)}, \nonumber \\
    &\quad P_{\text{CX}}^{(XY)} + P_{\text{CX}}^{(YX)}, \; P_{\text{CX}}^{(XZ)}, \; P_{\text{CX}}^{(IY)}, \nonumber \\
    &\quad P_{\text{CX}}^{(YX)} + P_{\text{CX}}^{(YY)}, \; P_{\text{CX}}^{(ZX)}, \; P_{\text{CX}}^{(XY)}, \; P_{\text{CX}}^{(YX)}, \nonumber \\
    &\quad P_{\text{CX}}^{(ZX)}, \; P_{\text{CX}}^{(IY)}, \; P_{\text{CX}}^{(IZ)} + P_{\text{CX}}^{(XY)} + P_{\text{CX}}^{(YY)} + P_{\text{CX}}^{(ZZ)}, \nonumber \\
    &\quad P_{\text{CX}}^{(ZI)}, \; P_{\text{CX}}^{(IZ)} + P_{\text{CX}}^{(ZI)}, \; P_{\text{CX}}^{(IZ)} + P_{\text{CX}}^{(ZZ)}, \; \ldots \Big].
\end{align}

\textbf{Type $P_{B3}^{(X)}$:} This formula has 62 XOR components. A representative subset:
\begin{align}
    P_{B3}^{(X)} &= \bigoplus \Big[ P_{\text{CX}}^{(XY)}, \; P_{\text{CX}}^{(YX)}, \; P_{\text{CX}}^{(XZ)}, \; P_{\text{CX}}^{(YI)}, \nonumber \\
    &\quad P_{\text{CX}}^{(IY)}, \; P_{\text{CX}}^{(YX)} + P_{\text{CX}}^{(YY)}, \; P_I^{(Y)}, \; P_{\text{CX}}^{(ZX)}, \nonumber \\
    &\quad P_{\text{CX}}^{(IY)}, \; P_{\text{CX}}^{(ZX)} + P_{\text{CX}}^{(ZY)}, \; P_{\text{CX}}^{(ZY)}, \; P_{\text{CX}}^{(XY)}, \nonumber \\
    &\quad P_{\text{CX}}^{(YY)}, \; P_{\text{CX}}^{(IZ)} + P_{\text{CX}}^{(ZI)}, \; P_{\text{CX}}^{(ZI)} + P_{\text{CX}}^{(ZZ)}, \; \ldots \Big].
\end{align}

\textbf{Type $P_{B4}^{(X)}$:} This formula has 68 XOR components arising from 34 distinct detector patterns. A representative subset:
\begin{align}
    P_{B4}^{(X)} &= \bigoplus \Big[ P_{\text{CX}}^{(IY)} + P_{\text{CX}}^{(XZ)}, \; P_{\text{CX}}^{(YX)}, \; P_{\text{CX}}^{(YI)} + P_{\text{CX}}^{(ZX)}, \nonumber \\
    &\quad P_{\text{CX}}^{(XY)}, \; P_{\text{CX}}^{(IY)}, \; P_{\text{CX}}^{(YX)} + P_{\text{CX}}^{(YY)}, \nonumber \\
    &\quad P_{\text{CX}}^{(IZ)} + P_{\text{CX}}^{(XY)} + P_{\text{CX}}^{(YY)} + P_{\text{CX}}^{(ZZ)}, \; P_{\text{CX}}^{(IZ)} + P_{\text{CX}}^{(ZI)}, \; \ldots \Big].
\end{align}

\textbf{Type $P_{B5}^{(X)}$:}
\begin{align}
    P_{B5}^{(X)} &= \bigoplus \Big[ P_{\text{CX}}^{(YZ)} + P_{\text{CX}}^{(ZY)}, \; P_I^{(Y)}, \; P_{\text{CX}}^{(YI)} + P_{\text{CX}}^{(ZX)}, \nonumber \\
    &\quad P_{\text{CX}}^{(IZ)} + P_{\text{CX}}^{(ZI)}, \; P_{\text{CX}}^{(ZZ)}, \; P_{\text{CX}}^{(IZ)} + P_{\text{CX}}^{(XY)} + P_{\text{CX}}^{(YY)} + P_{\text{CX}}^{(ZZ)}, \nonumber \\
    &\quad P_I^{(Z)}, \; P_I^{(Z)}, \; P_I^{(Z)}, \; P_I^{(Z)}, \; P_{\text{CX}}^{(XZ)} + P_{\text{CX}}^{(YI)}, \nonumber \\
    &\quad P_{\text{CX}}^{(IY)} + P_{\text{CX}}^{(ZX)}, \; P_{\text{CX}}^{(YX)} + P_{\text{CX}}^{(ZI)}, \; \ldots \Big].
\end{align}

\textbf{Type $P_{B6}^{(X)}$:} This formula has 57 XOR components. A representative subset:
\begin{align}
    P_{B6}^{(X)} &= \bigoplus \Big[ P_{\text{CX}}^{(YI)}, \; P_{\text{CX}}^{(YY)} + P_{\text{CX}}^{(ZY)}, \; P_{\text{CX}}^{(YZ)}, \nonumber \\
    &\quad P_{\text{CX}}^{(YX)} + P_{\text{CX}}^{(ZX)}, \; P_{\text{CX}}^{(ZY)}, \; P_{\text{CX}}^{(IZ)} + P_{\text{CX}}^{(ZI)}, \nonumber \\
    &\quad P_{\text{CX}}^{(ZI)} + P_{\text{CX}}^{(ZZ)}, \; P_{\text{CX}}^{(IZ)} + P_{\text{CX}}^{(ZZ)}, \; P_{\text{CX}}^{(ZI)}, \nonumber \\
    &\quad P_{\text{CX}}^{(IZ)} + P_{\text{CX}}^{(XZ)} + P_{\text{CX}}^{(YZ)} + P_{\text{CX}}^{(ZZ)}, \; P_I^{(Z)}, \; \ldots \Big].
\end{align}
 
\subsection{Z-stabilizer graph edge formulas}
\label{subsec:ZstabFormulas}

The $Z$-stabilizer matching graph detects $X$ and $Y$ errors on data qubits. Similar to the $X$-graph, it has 18 edge types: 3 spacelike (S1--S3), 4 timelike (T1--T4), 5 diagonal (D1--D5), and 6 boundary (B1--B6). The explicit formulas are obtained from the $X$-stabilizer formulas above by replacing all $Z$-type Paulis with $X$-type Paulis, exploiting the X/Z symmetry of the surface code circuit.
 
\subsection{Summary and verification}
\label{subsec:FormulaAccuracy}
 
The formulas were derived by systematically tracing error propagation through the syndrome extraction circuit for each possible Pauli error at each fault location. The methodology is:
\begin{enumerate}
    \item For each fault location (CNOT, idle, state preparation), activate a single Pauli error.
    \item Generate the detector error model (DEM) using Stim.
    \item Identify which DEM patterns contain the target edge's detector pair.
    \item Group contributions by pattern and sum Paulis from the same location.
    \item XOR-combine all pattern contributions to get the final formula.
\end{enumerate}

The formulas are \textbf{distance-independent}: the same formulas apply identically for $d = 5, 7, 9, 11, 13$ and beyond. This is because edge probabilities depend only on local stabilizer geometry, not global code size. Only the \textit{count} of each edge type changes with distance. For example, at $d=5$ the X-stabilizer graph has 8 type-S1 edges, while at $d=7$ it has 18, and at $d=13$ it has 72. The formulas enable gradient-based optimization through their differentiability, allowing neural networks to learn effective noise parameters by directly optimizing edge probabilities used by the MWPM decoder.

\bibliography{q}
 
\end{document}